\documentclass[article]{svmono}

\usepackage[greek,english]{babel}
\usepackage{marvosym}
\usepackage{graphicx,caption}        

\usepackage{booktabs}
\usepackage{threeparttable}

\def\begmarg{\par \begingroup  \leftskip2.5em \rightskip2em \large}
\def\endmarg{\par \endgroup }

\usepackage{cleveref}

\def\H{{1\over2}}
\def\3H{{3\over2}}
\def\simlt{\lower.5ex\hbox{$\; \buildrel < \over \sim \;$}}
\def\simgt{\lower.5ex\hbox{$\; \buildrel > \over \sim \;$}}

\def\magspt{$\buildrel{\rm m}\over .$}
\def\degs{$^{\circ}$}

\def\secspt{$\buildrel{\prime\prime}\over .$}

\def\degspt{$\buildrel{\circ}\over .$}
\def\mags{$^{\rm m}$}

\def\simlt{\lower.5ex\hbox{$\; \buildrel < \over \sim \;$}}
\def\simgt{\lower.5ex\hbox{$\; \buildrel > \over \sim \;$}}

\begin{document}

\pagestyle{plain}
\ \ 
\vspace{1cm}

\begin{center}
{\huge {\bf }}
\bigskip
  
{\huge {\bf Pieter Johannes van Rhijn,}}
\bigskip

{\huge {\bf Kapteyn's Astronomical Laboratory}}
\bigskip

{\huge {\bf and the Plan of Selected Areas}}
\vspace{1cm}

\noindent
{\LARGE Pieter C. van der Kruit},\\
{\Large Kapteyn Astronomical Institute, University of Groningen,}\\
{\Large P.O. Box 800, 9700AV Groningen, the Netherlands}.\\
{\Large email: vdkruit@astro.rug.nl}
\vspace{2cm}

\end{center}

{\large

\noindent
Version \today.
\vspace{2cm}

\noindent
This manuscript has been accepted for publcation
by the {\it Journal of Astronomical History and
Heritage}. This reprint version has been produced in \LaTeX.
The paper is scheduled for the September 2022 issue and some changes may occur
in the production process.

\newpage

\noindent
{\Large {\bf Abstract}}
\bigskip
  
In this contribution I discuss the Kapteyn Astronomical Laboratory during the
period of Pieter Johannes van Rhijn’s directorate, which lasted from 1921 to 1957.
It had developed under the founder Jacobus Cornelius Kapteyn into one of the leading
astronomical research institutes in the world. When van Rhijn took over at the
retirement of Kapteyn, it was in the process of coordinating Kapteyn’s {\it Plan
of Selected Areas} with the aim of determining the structure of the Sidereal
System. Kapteyn had just with the help of van Rhijn presented a first model of the
distribution of stars in space from existing observations and on this basis
constructed a `first attempt' model of the Stellar System
including dynamics, i.e. derived the
gravitational field from distribution of the stars and showed how the random and
systematic motions of the stars, the kinematics, could provide
equilibrium and stability of the
system. Under van Rhijn the work on the Selected Areas progressed well, but the
Groningen Laboratorium went into a decline, loosing much of its status and
prestige.

I conclude the following. There was very little
choice for Kapteyn’s successor and van Rhijn was appointed effectively by default.
Kapteyn himself must have seen that the future for astronomy in the Netherlands
was Leiden Observatory where already during his lifetime two of his prot\'eg\'ees,
Willem  de Sitter and Ejnar Hertzsprung, were in charge and to
which a third one was added not much later in the person of Jan  Oort.  Kapteyn
had his {\it Plan of Selected Areas} in addition to the scientific case
set up as a way of ensuring the future of the
Groningen laboratory at least for the time of this effort, arranging it to be
supplied with plate material to complete the provision of the data for the final
attach on the Sidereal Problem.

Van Rhijn’s research was solid and professional
work, but in his papers he invariably stopped before discussing
how his findings did fit into the larger scheme of things.
He maybe was unimaginative but it did lack the link to
the larger view towards the emerging picture of the structure of the Galaxy.
Comparison of his work in the 1930s on the distribution of stars as a function of
spectral types and Jan Hendrik Oort’s determination of his 1938 crosscut through
the Galaxy illustrates this. While this work by Oort constitutes the second attempt
following up on Kapteyn’s first, and may be seen as the actual completion of the
research for which the {\it Plan of Selected Areas} was designed, van Rhijn
continued the adding of more and more data in the context of the Plan without
further redefining the aims in the light of developments. The {\it Spectral
Durchmusterung} carried out in collaboration with the Bergedorf-Hamburg Sternwarte
and Harvard College Observatory was an enormous drain on the Groningen resources.
Van Rhijn’s successful attempt to obtain his own telescope resulted in very
relevant and correct determinations of the wavelength dependence of interstellar
extinction from photographic spectra, but these results became only available
in the 1950s when
the field had turned to photoelectric methods.

Kapteyn’s concept of an astronomical laboratory
lost its viability in van Rhijn’s time when observatories were no longer interested
in providing observational material for others to do research. Kapteyn’s model
on the other hand
of a laboratory-like institution, where research is based on material obtained
elsewhere on Earth or in space, is now the normal operating mode in astronomical
research. The only difference is access to observing facilities primarily,
which is now structural and guaranteed (subject to proposal acceptance of course),
through national and international organizations rather than access on the
basis of one’s fame or stature.

The cause of the loss of the Kapteyn Laboratorium of its prominent place in the
international context was largely van Rhijn’s
unwavering dedication to complete Kapteyn’s work. The {\it Plan of Selected Areas}
had been important for the progress of astronomy, but effectively reached the goals
Kapteyn had had in mind for it already in the 1930s with Oort ‘s 1932 work on the
vertical force field and local ‘Oort limit’ and his 1938 crosscut through the
Galaxy.

Van Rhijn was unfortunate to be hampered throughout  almost his complete
directorate by factors, that severely limited his attempts to obtain more
funding in spite of local support by his university. These were of course in the
first place the Great Depression of the 1930s and the Second World War and its
aftermath, while during most of the 1940s he suffered from tuberculosis.
But also the remote location of Groningen compared to
Leiden, where a major infrastructure led by three important prot\'eg\'ees of
Kapteyn was in place, and the governmental bias towards support for Leiden over
Groningen was an important factor.

Finally I examine the developments in the 1950s and the circumstances that
made  Adriaan Blaauw accept his appointment as van Rhijn’s successor in 1957 and
initiate the beginnings of the revival under his leadership.
\bigskip

\newpage

\ \ \ \\
\vspace{2cm}

\begin{tabular}{rlr}
  & {\bf {\LARGE CONTENTS}} &  \\
  && \\
 & Abstract & 2 \\
1 & Introduction & 5 \\
2 & Pieter Johannes van Rhijn and his early research & 10 \\
3 & Kapteyn’s Plan of Selected Areas & 16 \\
4 & The succession of Kapteyn & 23 \\
5 & Van Rhijn’s astronomical work up to 1930 & 27 \\
6 & Bergedorf(-Groningen-Harvard) Spektral-Durchmusterung & 34 \\
7 & Research based on the Selected Areas and the Spektral-Durchmusterung & 38\\
8 & Interstellar extinction & 42 \\
9  & Oort’s second approximation & 47 \\
10 & Van Rhijn’s plans for his own observatory & 53 \\
11 &  Realization of van Rhijn’s telescope & 58 \\
12  & Adriaan Blaauw and Lukas Plaut & 64 \\
13  & Van Rhijn and the Laboratorium during the War & 69 \\
14  & Recovering from the War & 78 \\
15 &  Staff and workers at the Laboratorium & 81 \\
16 & The late forties and early fifties & 88 \\
17 &  The Vosbergen conference & 93 \\
18 & Groningen-Palomar Variable-Star Survey & 97 \\
19  & Van Rhijn’s final research and seventieth birthday & 104 \\
20 & Van Rhijn’s succession & 111 \\
21  & The Laboratorium after van Rhijn & 122 \\
22  & Discussion & 128  \\
& Acknowledgments & 137 \\
& References & 137 \\
\end{tabular}

\newpage

\section{Introduction}\label{sect:Intro}

This paper concerns the `Sterrenkundig Laboratorium
Kapteyn’\footnote{\normalsize Because it concerns names I will as much as possible
keep the formal names of institutions and observatories in the language
of their location, so Sterrenkundig Laboratorium Kapteyn or for short
Kapteyn Laboratorium, the Sterrewacht
Leiden, Bergedorfer Sternwarte, Helsingfors Observatorion, Mount Wilson
Observatory, etc. But, since it would be impractical, I will
refrain from applying this to universities, but I  will not translate
`Rijks-universitei Groningen' into the sometimes used `State University of
Groningen' (a correct translation would be `National University',
since the term
`State University' refers to universities funded by states
in the USA rather than by a nation) but use the official University of
Groningen.}
in Groningen, the Netherlands, during the directorship of
Pieter Johannes van Rhijn (1886--1960)\footnote{\normalsize I add in general
for persons when they first appear
in this text their full set of first names and the years of birth an death,
but for those still alive I refrain from quoting their year of birth.},
which lasted from 1921 to 1957.

At the start of the academic year in 1921 Pieter van Rhijn, see
Fig.~\ref{fig:1921}, was
appointed professor of astronomy at the University of Groningen and director
of its astronomical laboratory. This laboratory had been founded by Jacobus
Kapteyn, see Fig.~\ref{fig:JCK},
and on the occasion the curators of the University and with
consent of the Minister had renamed it 
the `Sterrenkundig Laboratorium Kapteyn’. When Kapteyn's
attempts -- starting not long after he was appointed
professor of astronomy in 1878 without any facility to speak of for research
and teaching in astronomy -- to obtain a real observatory  had failed,
he had reverted to this ‘observatory without a telescope’. Under Kapteyn it
had risen to become one of the most prominent
institutions worldwide, overshadowing the Sterrewacht Leiden, which was in a
deep decline.  Under van Rhijn the Groningen Laboratorium slipped in turn to a 
lower rank on the ladder of prominence, status and prestige. 
I will look into the circumstances, context and background to this.
\bigskip

I will start with context and briefly consider astronomy
in the Netherlands up to roughy the end
of this period, i.e. the 1950s.
A first reference would be David Baneke’s history of astronomy
in the Netherlands in the twentieth century \cite{DB2015}, but this still awaits
being published in English translation. In recent years a
few further studies on particular aspects have been published. But also
a number of biographies of Dutch astronomers of this period
have appeared (a few unfortunately also in Dutch only).

\begin{figure}[t]
  \sidecaption[t]
\captionsetup{width=.36\textwidth}
\includegraphics[width=0.64\textwidth]{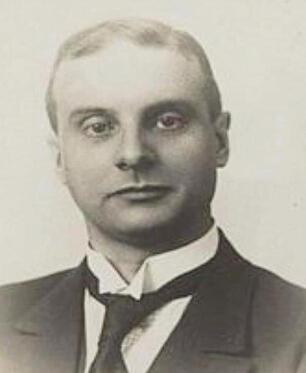}
\caption{\normalsize  Pieter Johannes van Rhijn in his late twenties around the time
  he obtained his PhD in 1915.   Kapteyn Astronomical Institute.}
\label{fig:1921}
\end{figure}

\begin{figure}[t]
\sidecaption[t]
\includegraphics[width=0.64\textwidth]{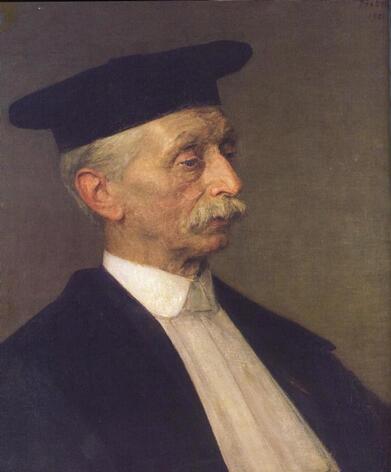}
\caption{\normalsize  Jacobus Cornelius Kapteyn. Painting by Jan Veth, now
located in the Senate Room of the Academy Building of the University of
Groningen. It was originally intended as a present to his wife at the 40-th
anniversary of his professorship, but she disliked it and a new one was
painted. Veth later over-painted it with gown, barret and jabot for its present
purpose. For the full story see \cite{JCKbiog} or \cite{JCKEng}.  
University of Groningen.}
\label{fig:JCK}
\end{figure}

Although astronomy in the Netherlands is generally considered to have started
with Christiaan Huygens (1629--1695)
and his ‘discovery’ of Saturn’s rings and satellite Titan (biography in
\cite{CDAn}), modern astronomy in the
Netherlands goes back to Frederik Kaiser (biography in \cite{RGent}).
Kaiser was the founder of the Sterrewacht
Leiden, the building modeled after Pulkovo Observatorium and opened in 1861.
Kaiser had made quite a name for
himself when Halley’s comet was due to return in 1835. He revised calculations
made earlier by renowned scientists
and predicted the time of passage of the perihelion to an astounding accuracy
of 1.5 hours, where many celebrated
astronomers had been wrong by much larger time intervals, sometimes
as much as 9 days.
Under Kaiser the Sterrewacht became well-known for its extremely accurate
astrometry. However, under his successor Hendricus Gerardus van de
Sande Bakhuyzen (1838--1923), it lost much of its influence because he decided
not to embrace photography as the new tool while in fact it
turned out  superior to visual
observing. Under his younger brother Ernst Frederik (1848--1918), who was
appointed director in 1908, Leiden lost even more of its
status of excellence.
Instead Groningen took over, where Jacobus Kapteyn  became one of the most
prominent astronomers worldwide with his studies of the
distributions of stars in space and stellar kinematics (biography in
\cite{JCKbiog} and \cite{JCKEng}).

The turning point for Leiden came in 1918, when after the sudden death of
Ernst van de Sande Bakhuyzen, Willem de Sitter (1872--1934) became director,
having been appointed professor of astronomy in Leiden
already in 1908. De Sitter
(biography in \cite{Guich}) was Kapteyn’s first student working on the
structure of the system of Galilean satellites
around Jupiter. In the reorganization of the Sterrewacht of 1918, in which
Kapteyn had had an important voice as well, a new department on stellar
astronomy was created next to de Sitter’s theoretical one (in this Einstein’s
General Relativity became a second focus). The new department was headed by
Ejnar Hertzsprung (1873--1967), who already was a world renowned astronomer
(no recent biography, but see \cite{Herrmann}), but the third
department on astrometry remained without a head, when the Prime Minister
refused to appoint Antonie Pannekoek (1873--1960) to the post because of his
communist sympathies. Pannekoek was appointed subsequently at the
municipal University of Amsterdam, where he founded an
Astronomical Institute and made a
name by mapping the Milky Way and more impressively, founding astrophysics of
stellar atmospheres, exploring ways to apply the Saha ionization
formula to study the physical conditions in these outer envelopes (see
\cite{Pannekoek}). In 1925 Jan Hendrik Oort was appointed at the
vacant Leiden position.
Utrecht under Marcel Gilles Jozef Minnaert (1893--1970) rose to prominence
particularly for producing a detailed solar spectrum and founding a line of
research in solar physics and astrophysics
(biography in \cite{Minnaert}). Later additions to the staffs
until 1957 (the end of van Rhijn's term in office)
were Hendrik Christoffel (Henk) van de Hulst (1918--2000)
(biography in \cite{Delft}), Pieter Theodorus Oosterhoff (1904--1978)
and Adriaan Blaauw (1916--2010) in
Leiden, Herman Zanstra (1894--1972) in
Amsterdam and Cornelis (Kees) de Jager (1921--2021) in Utrecht. Comprehensive
biographies of these last four remain to be written.

In the twentieth century, Dutch astronomy had become a major player worldwide
in the field of astronomy. Consider the large number of Dutch
astronomers that were born before WWII and rose to leading positions (usually
directors) abroad, as listed by Jan Oort \cite{JHO1982}: Jan Schilt, Willem
van den Bos, Dirk Brouwer, Peter van de Kamp. Willem Luyten, Gerard Kuiper,
Bart Bok, Leendert Binnendijk,
Maarten Schmidt, Lo Woltjer, Gart Westerhout, Tom Gehrels and Sydney
van den Bergh. To illustrate this point further I note that almost all
astronomers in the Netherlands mentioned in the previous paragraph
were awarded what was probably
the most prestigious honor, the Gold Medal of the Royal
Astronomical Society: Kapteyn 1902, Hertzsprung 1929, de Sitter 1931, Oort,
1946, Minnaert, 1947, Pannekoek, 1951, Zanstra 1961, de Jager 1981.
Most of these also received the Bruce Medal of the Astronomical Society of the Pacific.
Van de Hulst, Blaauw and Oosterhoff are missing in the Gold Medal list, but
the first two of them received the Bruce Medal in 1978
and 1988. 

In this success story van Rhijn is conspicuously missing. 
\bigskip

In his presentation during a symposium in 1999 in Groningen on the {\it Legacy
of Jacobus C. Kapteyn} \cite{Legacy}, Woodruff T. Sullivan put it as follows
(\cite{WTS99}, p.230):
\begmarg
\noindent
[...]. at Groningen after the death of Kapteyn in 1922, his eponymous
Laboratory went into a period of decline coinciding with the long directorship
[...] of his prot\'eg\'e Pieter van Rhijn. Van Rhijn did some excellent work
in the 1920s [...], but
both his research and leadership for the rest of his career were unimaginative
and stuck in a rut (Blaauw, 1984).[...]

Kapteyn’s Plan of Selected Areas, the study and organization of which
occupied a large portion of his successor van Rhijn’s career [...] , this
work was done in a routine and unimaginative manner, and contributed to
Groningen’s period of decline. 
\endmarg

In the 1984-paper of Adriaan Blaauw 
that Sullivan refers to to support his case, Blaauw actually is much milder
(\cite{Blaauw1983}, p.56; my translation):
\begmarg
[...] the foundation of an astronomical laboratory. But the complete dependence
on the cooperation of foreign observatories that went along with that 
was a very heavy burden on the future of the institute.
Could van Rhijn reasonably be expected to continue the special
fame that the institute had acquired through Kapteyn, and which was a
necessary condition for this kind of cooperation to maintain? [...]

[Van Rhijn] was a thorough scientific researcher [...].
His approach to research, however, was highly schematic,
strongly focused on further evaluation of previously
designated properties or quantities. This benefited the thoroughness of
the intended result, but came at the expense of flexibility and
reduced the opportunity for unexpected perspectives; as a result, it did not
lead to surprising discoveries. 
\endmarg

Or in Blaauw's interview for the American Institute of Physics Oral History
Interviews Project \cite{Blaauw1978}:
\begmarg
[...] the main lines at Groningen by van Rhijn, who followed very much in the
course set by Kapteyn, [...]. And in that time, he
accomplished a very large amount of, you might call it routine works. But it
is a very thorough compilation of photometry and proper motions in the
Selected Areas and also work on such problems as luminosity functions.
\endmarg
\bigskip

It may not be  surprising that no comprehensive biography of van Rhijn
has ever been written, but it remains remarkable that  shortly after his
death no extensive obituary concerning van
Rhijn has appeared in any major international
astronomical journal. The only somewhat extensive one
was written by Adriaan Blaauw and Jean Jacques Raimond (1903--1961)
-- both having obtained
their PhD's with van Rhijn as `promotor' (supervisor) -- in Dutch in the
periodical of the association of amateur astronomers and meteorologists
\cite{BlaRaim} (the contribution by Adriaan Blaauw in the {\it Biographical
  Encyclopedia of Astronomers} \cite{Bea}, is modeled on this
article by him and Raimond). But they did not think it worthwhile,
necessary or appropriate to publish an English language article
to honor the life of an important astronomer.
Bartholomeus Jan (Bart) Bok (1906--1983),
who also obtained his PhD under van Rhijn,
published obituaries in {\it Sky \&\ Telescope} on Pannekoek
and van Rhijn \cite{BokST}. And
in 1969 Ernst Julius \"Opic (1893--1985), catching up on a suspension of the
{\it Irish Astronomical Journal}, listed van Rhijn among a set of over
twenty astronomers that had died in previous years \cite{IrAJ}. 
But that is all.

In fact, there is not a single article on van Rhijn and Groningen astronomy
under him in the English literature other than 
Blaauw’s article in the {\it Biographical Encyclopedia}, and
in Dutch there is only the one by Blaauw \cite{Blaauw1983}
that Sullivan referred to. 
At van Rhijn’s seventieth birthday in 1954 a kind of festschrift appeared
(the Dutch title translates into English as `in 
Kapteyn’s footsteps'; see below) with articles by his colleagues and students,
but the articles are mostly in Dutch and refer to 
then current research by the authors themselves and do not address van Rhijn’s
contributions in any detail. 
\bigskip

In this contribution I will critically examine the Kapteyn
Laboratorium during the period of van Rhijn’s directorate 1921-1957. 
Questions addressed include: Was van Rhijn the successor Kapteyn had wished
for or had he preferred someone else? What future might Kapteyn have had
in mind for his Laboratorium? Was van Rhijn's work really so unimaginative
and how important has it turned out to be?
Was Kapteyn’s concept of an astronomical laboratory,
an observatory without a telescope, a viable one in the long run?
What precisely caused the Kapteyn Laboratorium to loose its prominent place
in the international context? How important was the
{\it Plan of Selected Areas} for the progress of astronomy and did
it ever reach the goals Kapteyn had in mind for it? Did van Rhijn get the
support he deserved from his university and the government? How influential
was van Rhijn and what was his role in the prominence of Dutch astronomy
that followed during the twentieth century? What made Adriaan Blaauw accept his appointment as van Rhijn’s successor?

I rely heavily on Adriaan Blaauw's 1983 chapter \cite{Blaauw1983}, and
I will when appropriate quote verbatim from this in translation.
\bigskip

A few words on archival sources. The van Rhijn Archives are now part of the
Regionaal Historisch Centrum  `Groningen Archieven', founded in 2001 by the
Ministry of Education, Culture and Science and the Municipality of Groningen,
where the University of Groningen has deposited its Archives. The van Rhijn
Archives were transferred from the Kapteyn Astronomical Institute through the
`Archief Depot' of the University. Much of the correspondence is missing in
the Groninger Archieven under the corresponding heading, including that
between van Rhijn and the Curators of the university. Why these letters have
been culled from the collection is unknown, but probably to avoid duplication.
This, unfortunately, has made this very much less accessible.

Van Rhijs corresponded quite extensively with Jan Hendrik Oort (1900--1992)
in Leiden. These letters are missing in Groningen
also, but quite a collection is available in the Oort Archives,
which contains over 400 letters between van Rhijn and Oort. These are mostly
handwritten letters, of which usually no copies were kept,
so  most are from van Rhijn
to Oort. Also a few letters of Adriaan Blaauw and Lukas Plaut to Oort
proved to be relevant to this article. These letters are  almost entirely in
Dutch of course. An extensive inventory of the Oort
Archives has been published by Jet Katgert-Merkelijn \cite{JKM97}. No
professional scans are available for the Oort Archives, but over 27,000
reasonable to good quality photographs, taken by me,  of mainly letters and
notes are available on my special Oort homepage accompanying my biography of
him \cite{JHObiog}. 

Much information on the University of Groningen during van Rhijn’s days
is available in part II of Klaas van
Berkel’s monumental history of the university \cite{KvB2017}. Again this is in
Dutch unfortunately, and of not much use to English speaking researchers. An
extensive English summary has been announced.

\section{Pieter Johannes van Rhijn and his early research}\label{sect:Early}

I start this section with some biographical notes on van Rhijn. As mentioned,
at present no  comprehensive biography on him  has been written.

Pieter Johannes van Rhijn was born on March 24, 1886 in the city of Gouda, some 20
km northeast of Rotterdam. His father, Cornelis Hendrikus van Rhijn (1849--1913),
was a clergyman and not long after Pieter’s birth he was appointed professor of
theology at the University of Groningen. In \cite{Nijen} we read about C.H. van
Rhijn (my translation):
\begmarg
Van R. had a hard time fitting into the directional scheme of his time. The
liberal, tolerant upbringing, received in the parental home, continued to
influence him throughout his life. He wanted to be an unapologetic biblical
theologian.
\endmarg
Pieter’s upbringing therefore was undoubtedly relatively liberal as well. Cornelis
van Rhijn had had a first marriage, which however lasted only three years after
which his wife died. This marriage produced no children. Cornelis married again
in 1882 with Aletta Jacoba Francina Kruijt (1859--1945), and Pieter Johannes was
their second child. He had one older sister and three younger brothers, of which
one, Maarten  van Rhijn (1888--1966), eventually like his father became a professor
of theology, but then at the University of Utrecht. On the Web there is a long and
detailed genealogy of the family van Rhijn (earlier often spelled van Rijn),
\cite{PJgenea}. The keeper of this, Gert Jan van Rhijn, is a grandson of van
Rhijn’s brother Maarten, has provided some very useful biographical material
to me.
The genealogy goes back to the year 1510 and to ancestors who lived very close to
Leiden, but the famous seventeenth century painter from Leiden, Rembrandt with
the same surname, turns out to be not part of his ancestry. Pieter grew up in
Groningen, where eventually he studied astronomy with Kapteyn. In 1915
obtained his PhD with Kapteyn as supervisor (`promotor').

\begin{figure}[t]
\sidecaption[t]
\includegraphics[width=0.49\textwidth]{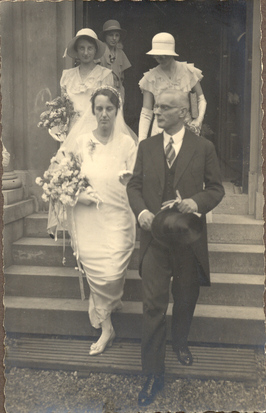}
\includegraphics[width=0.49\textwidth]{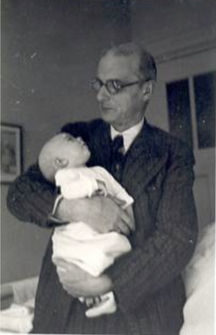}
\caption{\normalsize  Left: Pieter J. van Rhijn and Regnera (`Rein') L.G.C. de
Bie at their wedding in 1932. Right: Van Rhijn with his son, who was born in
1934. From the Website of the `Stichting Het geslacht van
Rhijn' \cite{PJgenea}, with permission.}
\label{fig:wedding}
\end{figure}

In 1932, at the age of 46, he married Regnera Louise Geertruid Constantia de Bie
(1906--1997), twenty years his junior (see Fig.~\ref{fig:wedding}, left).
In his letters
van Rhijn referred to her as `Rein’ (interestingly pronounced the same as ‘Rhijn’).
She had a degree in Law. The marriage certificate says she was ‘without occupation’.
Her father was a juvenile judge and later president of a district court
(Arrondissementsrechtbank) in Rotterdam. Van Rhijn and his wife eventually would
have two children, a son and a daughter (see Fig.~\ref{fig:wedding}, right).

Before his marriage van Rhijn was well occupied by the fact that his elder
sister Adriana Josephine van Rhijn (1884--1976),
whose nickname was Joos, lived in Groningen after being divorced in 1921.
She had four children, at the time of the divorce between 2 and 10
years of age and van Rhijn helped her extensively in raising the children.
At the time of his own marriage these children were
eleven years older and his involvement with them must slowly have become
less time consuming.

During the Second World War van Rhijn contracted tuberculosis and his
recovery took a fair number of years.  More about this is covered in the later
sections of this paper.
It did have a major impact on the rest of his life, for his vigor never fully
returned.

Not too much has been written about van Rhijn as a private person. Through
Gert Jan van Rhijn, the keeper of
the van Rhijn genealogy \cite{PJgenea}, I have been able to get answers to some
questions about Pieter van Rhijn from the latter’s  daughter, who is still alive
(contrary to his son). With his background of being the son of a clergyman and
theology professor, the question arises how religious Pieter van Rhijn really was.
According to his daughter 
he went to church but not every week and the church he
went to was that of the Remonstrants, because `he could not believe the creation
story’.

The Remonstrant movement had started with the liberal Leiden professor Jacobus
Arminius ($\pm$1555--1609). The view of Calvinism was that humans were predestined
for Heaven or Hell and had no free will to choose to live either in virtue or in
sin. The followers of Arminius (sometimes called the `Rekkelijken’ or Flexibles)
opposed this and a few related points of view, which were collected in the
`Five Articles of Remonstrance’ (objection or statement or defence)
in 1610, which caused them to be designated
Remonstrants. Their views were particularly disputed by Franciscus Gomarus
(1563--1641), whose followers were called `Preciezen’ (Strict Ones) or
Counter-Remonstrants. Gomarus actually already taught in Leiden when Arminius
arrived there. During the Synod of Dordrecht (1618--1619) the Remonstrants were
expelled by the Protestant Church and the movement separated from the
Calvinists.
The movement still exists in the form of the Remonstrant Brotherhood
and has somewhat over 5000 members and `friends’ (sympathizers) in the Netherlands.
It stands for a liberal interpretation of Christianity; their faith is characterized
by a non-dogmatic attitude, in which freedom of thought and belief based on one's
own insights is central and developments in modern science and insights in general
are taken into consideration. There was therefore room for van Rhijn to adhere
modern scientific developments and to adopt evolution over biblical creation.

Van Rhijn's daughter (who like het aunt was called `Joos' in the family)
stressed that (my translation)
\begmarg 
van Rhijn did not like dogmatic beliefs, but his wife had a more orthodox
upbringing. At home they read from the Bible, but  Pieter Johannes never led in
prayer. He had a broad view and was averse to strict rules. His wife had had a
strict upbringing. If something had to be done this way or that, Pieter Johannes
van Rhijn would say: So, who stipulates that [`Wie stelt dat vast', stressing
that no other authority than himself would decide on his behavior, action or
options]?

They [he and his wife] went to church together. The faith was also passed on, so
daughter Joos went with them to catechism. Mother found it liberating that Pieter
Johannes thought more freely about religion. 
\endmarg

Van Rhijn had no special functions in the church  and did not participate in
activities other than the Sunday service. In the Netherlands, people had (and have)
attached small signs to their front doors, traditionally with the name of the man,
as head of the family, imprinted on it. In van Rhijn’s case his wife’s maiden name
was added to it. 

Van Rhijn was deeply infuriated about the German occupation during Worl War II.
He was hurt and outraged by the supposed German superiority and breach of Dutch
sovereignty. In a
in a letter of December 1945 to his sister-in-law\footnote{\normalsize This is
Lies de Bie, a
younger sister of his wife, who and her preacher husband Foppe Oberman  at the time were still in Java. A copy this letter was kindly provided by Gert
Jan van Rhijn.} he expressed this as follows (my translation):
\begmarg
What has struck me most in the Germans is their unsparing arrogance (`das
Herrenvolk') and their virtuosity in lying. They can lie brilliantly. Therein lies
a certain merit.  The Krauts only did it stupidly. And sometimes I almost burst at
the pharisaism of e.g. Seys Inquart (the German chief in the Netherlands), who
said he had come to the Netherlands to protect us from Bolshevism. But those are
echos from a distant past. The gentlemen Nazi leaders including Seys Inquart are
now on trial in Nuremberg and within a few months they will probably no longer be
among the living. 
\endmarg
\noindent
Arthur Seyss-Inquart (born Artur Zajtich; 1892--1946) actually was not German, but Austrian.
Pieter van  Rhijn  played the viola, but there was no frequent playing together in
the household. Adriaan Blaauw has recorded \cite{Blaauw1983} his fond recollection
of (my translation):
\begmarg
\noindent
the warmth of his domestic environment, where playing music together -- van Rhijn
was an excellent viola player -- was a bright spot in the dark War years. 
\endmarg
\noindent
Blaauw himself played the flute. Van Rhijn's  daughter has remarked that
\begmarg
\noindent
as a younger man in his student days he had played in ensembles. He was not
good enough to play first violin. [...]  Eventually he found it more
interesting to play the viola. Later in life he did not have the strength [to
play]. 
\endmarg

She also recorded that she went to performances of Bach’s  Matth\"aus Passion
together with her father. Mrs. van Rhijn sang in a choir. Van Rhijn was interested
in Mozart and Schubert quartets. When recovering from tuberculosis after the War
he noted in the letter to his sister-in-law referred to above that he was allowed
out of bed eight hours a day and walking 30 minutes, and added that
in addition to reading he
listened to concerts on the radio, where Mozart was an obvious favorite. He
wrote (my translation)
\begmarg 
Mozart is a musician who really doesn't belong on this Earth. When hearing
his music, it is as if the heavens break open over this cursed
world. My colleague
van der Leeuw professor of theology (now a Minister) says that after hearing a
Mozart concert he no longer believes in the Original Sin for three days.
\endmarg 
Gerardus van der Leeuw (1890--1950), a good friend of van Rhijn,
was historian and philosopher of religion, and for a short period
Minister of education immediately after the War.

Van Rhijn
read a lot throughout his life, but particularly when he was recovering from
tuberculosis, especially English novels.  British authors that he liked
included Howard Spring, John Galsworthy, Nevil Shute, J.B. Priestley, Graham
Greene, W. Somerset Maugham, Duff Cooper, C.L. Morgan, Jane Austin and the
Bronte sisters. American writers were fewer, but included J.P. Marquand, Louis
Bromfield and Pearl Buck. He mentioned in the letter quoted that he also had read
some excellent poetry.

As for outdoors activities, he also often went hiking to south of Groningen, one
would think with his family.

Pieter Johannes van Rhijn died in Groningen on May 9, 1960 at age 74.
\bigskip

Van Rhijn obtained his PhD in 1915, on a thesis consisting of two quite
separate studies. For this he had spent an extended period at Mount Wilson
Observatory (1912--1913), which of course had been arranged 
by Kapteyn with George Ellery Hale (1868--1938).
The title of the thesis was {\it Derivation of the
change of colour with distance and apparent magnitude together with a new
determination of the mean parallaxes of the stars with given magnitude and
proper motion}. The thesis was never fully published
in a journal; a shorter paper appeared with the same title as the thesis itself
\cite{PJvR1916}, although it treated really
only the first part, which concerned the question
of reddening of starlight with distance from the Sun. In 1909, Kapteyn 
 had hypothesized that extinction of starlight would be primarily 
scattering rather than absorption, and that this could be expected to be 
stronger at shorter wavelengths, giving rise to reddening of starlight with 
distance \cite{JCK1909}. Van Rhijn addressed this issue using a study
called the {\it Yerkes Actinometry}, which was published in John
Adelbert Parkhurst (1861--1925) in 1912.
With this data-set, together with proper motions from other places, van Rhijn
determined the change in color of stars with distance. Kapteyn has assumed
(as we now know incorrectly)
that the wavelength dependence was like that of Rayleigh
scattering in the Earth's atmosphere, so that reddening could be related to
total extinction. Van Rhijn thus found that it
amounted to ‘{\it ceteris paribus} [all other things being equal] only
$0.000195 \pm 0.00003$ magnitudes per parsec’, only about half that what Kapteyn
had found in 1909. Interestingly -- and disappointingly --
there was absolutely no
discussion of this difference, neither in the paper, nor in the thesis itself.
We will see in what follows that this is the first instance of an emerging pattern
of van Rhijn never
adding a discussion in a paper to put things in a larger context.
Van Rhijn’s value is much lower than the currently adopted roughly one
magnitude per kiloparsec. However, this region of {\it Parkhurst Actinometry}
is centered on the celestial pole and that means almost $30^{\circ}$ Galactic
latitude. So in hindsight
it is not surprising that the inferred extinction was smaller
than in all-sky determinations or regions exclusively  in the Milky Way.

The second part of the thesis
formed an important step in the preparation to derive the
distribution of stars in space, the ultimate aim of Kapteyn’s scientific work,
and was published much later after more work and additional data.
Because it is relevant to the discussion of van Rhijn’s work for the rest
of this paper, I summarize
first the procedure Kapteyn had designed for this very briefly. It begins with
two fundamental equations of statistical astronomy.
They link the counts of stars and the average parallax as a function of apparent
magnitude to the distribution of the total density of stars as a function of
the position in space and to the luminosity function. I will
not present these here; if interested see the full, online version of Appendix A
\cite{AppA} of \cite{JCKEng}\footnote{\normalsize Note that in Box 15.3 or
\cite{JCKbiog} the function
$D(r)$ has been erroneoously omitted from the integrand of both equations.}.

Distances of stars can be measured directly from trigonometry using as base the
diameter of the Earth’s orbit. But Kapteyn and others had concluded that it
was not possible to do this in a wholesale manner. So he decided to use the
motion of the Sun through space as baseline. Already in one year this amounts
to four times larger a baseline and furthermore only increases with time.
This method of so-called secular parallaxes uses proper motions which contain
a reflection of the motion of the Sun through space in addition to the
peculiar motion of each star itself. His assumption was that these latter
components were random and isotropic. However because of these random
motions these secular parallaxes  can only be derived statistically.

Kapteyn developed the method to solve these
fundamental equations basically in two papers,
\cite{JCK1900} and \cite{JCK1902}. The first step is to determine, with the
secular parallax method, the average parallax of the stars of any apparent
magnitude $m$ and proper motion $\mu$. We can also estimate the average absolute
magnitude when we know the average parallax. Of course for an individual
star of an observed $m$ and $\mu$  the parallax will be different from the mean.
But for the problem of the distribution  of the stars in space it is not
necessary to know the distance of each individual star, only how many stars
there are at a certain distance. Furthermore, we need to know in principle
the distribution law of the parallaxes around the average value.
However, Kapteyn showed that widely
differing assumptions of this law led to results that differ but little. Now,
since it thus becomes possible to find, for any given distance, the number of
stars of each absolute magnitude, we can evidently derive the two fundamental
laws which determine the arrangement of the stars in space, viz. the
distribution of the stars over absolute magnitude, which Kapteyn called the
‘luminosity curve’ and for each line of sight
the law of total stellar density with distance from the Sun.

Kapteyn developed a
convenient numerical procedure of conducting these computations, after he had
introduced two assumptions: (1) there is no appreciable extinction of light in
space; (2) the frequency of the absolute magnitudes (the luminosity
curve) does not change with the distance from the Sun.
The point then is that this procedure
gives indeed both the luminosity curve in the solar
neighborhood {\it and} the density law, but the latter
up to only small distances, since stars
with measured proper motions are involved. So this really
only constitutes a first step, namely determining the local luminosity
function. The next is to use deep star counts to find the
star density out to larger distances, but that is a separate problem of
`inverting' these two fundamental equations. This summary 
illustrates how the second part of van Rhijn’s thesis, {\it a new determination
of the mean parallaxes of the stars with given magnitude and proper motion},
fitted into the grand scheme of things, and that it constituted the
important preliminary step before one could derive a model for the distribution
of stars in space. 
\bigskip

After his return from Mount Wilson, van Rhijn had been appointed assistant at
the Groningen Laboratorium in 1914.
In addition to completing his PhD thesis, van Rhijn was heavily involved
in Kapteyn’s activities to prepare for his first attack on the problem of the
structure of the Sidereal System. Much of this work had been done
done with the assistants, first Willem de Sitter,
Herman Albertus Weersma (1877--1961) and Frits Zernike (1886--1966), 
and later van Rhijn. The program led to major contributions in the
{\it Publications of the Astronomical Laboratory at Groningen}:
van Rhijn in 1917 
\cite{PJvR1917}, Kapteyn, van Rhijn \&\ Weersma in 1918 \cite{KRW1918},
Kapteyn \&\ van Rhijn in 1920 \cite{KvR1920a}. All this work was part of the
necessary preparatory effort to bring
together information and data etc. available into
a form that could be used to study the ‘Construction of the Heavens’. This was
finally done for the first time by Kapteyn and van Rhijn in 1920 \cite{KvR1920b}, in
which they presented a crosscut through the Stellar System, which
came to be known as the Kapteyn
Universe. Since the analysis was done with count
averages over all longitudes, this
schematic model had the Sun by definition
in the center. It was the basis for the {\it First
attempt at a theory of the arrangement and motion of the Sidereal System} by
Kapteyn \cite{JCK1922},  which was the
start of  observational galactic dynamics and moved
the Sun out from the center by 0.65 kpc. 

Under and in collaboration with Kapteyn, van Rhijn had been extremely
productive. He even found time to complete another fundamental study {\it On
the brightness of the sky at night and the total amount of starlight}
\cite{PJvR1921}. This was another old favorite of Kapteyn, who wanted to find a
constraint on the distribution of stars in space by determining an observational
value for the amount of integrated starlight.
It had been attempted before in a PhD thesis under Kapteyn by a
student with the name Lambertus Yntema (1879--1932),
who tried to measure this from a dark
site not too far from Groningen by two ingenious methods developed by Kapteyn,
comparing an illuminated small surface to the sky viewed next to it and by
photographing the sky through a small hole
and comparing that to an exposure of the full moon
\cite{LY1909}. He found that there was a
significant amount of  ‘Earth light’,  a sort of  ‘permanent aurora’, but he
speculated that part of his observed light would be diffuse starlight.
Yntema left astronomy ending up as rector of a lyceum or grammar school in Bussum.
Van Rhijn, during his stay at Mount Wilson, had repeated this study from the
mountain top. Zodiacal light is another factor; it is in fact comparable in
brightness to integrated starlight. Actual direct measurements
of integrated starlight have only been obtained in recent times, particularly
by the Pioneer 10 spacecraft on its way to
Jupiter, when in 1973 it had passed the asteroid belt. The zodiacal light
there was negligible and the spacecraft
produced maps of the integrated star light that could be used to study the
properties of the stellar distribution in the
Galaxy; for details see \cite{PCK86}.
Both Yntema and van Rhijn claimed values that compared to
our current
knowledge are not bad at all, but this must be fortuitous. For a more
thorough and detailed discussion see \cite{JCKbiog} or \cite{JCKEng}.

All of this goes to show that van Rhijn
produced quite a lot of high
quality work, which however was initiated by Kapteyn and executed under
his supervision.

\begin{figure}[t]
\begin{center}
\includegraphics[width=0.98\textwidth]{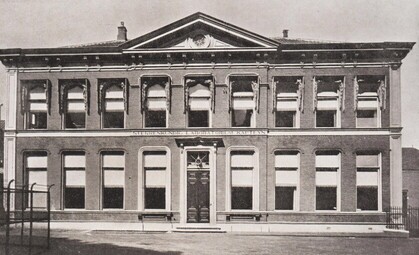}
\end{center}
\caption{\normalsize The `Sterrenkundig Laboratorium', when it had been
named after Kapteyn on the occasion of his retirement.  
Kapteyn Astronomical Institute.}
\label{fig:lab}
\end{figure}

\section{Kapteyn's Plan of Selected Areas}\label{sect:PSA}

Kapteyn launched the {\it Plan of Selected Areas} in 1906 \cite{SA},
after a few years of planning and
discussions. A few relevant pieces of the text are important to cite here for
background to what is discussed below.

The Plan was designed to address a particular problem:
What is the distribution of stars in space and what is the structure of the
Sidereal System? But there is more to it, as we can read in Kapteyn's
Introduction in \cite{SA} (p.3):
\begmarg
[...] to make at last definite plans for the future work of the laboratory.
\endmarg
So, it was more than an endeavor for the interest of answering a scientific
question, it was also Kapteyn’s way of providing a future for the laboratory
he had founded (Fig.~\ref{fig:lab}). The text continues:
\begmarg
After much consideration the most promising plan seems to me to
be some plan analogous to that of the gauges of the Herschels.

In accordance with the progress of science, however, such a plan
will not only have to include the numbers of the stars, but all the data
which it will be possible to obtain in a reasonable time. My intention had 
been all along to give to this plan such an extension that, with the
exception of the work at the telescope, the whole of it could be undertaken
by our laboratory.

In working out details, however, I soon found out that, with a plan
any ways on a scale meeting the requirements of the case, such would be
impossible unless the funds of our institution were materially increased.
\endmarg

\begin{figure}[t]
\sidecaption[t]
\includegraphics[width=0.64\textwidth]{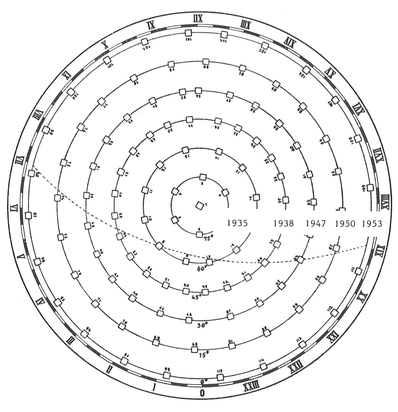}
\caption{\normalsize The \textit{Selected Areas}
in the \textit{Systematic Plan} of the northern hemisphere; the
numbers run from  Area no. 1 in the center (the North Pole) to No. 115 on the
equator. The scale on the border is Right Ascension in hours or 15\degs\ 
on the equator.
Declinations are indicated next to the circles, which are spaced by 15\degs.
The dashed line indicates the center line of the Milky Way,.
Year numbers indicate when the volume of the
\textit{Bergedorf(-Groningen-Harvard)
Spektral-Durchmusterung} for the relevant declination range has been published.
Kapteyn Astronomical Institute.}
\label{fig:SANorth}
\end{figure}

Kapteyn presented the {\it Plan of Selected Areas}
as a refinement of the star gauges of William
Herschel (1738--1822) and his son John Frederick William Herschel (1792--1871).
Kapteyn quoted that they covered 145 square degrees (of about 41000 for the
full $4\pi$ steradians of sky) and counted about 117,600 stars. But this work
constituted only counts, no magnitudes or other properties. Kapteyn set an aim
of  200,000 stars, or almost twice the number of stars in the Herschel gauges.
This number for stars down to magnitude 14 required 400 square degrees, for
which Kapteyn then defined 252 areas, of dimensions either rectangular with
sides of order 75 arcmin or circular with radius 42 arcmin
(in both cases an area of about 1.5 square degrees).
This in regard of a reasonable field of view
for plates in photographic telescopes. The size eventually was not a fixed
property, areas could be smaller in crowded parts of the sky or larger in
sparse ones. The program envisaged counts to as faint a level as possible,
photographic and visual magnitudes (and thus colors),
and as far as possible parallaxes,
proper motions, spectral types and radial velocities. Kapteyn estimated that this
(not counting `additional short exposures') would require in total 9710
exposures on 2620 plates plus an unknown number for (hopefully) some 6300
radial velocities. The Plan also provided for determinations of the
total sky brightness to set a further constraint.

At the insistence of Edward Charles Pickering (1846--1919), director of the
Harvard College Observatory, the Plan consisted of two parts,
a {\it Systematic Plan} of 206 areas distributed evenly across the sky,
as envisaged by Kapteyn, and an
additional 46 for a {\it Special Plan} chosen to be particularly
suitable for more
detailed studies of the structure of the Stellar System.
The  {\it Systematic Plan}
aimed for a mean separation of the areas of order 15\degs. The distribution
of the regions 1 to  115 in the northern hemisphere
from the pole to the equator is shown in
Fig.~\ref{fig:SANorth}. This is a projection of a sphere onto a flat plane and
distortions resulting from this hide the approximate regularity
on the sky. The
declinations are as required for the desired spacing in rings of 15\degs\
difference. The 24 regions on the equator are spaced by one hour or
15\degspt0. The rings at declinations 15\degs, 30\degs, and 45\degs\ also
contain 24 regions that have spacings of respectively 14\degspt5,  13\degspt0,
and 10\degspt6.  For declination 60\degs\ the 12 regions are spaced by
15\degspt0, while for the 6 at declination 75\degs\ it is 15\degspt5. This is
indeed a rather uniform distribution, except that it is not evident
why Kapteyn did not restrict
the declination zone 45\degs\ to 18 regions spaced by 14\degspt1
instead of 24 of 10\degspt6.
The distribution is not {\it precisely} uniform because for calibration
purposes they were required, if at all possible, to be centered on a star of
magnitude 8.0 to 9.0. But there also should be no star brighter than this --
as Pickering had stressed --, since its light would scatter
over much of the photographic emulsion and affect the determination of
magnitudes of the faint stars.

This sounds like a straightforward selection procedure,
but in fact is less trivial than it may seem, as
a quick calculation shows. The density of stars between 8-th or 9-th magnitude
in photometric band V (roughly visual) averages about 2 per square degree (in
round numbers 90,000 stars for the full sky of 41,000 square degrees).  A
Selected Area covers about 1.5 square degrees. The density varies of course
substantially with Galactic latitude, ranging from somewhat less than one in
the poles to up to ten or more stars
per square degree at low latitude (for these
numbers see Fig.~1 in \cite{PCK86}, which is produced with the export version
of the Galaxy model computer code of \cite{BS80}). At high latitude this is a
serious limitation for the choice of fields, since fluctuations around an
average of one star per field yields a significant number of field with none.
The further condition that there should be no bright star -- of, say,
magnitude 7 or brighter -- in the Area excludes a  significant part of the sky;
there are about 16,000 of such stars over the full sky or one star per 2.6
square degrees. This would imply that a randomly chosen field of 1.5 square
degrees has a chance of one in two or so to have a star brighter than magnitude
7 in it. Kapteyn did not comment on this but his combined conditions must have
been a serious constraint and have made the choice of the positions of his Areas
far from simple.

The selection of the Areas for the {\it Special Plan} was done along a
number of lines, that I will summarize as follows with excerpts from Kapteyn’s
original 1906 proposal \cite{SA}:
\begmarg
A. Quite a number of areas (19) in which […]  maps show a black opening
surrounded by rich parts [...]; or a rich part, or rich parts, between dark
spaces  [...]; or where there is at least a sudden change in star density
[...]. \\
B. A series of areas on the great branches of the Milky way and the rift
between them.    [...] Moreover several plates of the series A will very
usefully serve the study of this division of the Milky way.\\
C. Further: very rich or very poor parts of the Galaxy [...]: Here too, many
of the plates of series A will be useful, so for instance [the one]  on the
border of the southern Coalsack.\\
D. Two Milky way areas   [...] for which de Sitter finds a marked difference,
in opposite direction, between photographic and visual magnitude.\\
E. For the rest I have included, of {\it extra}-galactic regions: two
areas
[..]  coinciding with the richest parts of the {\it Nubecula minor};
{\it two} areas
[...] coinciding with the parts richest in stars and nebulae of the {\it
Nubecula major}; {\it one} area  [...]  coinciding with the part of the Orion nebula
with strongest contrast in star density; one plate [...] covering a part of
the sky, near the North Pole of the Milky way, exceptionally rich in small
nebulae. [Italics from Kapteyn]
\endmarg
It is interesting to consider the last category (E.) in somewhat more detail.
The term {\it extra-}galactic means of course away from the Milky Way on the
sky, while eventually it became to mean outside our Galaxy.
The {\it Nubecula major} and {\it Nubecula minor} are the Large and Small
Magellanic Clouds. At the time the Magellanic Clouds were generally considered
to be detached parts of the Milky Way, although the suggestion that they were
separate stellar systems had been made \cite{CA1867}. The brightest stars in
them are of apparent magnitude 9 or 10 and this by
itself would also suggest relatively large distances. Kapteyn noted in a
footnote that there is no field in the {\it Systemstic Plan} that covers
either of the Clouds.

The exceptional density of nebulae in the final Area refers to the Virgo
Cluster. It is one of the interesting coincidences in galactic and extragalactic
astronomy that the relatively nearby (some 20 Mpc)  Virgo Cluster of galaxies
is on the sky located close to the Galactic north pole and consequently that
our Galaxy would be seen close to face-on for astronomers there.  The dense
center of the Virgo Cluster is some 15\degs\ from the pole,
and there is an only about 3\%\ chance to see another spiral
galaxies that close to face-on
\cite{PCKMWG}\footnote{\normalsize
See this reference also for some more remarkable
coincidences related to the orientation of nearby galaxies and the position of
the Sun in the disk of the Galaxy.}. The region around the Galactic
North Pole is indeed
rich in nearby galaxies belonging to the Virgo Cluster and its extensions into
the Local Supercluster. Kapteyn’s Area No. 32 in the {\it Special Plan} is
about 1\degs\ away from the actual Galactic north
pole as defined now.  Kapteyn has not specified what he proposed to about
the small nebulae. In the catalogue
resulting from the {\it Special Plan} no special attention has been given to
the nebulae in this Area.
\bigskip

There are two aspects that I want to stress. Although Kapteyn
may have considered his plan a work of service to posterity,
much like the {\it Carte du Ciel}, and the great star catalogues {\it Bonner
Durchmusterung}, his own and David Gill's (1843--1914)
{\it Cape Photographic Durchmusterung}, and
the {\it Astronomische Gesellschaft Katalogs}, he had in the first place, and in
clear distinction to these works, a well-defined scientific goal in mind.
He might have had some hope in
1906 that he himself would be able to use at least a major part of the data
to construct a model of the Sidereal System. That
hope turned out to be unrealistic, 
and he was forced to replace this desire with his {\it First attempt at a
theory of the arrangement and motion of the Sidereal System} \cite{JCK1922},
based on
data before the bulk of the Selected Areas had been surveyed.
But he would lay the foundations for future
astronomers to do so. He expected though that his
successor would surely complete what he had been unable to accomplish.

Secondly, he saw it as a program for his Groningen Laboratorium to carry out
completely by itself except for the data collection at the telescopes.
And that this would
require extra investment. However, it turned out that he should not have
counted for this on his university, and certainly not on his government.
Just as when in 1886 he set out
to measure up the plates for the CPD and reduce the date to positions and
magnitudes, he had to rely for the largest part on private funds. He now had
his own Laboratorium with, since 1911, adequate housing, contrary to the 1880s
when physiology professor Dirk Huizinga (1840--1903)
generously provided some working space (as it happened in the same building
that later housed the Laboratorium). Luckily,
he was saved then by George Hale, when in 1908 he adopted the {\it Plan of
  Selected Areas} as the prime observing program for his brand new giant 60-inch
telescope on Mount Wilson. Kapteyn was appointed Research Associate by the
Carnegie Institution of Washington, which enabled him to visit California
annually (at least until the first World War broke out) and take charge of the
execution of the program at the big telescope.

But is turned out unrealistic that all work could be done in Groningen.
For example, the following was an unforeseen complication. It turned out
much less straightforward to derive stellar magnitudes from photographic plates
taken with a large reflector than a small photographic telescope.
To illustrate this the following. The field of view of the 60-inch telescope was
only about 23 arc minutes and many of the fields had no stars brighter than
12-th magnitude.
Using the same Fig.~1 of \cite{PCK86} as above it can be deduced that
averaged over the
whole sky  the number of stars of 12-th magnitude or brighter is
about 70 per square degree, which means on average 10 or
so per  23 arcmin field of the 60-inch. But at high latitudes it is on
average one or a few, so that fluctuations will assure that a
non-negligible fraction has no stars brighter than magnitude 12.
This complicated matters as calibration of the zero-point of the magnitude
scale relative to the standards around the celestial North Pole had to be
permormed in such Areas without making use of stars brighter than magnitude 12.
Furthermore stars away from the center of the field had
distorted images from optical aberration's such as coma
due to the parabolic shape of the mirror.

All this required
special attention and much work was taken off Kapteyn's  hands
by Mount Wilson astronomer Frederick Hanley Seares (1871--1964), who took
this upon himself to solve. This was an enormous effort and it lasten until
1930 until the project was concluded with the publication of the
catalogue \cite{MWilson}.  For manpower in Groningen Kapteyn
relied on funding from various sources rather than the university and
significant parts of the manpower was actually paid for by the Carnegie
Institution. The university and government provided only one
computer (`rekenaar', a person whose primary task was to perform calculations)
and one clerk in addition to the professorship and assistant. When
Kapteyn retired the pay-roll of the university staff consisted of the
professor, one assistant, three computers and a clerk (see Table 1 in section
\ref{sect:staff} below), but that increase had only occurred in 1919. 

The University of Groningen had committed to the {\it Plan of Selected Areas}
when it agreed with Kapteyn’s appointment by the Carnegie Institution (it was
less generous than this suggests, since 
he had to make up in teaching that he failed to provide during his Mount Wilson
visits over the remaining months). To some extent the lack of financial
support is understandable as the funding for the universities in general was
tight. However, due to its remoteness from `Den Haag'
(the government seat in the Hague) and the difficulty
of effective lobbying, and due to low student numbers in Groningen, it was
particularly tight there.With the appointment of Willem de Sitter 
as director in 1918 a
major reorganization had taken place at the Sterrewacht Leiden, and its staff
was three or four times larger than at the Groningen Laboratorium. 
\bigskip

The execution of the {\it Plan of Selected Areas} during the remaining years
of Kapteyn was restricted,
as far as the work in Groningen is concerned, to the cataloging
of stars in the Areas from the plates it received from Harvard and
Mount Wilson with the vital support of
the directors Edward Pickering and George Hale.
The plates of all 206 Areas Pickering had committed to
were taken at the Harvard College Observatory in the north and at
its Boyden Station in Arequipa, Peru, in the south,  respectively
between 1910 and 1912 and between 1906 and 1916, and from all of these plates
stellar positions and magnitudes had to be derived.
All those measurements were performed in Groningen.

The catalogue resulting from the Harvard plates, which went down to about
magnitude 16, was published in three installments by Pickering \&\ Kapteyn in
1918 and by Pickering, Kapteyn \&\ van Rhijn in 1923 and 1924 in the
{\it Annals of the Astronomical Observatory of Harvard College}
\cite{Harvard1}\cite{Harvard2}\cite{Harvard3}.
The work had been a joint undertaking between Harvard and Groningen. As 
Pickering wrote in the Preface to the first volume:
\begmarg
It was, therefore, a pleasure to aid the plans of Professor Kapteyn in his
work on the Selected Areas. We have accordingly taken the photographs needed
to show the faint stars, determined the magnitudes of a sequence of standard
stars in each area, and provided the means of publication of the results.
Professor Kapteyn has undertaken the laborious work of measuring the plates,
reducing the measures, and determining the positions and magnitudes of large
numbers of stars.
\endmarg

The Durchmusterung became known as the {\it Harvard Durchmusterung} in spite of
the fact that while the plates were indeed taken at Harvard, the bulk of the
work was done in Groningen. I will therefore refer to it as the {\it
Harvard(-Groningen) Durchmusterung}.

According to the {\it Third Report} of the progress of the Plan
\cite{PJvRrep3}, the number of stars in the
{\it Harvard(-Groningen) Durchmusterung} was some 250,000, for all of which
positions and magnitudes had been measured in Groningen. An examination of the
three volumes shows accurate quotes for the total number in the first two
volumes, while adding up numbers in Table 1 of the third volume is easily
performed. This results then is a total of  84,002 + 65,753 + 82,226 =  231,981
stars, so it is actually somewhat less than what had been advertised.

The work in Groningen was quite extensive and, since great care had to be taken,
very time consuming. Most of it was done by dedicated staff,
but had to be overseen by
an astronomer.  In Kapteyn’s Introduction to the first installment in 1918
on the northern hemisphere, he stressed that his own contribution was minor,
but that the supervision of the work was left to the assistants, which he
listed as Willem de Sitter, Herman Weersma, Frits Zernike and Pieter van Rhijn.
The second and third volumes covering the south were authored
by Pickering, Kapteyn (posthumously) and van Rhijn, and the latter added Jan
Oort to this list of assistants.
\bigskip

The primary plates at the Mount Wilson 60-inch of the 139 Selected Areas
north
of declination -1\degs\ were taken between 1910 and 1912, plates with shorter
exposure time and/or reduced aperture were taken from 1913 to 1919, and a set
of exposures of the Selected Areas and a field with stars in the {\it
North Polar Sequence} with a 10-inch telescope in the same years.
Overall more than one
thousand plates were taken. The project was executed at Mount Wilson under
the leadership of Frederick Seares.

For the {\it Mount Wilson(-Groningen) Catalogue} the work was more evenly
divided than for the {\it Harvard(-Groningen) Durchmusterung}. In addition to
the ‘long-exposure plates’ together with short exposure plates, that were all
sent to Groningen, to facilitate the transition from faint to brighter stars,
a long and thorough investigation had been conducted at Mount Wilson to learn
how to do photographic photometry with large reflectors
and to define a magnitude
scale and zero-point in each of the Selected Areas. For this purpose
another set of exposures, using the 60-inch telescope stopped to
smaller apertures  was obtained.  As a consequence two sets of stars were
catalogued: the long-exposure plates were only studied over 15 arcmin areas at
low Galactic latitudes (below 40\degs) and 20 arc\-min at high latitude. The
plates measured for magnitude scales and zero points were measured over a
larger area and this yielded ‘supplementary’ stars outside the ‘Groningen
areas’. The {\it Mount Wilson Catalogue of photographic magnitudes in Selected
Areas 1–139} by Seares, Kapteyn, van Rhijn, Joyner \&\ Richmond appeared in
1930. It reached magnitude 18.5 or so. The final catalogue contains 67,941
stars, of which 45,448 had been measured in Groningen.
\bigskip

The measurements and reductions for the plates for the {\it Special Plan}
\cite{vRK1952} were seriously delayed by the economic depression and associated
shortage of manpower and then the Second World War and in the end were only
published in 1952. It was no longer a joint publication between Groningen and
Harvard, but solely by the Kapteyn Laboratorium. The work concerned all the
46 Areas of the {\it Special Plan} and the total number of stars measured
was 140,082 stars down to 15 to 16 photographic magnitude. Measurements and
reductions were supervised by van Rhijn’s assistants starting with Willem
Jan Klein Wassink, then Bart Bok, Petrus Bruna, Jean-Jacques Raimond, Adriaan
Blaauw and ending with Lukas Plaut (and van Rhijn himself did some of this
also of course). The plates were taken decades before the
publication, most of them in 1911 and 1912, but some as long ago as in 1898
(so before the {\it Plan of Selected Areas} was even defined and at that time
for some other purpose). The measurements extended only over a restricted part
of the plates, in squares of roughly 50 to 100 arcmin on a side,
in general depending on the structure in the star field in the
particular Area or the richness.

The production of this catalogue constituted a very large
expenditure in terms of time and manpower. The use that has been made of
it is very limited it
seems, to some extent of course since its appearance was long overdue, decades
after the plates had been taken. The NASA  Astrophysics Data System
ADS\footnote{NASA/ADS (ui.adsabs.harvard.edu) provides some information in
terms of numbers of citations
and of reads as measures of usefulness or at least one aspect of this.
Citations to other publications are reasonably complete in ADS
since the mid-fifties, when journals started to
collect references at the end of papers
rather than in footnotes in the text, so that it was practical for
ADS to collect them (see section 14.6 of \cite{JHObiog} for
some more details). Reads, when the full record in the ADS has
been downloaded by a user, are available since roughly the year 2000.}
records no
(zero!) citations after the publication of the {\it Durchmusterung},
so it is unlikely it has ever been used as source material
for a study as envisaged initially.  The number of
reads is only a few (except in 2021, which are due to me for this study).
But then finding it on ADS merely yields the information
that the publication is not available online. As we will see  below
(section~\ref{sect:extinction}), the results have actually been used
in a preliminary form  in the 1938 PhD thesis of Broer Hiemstra \cite{BH38},
who refers to the unpublished counts of the {\it Special Plan}. This thesis
concerned distances and total extinction of dark clouds, and contained
extensive determinations of proper motions from other plates
than the Harvard ones for the {\it Special Plan}. As far as I am aware no
extensive use is made of the counts in the fields in the Magellanic Clouds;
these amount tot 9777 stars in the richest part of the SMC and 7818 in that
of the LMC.
It would not be surprising if more than maybe a few researchers of the Clouds
are or were actually aware of  the existence of these uniform data sets of
magnitudes and star counts.

\begin{figure}[t]
\sidecaption[t]
\includegraphics[width=0.51\textwidth]{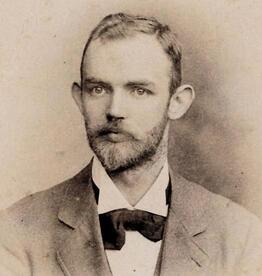}
\includegraphics[width=0.47\textwidth]{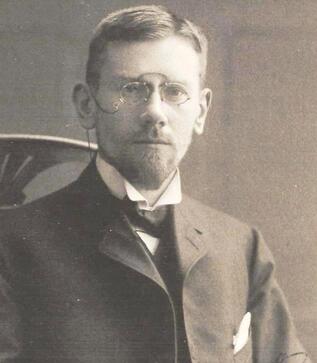}
\caption{\normalsize  Willem de Sitter and Herman A. Weersma.
These pictures come from the {\it Album Amicorum}, presented to H.G. van de
Sande Bakhuyzen on the occasion of his retirement as professor of astronomy
and director of the the Sterrewacht Leiden in 1908 \cite{Album}.}
\label{fig:dSW}
\end{figure}

\section{The succession of Kapteyn}\label{sect:Succession}

In 1908 there had been  a vacancy in the directorship of the Sterrewacht Leiden
after the retirement of Hendrikus van de Sande Bakhuyzen.
It had of course been offered to Kapteyn, but -- as actually
he had done earlier in the case of 
Utrecht -- he had declined. Ernst van de Sande Bakhuyzen
(1848--1918) had been appointed director and de Sitter (see Fig.~\ref{fig:dSW})
professor. Herman Weersma (Fig.~\ref{fig:dSW})
in Groningen in 1908 took de Sitter's place and became Kapteyn's
assistant. Weersma had obtained his PhD in that year, the thesis title was
{\it A determination of the apex of the solar motion according to the methods
of Bravais}.
Kapteyn considered Weersma his probable successor, but in 1912
Weersma left to take up a position as secondary school teacher, because he
wanted to have more time to devote to his `real interest', philosophy.

Kapteyn was at a loss. There was no suitable candidate in the Netherlands.
After I had written my biography of Kapteyn \cite{JCKbiog}, Prof. Kalevi
Mattila of Helsinki University informed me of a development. The story is as
follows:

On November 3, 1912, not long after Weersma's departure, Kapteyn wrote to his
longtime friend and collaborator Anders Severin Donner (1854--1938) of
Helsingfors Observatorion (Helsingfors is Swedish for Helsinki) as follows
(provided by Mattila, my translation from the German):
\begmarg
At present my greatest worry is to find a successor that will keep the
laboratory from going into a decline. My assistant leaves me. Do you know
anyone -- assiduous, practical, reliable -- who without too much pay (he will
get immediately 2000 German Mark, but I see a possibility, if he is really good,
that next year he might be paid and extra 1200 Mk)
could develop in the next 9 years to
become my successor?
\endmarg

\noindent
Prof. Matilla continued his email to me:
\begmarg
Donner's pupil Yrj\"o V\"ais\"al\"a had just completed his Masters degree the same
month and Donner approached him with this proposal. Prof. Olli Lehto published
in 2005 a triple biography (in Finnish) of the three V\"ais\"al\"a
brothers, who all became professors: Vilho (meteorologist), Yrj\"o
(astronomer, geodesist), and Kalle (mathematician). Having heard from me
of this letter by Kapteyn, Vilho got access also to some private
correspondence of Yrj\"o V\"ais\"al\"a to his bride Martta. Yrj\"o V\"ais\"al\"a
was very seriously considering the great scientific opportunities offered by
this position and how pleasent it would be to live among the nice and decent
people of Holland [the Netherlands]\footnote{\normalsize Holland is sometimes used as the
name of the nation, although it really is only two of the western provinces
of the Netherlands bordering the North Sea, where Amsterdam, Rotterdam
and other cities thrived as centers
of commerce and culture in the seventeenth century. The use of Holland instead
of the Netherlands is much the same as the often heard incorrect referring  to
Great Britain as England, and both should be avoided.}.
However, as he wrote to Martta: `Away from Finland for many
years, perhaps forever! I think that we would loose much when doing that. And
what about my little Martta in that foreign country! What if we would become
parents of even smaller Marttas and Yrj\"os! Would they become Finns or Dutch?
Or neither of them.' After such deliberations he finally asked Donner to
answer with `no' to Kapteyn. Donner motivated V\"ais\"al\"a's refusal in a
somewhat different, perhaps more diplomatic way: `We Fins are passionate
patriots'. In fact, this was a period in the
history of Finland with strong national feelings while the Russian tsar tried
to suppress the autonomous status of Finland. Five years later, in 1917,
Finland declared its independence.
\endmarg
Yrj\"o V\"ais\"al\"a (1891--1971) went on to found the astronomy department
at Turku University, where he designed a Schmidt-like telescope with which he
and his collaborators discovered 807 asteroids and 7 comets.
\bigskip

Kapteyn's next 
hope had been Frits Zernike,
who was appointed assistant, but he also left
already in 1914 to pursue a career in physics, which earned him eventually (in
1953) a Nobel Prize for his invention of the phase-contrast microscope. 
In \cite{Blaauw1983}(p.61),
Blaauw confirmed this. He quoted Zernike in a speech
in 1961 at
the occasion of the donation of the well-know  Veth painting of Kapteyn to the
university that Kapteyn had told him he would have liked to designate him as
the next director of the Laboratorium.\footnote{\normalsize
I have not been able to locate
a copy of this speech that Adriaan Blaauw must have seen, to confirm this.}

In 1916 Kapteyn arranged for van
Rhijn to be appointed his assistant, who became then the probable candidate to
succeed him as professor of astronomy and director of the 
Laboratorium. We may have a brief look at other young astronomers in the
Netherlands at the time as possible candidates,
say those who obtained a PhD between 1910 and 1920. Kapteyn
had four students in that period: Etine Imke Smid (1883--1960, PhD thesis
{\it Determination of the proper motion in right ascension and
declination for 119 stars} in 1914),
Samual Cornelis Meijering (1882--1943; 
PhD thesis {\it On the systematic motions of the K-stars} in 1916), Willem
Johannes Adriaan Schouten (1893--1971; PhD thesis {\it On the
determination of the principal laws of statistical astronomy} in 1918) and
Gerrit Hendrik  ten Brugge Cate (1901--1961; PhD thesis {\it Determination and
discussion of the spectral classes of 700 stars mostly near the north pole}
in 1920). These were not obviously extraordinary theses, so they were
apparently not considered suitable candidates.
Etine Schmid was the first female astronomy PhD in the Netherlands and
may not even have been eligible for a professorship or the  directorate.
The thesis was written in Dutch; I translated the title.
She went to Leiden where she worked with among others
physicist and later Nobel laureate Heike Kamerlingh Onnes (1853--1926),
but she left scientific research in 1916, when she got married.
Meijering became a
teacher at a high-school, Schouten had alienated Kapteyn because of his
unnecessary rude remarks on von Seeliger's work and ten Brugge Cate (who later
changed his name into ten Bruggecate) also became a teacher at a high-school.

In Leiden new PhD's in this period were Christiaan de Jong
(1893--1944; PhD thesis {\it Investigations on the 
constant of precession and the systematic motions of stars} in
1917), Jan Woltjer (1891--1946; PhD thesis {\it Investigations in the theory of
Hyperion} in  1918) or August Juliaan Leckie (1891--19??; PhD thesis
{\it Analytical and numerical theory of the motions of the orbital planes of
Jupiter's satellites} in 1919). De Jong, whose supervisor
was Ernst van de Sande Bakhuyzen, already became a gymnasium teacher  of
mathematics in Leiden before completing the thesis, which was
written in Dutch (title above is my translation). He may not have been
interested in a career in astronomy in the first place. De Jong was executed
by the Germans as retaliation of an act of sabotage he had not been concerned 
with. Woltjer became a Leiden staff member and worked on stellar pulsation 
theory until his death in 1946 as a result of undernourishment in the last
year of the War. His interests were remote from the ongoing investigations in
Groningenand anyway. Leckie moved to the Dutch East Indies, where be worked at
the Bandung Institute of Technology. The last two were students of Willem de
Sitter and worked on theoretical subjects involving planetary satellites.
In Utrecht Adriaan van Maanen
(1884--1946) had defended his thesis {\it The proper motions of 1418 stars in
and near the clusters h and $\chi $ Persei} in 1911 under Albertus Antonie
Nijland (1868--1936) and Kapteyn; he might have been a potential candidate,
but had moved permanently to Mount Wilson Observatory.

This list is not very long, so candidates were difficult to find.
Among these persons that
obtained their PhD’s before or in 1916, there is no obvious other
candidate that could
have been considered as being preferred over van Rhijn for the position of
assistant in Groningen. After van Rhijn had
been appointed, a further appointment would
have been difficut to make so especially after that only exceptionally promising
candidate could have been considered. There is no such person obvious in
the list Nor is there anyone among these that would be so
outstanding to be appointed as Kapteyn’s
successor in 1921 in place of van Rhijn.

Before 1921 van Rhijn  had done relevant and excellent
research. He had extensively worked on
proper motion observations and with Kapteyn  had used simulations to show that
with counts down to magnitude 17, which the plates taken with the 60-inch
would certainly be able to provide,
`the densities should become pretty reliable for the whole of
the domain within which the density exceeds 0.1 of that near the Sun'.
In Kapteyn’s last paper before retirement that van Rhijn co-authored, they
examined the matter of the Cepheid distances and the consequences for those of
globular clusters, concluding that Shapley’s distances were much
too large; Shapley's result was indeed
incorrect due to the now known concept of Stellar
Populations and the unknown differences between pulsating variables in these
two Populations (for more see \cite{JCKbiog}, p.592).
Van Rhijn before he assumed
the directorship definitely had been quite productive and had
produced excellent astronomy, but under the direction of
and on research programs initiated by Kapteyn.

The conclusion must be that van Rhijn with his dedication to the work of
Kapteyn and the Laboratorium
effectively by default became Kapteyn’s successor. He had done
very good research, although he never seemed to have had a
chance to define his own projects. This discussion serves to illustrate the
problem of finding suitable candidates for leading positions in astronomy at
the time. There simply was no other option.
Van Rhijn had certainly 
shown himself to be a competent researcher, so he was a man
Kapteyn could expect to dedicate himself to the continuation of his life's work.
There is no evidence that Kapteyn looked for other possibilities
abroad, being confident apparently that van Rhijn was the person to whom he
could entrust the future of his Laboratorium and the further execution of his
{\it Plan of Selected Areas}. 

\begin{figure}[t]
\begin{center}
\includegraphics[width=0.98\textwidth]{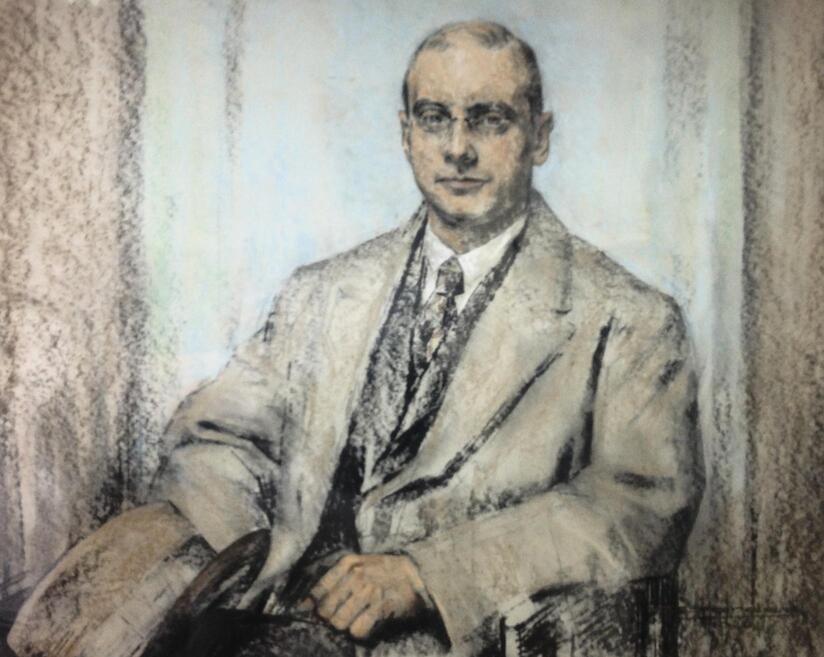}
\end{center}
\caption{\normalsize Pieter Johannes van Rhijn. Reproduction of a crayon
drawing donated by his relatives to the Kapteyn Astronomical Institute.
It dates from 1926. It has been produced by Eduard Gerdes (1887--1945),
who was a Dutch painter and art teacher. Later, in the 1930s, he would sympathize
with the national socialist party, and hold important functions in art policies
during the War. Gerdes died soon after the War ended, but the cause of his
mysterious death has never been
determined. He was posthumously found guilty for collaborating with the German
occupation.  This drawing decorates the Kapteyn Room.}
\label{fig:draw}
\end{figure}

\section{Van Rhijn's astronomical work up to 1930}\label{sect:1920s}
When Kapteyn retired from his positions of professor and director at the end
of the academic year in 1921, he
actually left Groningen, never to return, and completely terminated all
involvement in astronomy in Groningen.
During the first decade of van Rhijn’s directorship, roughly the 1920s, the 
Laboratorium still flourished.
A very important part of the {\it Plan of Selected 
Areas}, the star counts in the {\it Harvard(-Groningen) Durchmusterung} and the 
{\it Mount Wilson(-Groningen) Catalogue} progressed vigorously.
The publication of the 
first was completed in 1924, encompassing  the {\it Systematic Plan} of the 206 
all-sky Areas. This had been a major investment of effort on behalf of 
the Laboratorium, which in the mean time had been named
`Sterrenkundig Laboratorium Kapteyn' or just `Kapteyn Laboratorium'.
The {\it Special Plan} of 46 areas in the Milky Way on which Pickering had 
insisted so much and which Kapteyn had been forced to accept to ensure
his collaboration, went along at a very slow pace 
and was published only in 1952. The Groningen part of the work on the Mount 
Wilson program was also completed expeditiously, but the problem of deriving 
accurate magnitudes using large reflectors took much longer to solve and
as we have seen, was a 
responsibility of Seares at Mount Wilson; still it was completed and published 
in 1930. 

In 1923 van Rhijn had followed up the first and second reports,
published by 
Kapteyn \cite{JCK1911}
on the progress of the {\it Plan of Selected Ares}, by a third 
report \cite{PJvRrep3}.
It was an almost one hundred page booklet, published by the 
Laboratorium. It was followed up with a fourth report in 1930 in the {\it 
Bulletin of the Astronomical Institutes of the Netherlands} \cite{PJvRrep4}.
So van Rhijn had picked up his responsibilities with vigor. 

But he did also some very significant research, some in providing additional 
observational material to the Plan, some in analysis relevant to the 
aim of deriving a model for the Stellar System. In this he was assisted by 
young astronomers. Upon his appointment the then vacant position of assistant 
had for a short period been filled by Jan Oort, who soon after this 
appointment had been promised a staff position in Leiden by de Sitter, and  
had left for a period of two years to work with Frank Schlesinger (1871--1943)
at Yale Observatory in New Haven to learn astrometry.
It had been filled subsequently by students preparing a PhD 
thesis: one year by Peter van de Kamp (1901--1995)
who defended a thesis on  {\it De zonsbeweging met 
  betrekking tot apparent zwakke sterren} (The solar motion with respect to
apparent faint stars) in 1924, and subsequently by Willem Jan 
Klein Wassink (1902--1994),
who submitted a thesis on {\it The proper motion and the 
distance of the Praesepe cluster} in 1927. Klein Wassink left astronomy and 
like so many young PhDs in astronomy became a physics and mathematics teacher, 
in this case in Zierikzee in the province of Zeeland in the southwest.
After him the 
position was filled by students that would obtain their doctorate after 1930, 
Bart Bok and Jean Jacques Raimond (for details see below).

\begin{figure}[t]
\sidecaption[t]
\includegraphics[width=0.64\textwidth]{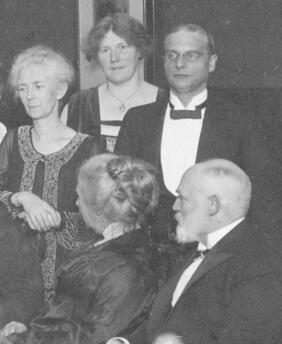}
\caption{\normalsize Pieter van Rhijn in 1926 at the dinner after the PhD thesis
defense of Jan Hendrik Oort, for which he acted as supervisor. Sitting at the
table are Willem de Sitter and his wife. Standing next to van Rhijn is Oort’s
(then future) mother-in-law and on the left Oort’s mother. From the Oort
Archives, see \cite{JHObiog}.}
\label{fig:JHOdiner}
\end{figure}

In addition to this van Rhijn (see Fig.~\ref{fig:JHOdiner})
had been `promotor' (thesis supervisor) of
three students, that had actually started under Kapteyn. The
first is Egbert Adriaan Kreiken (1896--1964; for an obituary see \cite{EAK}),
who completed a PhD thesis {\it On the color of the faint
stars in the Milky Way and the distance to the Scutum group}, in 1923.
Kreiken had been born in exactly the same house as Kapteyn in Barneveld,
where his parents had taken over the boarding school of Kapteyn's
parents after  the Kapteyn Sr. had died. After many peregrinations, Kreiken 
ended up in Ankara, Turkey were he founded the observatory which is now named 
after him. Jan Schilt (1894--1982; for an obituary by Oort see \cite{JHO1982}),
who had defended his thesis
{\it On a thermo-electric method of measuring photographic magnitudes}
in 1924, had moved to Leiden. Next
moved to the USA, where eventually he became chair of 
the Columbia astronomy department. Jan Oort had already moved to Leiden when 
he defended his thesis in 1926  on {\it The stars of 
high velocity}. 
\bigskip

In addition to work on the {\it Plan of Selected Areas} catalogues and 
associated matters, the coordination of the Plan’s execution and writing 
reports, and in addition to his teaching and supervision of students, van 
Rhijn did some very important and fundamental work. There are actually three 
papers in this category that I will very briefly introduce. The first concerns 
what we have seen was the first step to derive a model for the distribution of 
stars in space an that is the mean parallax of stars as a function of proper 
motion and magnitude in various latitude zones. In 1923 van Rhijn published an 
improved version of what he had published with Kapteyn in 1920, but now for 
spectral classes separately \cite{PJvR1923}. Secondly, after first having
discussed the relative merits and comparisons of trigonometric, statistical 
and spectroscopic parallaxes \cite{PJvR1925a}\footnote{\normalsize
Van Rhijn used ‘mean
statistical parallax’ in the title (note that ADS has omitted the ‘mean’).
Statistical parallax is different from secular parallax and this may be
confusing to some readers. I use the following. Usually 
statistical parallax is that for a group or cluster 
of stars for which both proper motions and radial velocities are known, where 
then the average parallax is derived from comparing the scatter in the 
radial velocities to that in the proper motions. It is therefore fundamentally
different from secular parallax from proper motions relative to the solar apex.
Van Rhijn has in the titles of the (sub-)sections in
the paper been careful to avoid confusion,}, van Rhijn
produced in 1925 an improved version of the luminosity curve of stars
\cite{PJvR1925}.

\begin{figure}[t]
\sidecaption[t]
\includegraphics[width=0.64\textwidth]{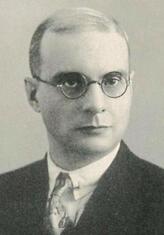}
\caption{\normalsize Pieter van Rhijn around 1930. Kapteyn
Astronomical Institute.}
\label{fig:vRhijn2}
\end{figure}

This was a very profound paper cited for a long time. It presented
what became known as the 
‘van Rhijn luminosity function’. Its importance today lies with studies of 
the Initial Mass Function (IMF) of star formation, where for stars below about 
one solar mass (that live longer than the age of the Galaxy)
it directly follows from this local luminosity function. ADS, 
the NASA Astrophysics Data System, still shows of order 3 reads per year over 
the last 15 or so. It has been of major importance. The ADS is highly 
incomplete in citations before the mid-1950s, so we cannot illustrate its 
impact using these.

And thirdly, in 1929 van Rhijn (Fig.~\ref{fig:vRhijn2})
presented a comprehensive set of star counts 
\cite{PJvR1929}. It was the final product of the {\it Harvard-(Groningen) 
Durchmusterung} and the {\it Mount Wilson(-Groningen) Catalogue},
the latter before it was 
actually published. It was presented as a table with positions at intervals of 
$10^{\circ}$ in Galactic longitude and latitude with the number of stars per 
square degree at integer values of photographic magnitude down to magnitude 18. 
Alternatively it was presented as a table of coefficients as a function of 
position for an analytical fit to the counts. The counts in the {\it 
Selected Areas}  in the Harvard and Mount Wilson data were for magnitudes 
larger than 10, between 6 and 10 van Rhijn used counts derived by Antonie 
Pannekoek using the {\it Bonner} and {\it Cordoba Durchmusterungs} and still 
brighter the {\it Boss Preliminary General Catalogue}. This together
constituted  a  major milestone in Kapteyn’s Plan. 

This would have been a point in time for van Rhijn to redo the analysis in the 
paper of Kapteyn and van Rhijn in 1920 \cite{KvR1920b} and derive an improved
version of the distribution of stars in space.
But in the mean 
time it had become a general feeling among astronomers that indeed interstellar 
extinction did exist, see for example Oort (1927, 1931),
waiting only for the final blow to the notion of the 
absence of absorption to be delivered by Robert Julius Trumpler
(1886--1956) in 1930. In any case, van Rhijn postponed 
such an exercise, at least felt it was premature to do so.
That the matter of interstellar extinction was definitely on 
his mind is clear from his paper in 1928, in which he studied the matter 
using diameters of globular clusters \cite{PJvR1928}. This was a clever and 
timely exercise. Remember that the notion of an  absence of absorption was 
primarily due to Harlow Shapley's (1885--1972)
study of colors of stars in globular clusters. Shapley
had found that, if Kapteyn's value were correct, the bluest
stars in globular clusters had 
to be intrinsically bluer by of order two magnitudes than those in the solar 
neighborhood and therefore had concluded that interstellar extinction was
negligible. Now, Arthur Stanley Eddington (1882--1944)
had proposed a mechanism in which no 
reddening would accompany extinction \cite{ASE26}. Eddington considered
`molecular absorption’, by which he meant dissociation of
molecules by starlight, which are continuously `disrupted and reformed’.
This is now known to
be much too inefficient. Van Rhijn decided to use diameters of 
globular clusters to investigate this. The formal solution for the amount of 
extinction he obtained was $.036 \pm .021$ magnitudes per kiloparsec,
which means very little if any absorption. Van 
Rhijn noted that this formal value would change
the distance of Shapley’s most distant globular 
cluster from 67 to 35 kpc, and then mentioned but did not discuss the large 
uncertainty of this.  With modern eyes this result would mean that he had 
found that indeed space was probably mostly  transparent towards the globular 
clusters. It is interesting to note that this approach of diameters of globular
clusters foreshadows Trumpler’s seminal analysis of open clusters two years 
later. 
\bigskip

\begin{figure}[t]
 \begin{center}
\includegraphics[width=0.98\textwidth]{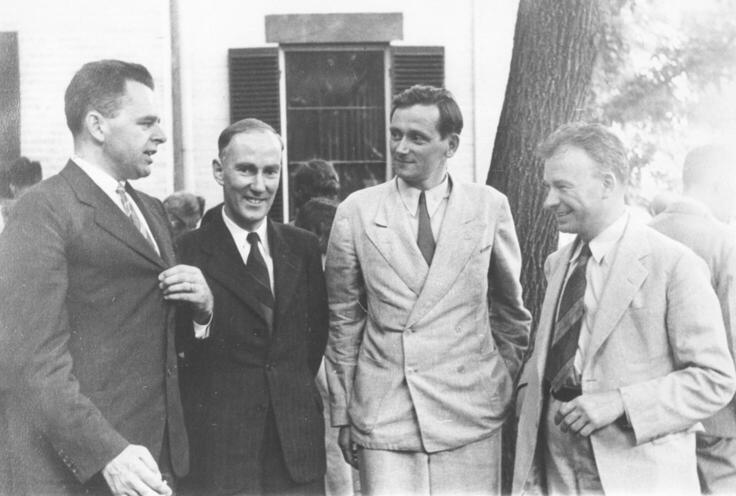}
\end{center}
\caption{\normalsize  Four students of Pieter van Rhijn in 1939. From left to
right Bart Bok, Jan Oort, Peter van de Kamp and Jan Schilt.  
Kapteyn Astronomical Institute.}
\label{fig:4studs}
\end{figure}

At the close of the decade of the twenties van Rhijn could look back onto a
productive period. The important  Harvard and Mount Wilson
catalogs in Kapteyn’s  {\it Plan of Selected Areas} had successfully been
completed and important first steps for an analysis towards a model of the
Stellar System, namely
star counts, mean parallaxes for the different spectral types
and a local luminosity function had been derived. Under his coordination the
Plan was progressing well. 

While during Kapteyn’s directorship 27 numbers of the {\it Publications of the
Astronomical Laboratory at Groningen} had appeared (in 22 years, but he
had started in 1900
with  backlog of a few), during the first nine years of van Rhijn’s
directorate (1921-1939) fourteen had been published.
In addition to this and the
Harvard and Mount Wilson Durchmusterungs, he had (co-)authored eight papers
in refereed journals, a progress report on the Plan (and prepared another one
to appear in 1930), and an obituary
about Kapteyn. In addition to the three PhD students he inherited from Kapteyn,
three PhD students had produced theses entirely under his supervision
and two more were on their way:
Bart Bok would finish in 1932 on {\it A
study of the $\eta$ Carinae region} and Jean Jacques Raimond on {\it The
  coefficient of differential Galactic absorption} in 1933 (see
Fig.~\ref{fig:4studs}).

There might have been other students as well that had an interest in astronomy
but did not pursue that, at least not in Groningen.
One was Helena Aletta (Heleen) Kluyver (1909--2001).
On March 15, 1929, van Rhijn wrote to
Jan Oort in Leiden (Oort Archives, website accompanying \cite{JHObiog},
Nr. 149; my translation):
\begmarg
During the Easter vacations Helena Kluyver, a very interested astronomy
student, will visit
Leiden. She wishes to see the instruments and to get an idea of the work
done at the Sterrewacht.
Can you help her a little?
\endmarg
Heleen Kluyver ended up moving to Leiden, where she remained in various
functions at the Sterrewacht, eventually obtaining a PhD (see \cite{JHObiog}
or \cite{JHOEng} for more on her).

And on January 12, 1931 Oort wrote (same source):
\begmarg
You will have heard that Heleen Kluyver has become half-assistant at the
Sterrewacht. She did obtain her Candidaats
[Bachelor nowadays] here. She is quiet and not very accessible, but when you
get to know her she is really a very nice
girl. In the beginning she may have been lonely in Leiden, so if the
occasion arises you may make some contact with her.
\endmarg

During the 1930s van Rhijn's  productivity in terms of number of
publications took a downward
turn and no new initiatives were taken. To examine the background and the
reasons for this more closely we have
to look at the program van Rhijn started soon after assuming the directorate
with the Bergedorfer Sternwarte at Hamburg and his attempts to obtain his own
telescope, an undertaking that Kapteyn had failed to successfully complete.
\bigskip

But before that I close this section by noting a remark Adriaan Blaauw made in
his 1983 chapter \cite{Blaauw1983}, where he points out that
(p.51; my translation):
\begmarg
\noindent
[...] in these days the Groningen laboratory still internationally was held in high
regard, is evident from the 1933 book {\it Makers of astronomy} by Hector
MacPherson, in which a prominent place was reserved for van Rhijn and his
institute. 
\endmarg
Hector Carsewell MacPherson (1888--1956), in his day, was
a highly respected astronomy teacher and
writer, especially of biographies of astronomers, see \cite{BG2011}. In the
final chapter of the book  \cite{Makers}, {\it Explorers of the Universe},
the persons presented by MacPherson are David Gill, Jacobus Kapteyn,
Willem de Sitter, Vesto Melvin Slipher (1875--1969), Arthur Eddington,
Harlow Shapley  and Pieter van Rhijn.
The latter, in MacPherson’s view,
is in the same class as Gill, Kapteyn, Eddington or Shapley. He
discussed in detail and praised the work leading up to the ‘Kapteyn Universe’,
giving part of the credit for it to van Rhijn, even designating it the
‘Kapteyn-van Rhijn Universe’. The Laboratorium in Groningen was staged as
‘the world-famed institution’. Maybe van Rhijn here gets more credit than he
deserves, but the high regard of Groningen astronomy is obvious.

\section{Bergedorf(-Groningen-Harvard) Spektral- Durchmusterung}\label{sect:Spektral-D}
The work in the preceding section is quite substantial. But another major
project had been started by van Rhijn, which however would turn out a major
investment of time with in comparison very little return. This is called
in full the {\it Bergedorfer Spektral-Durchmusterung der 115 n\"ordlichen
Kapteynschen Eichfelder enthaltend f\"ur die Sterne bis zur 13. photographischen
Gr\"osse die Spektren nach Aufnahmen mit dem Lippert-Astrographen der Hamburger
Sternwarte in Bergedorf und die Gr\"o{\ss}en nach Aufnahmen des Harvard College 
Observatory bestimmt durch das Sterrenkundig Laboratorium Kapteyn}. 

At the start of the {\it Plan of Selected
Areas} it was foreseen that spectral types would be determined
for as many stars as possible. Now, stars brighter than magnitude 8.5 were
classified as part of the {\it Henri Draper Catalogue}, published by Annie Jump
Cannon (1863--1941) and Edward Pickering in nine volumes of the
{\it Annals of Harvard
College Observatory}  between 1918 and 1924, containing about 250,000 stars
(and later extended with the {\it Henri Draper Extension}). As van Rhijn
remarked in the {\it Third progress report} on the
{\it Plan of Selected Areas} \cite{PJvRrep3}:
\begmarg
We thus have only to provide for the
classification of selected areas stars fainter than 8\magspt5.
\endmarg
The enormous
tasks of classifying a quarter of a million stars had been accomplished almost
entirely by Annie Cannon, which van Rhijn had used in \cite{PJvR1923} and which
had earned Cannon at Kapteyn’s proposal a
doctorate {\it honoris causa} from the University of Groningen in 1921
\cite{PCK2021}. In addition to this Milton Lasell Humason (1891--1972)
at Mount Wilson Observatory obtained
spectral types of at least ten stars per Area between magnitudes 11 and 12
with the 60-inch telescope of all 115 northern Areas, in total 4066 stars
(published in 1932, \cite{Hum32}). The aim, however, was to obtain many more spectral types.

Soon after he had assumed the directorship of the Laboratorium, van Rhijn had
approached Richard Reinhard Emil Schorr (1867--1951),
the director of the Bergedorfer Sternwarte near Hamburg,
and proposed to set up a joint project to fill the gap. Schorr accepted and
assigned Friedrich Karl  Arnold Schwassmann (1870--1964) to carry out the
program. Arnold Schwassmann had, after a PhD from G\"ottingen in 1893,
worked at a number of places, until he was appointed in 1902 on the staff at
Hamburg and was very much involved in moving the observatory out of Hamburg to
Bergedorf (completed 1912). Schorr had convinced
local businessman Eduard Lippert to finance an astrograph for the new site
rather than setting up a private observatory. It was a triple
telescope, consisting of a double astrograph with
in addition a telescope according to the {\it Carte du Ciel} specifications on
the same mount. Schwassmann, later together with Arno Arthur Wachmann
(1902--1990), used it for
hunting comets and asteroids. The instrument included an object prism that 
could be fitted to any of the three astrographs (see Fig.~\ref{fig:Lippert}).

\begin{figure}[t]
\begin{center}
\includegraphics[width=0.98\textwidth]{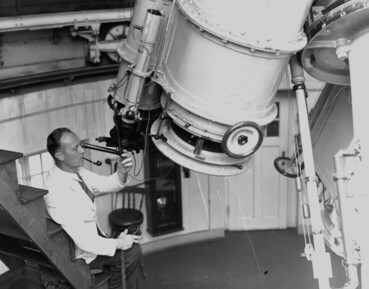}
\end{center}
\caption{\normalsize Arthur Arno Wachmann at the Lippert Astrograph, with
which the objective prism (and accompanying direct) plates for the
\textit{Bergedorf(-Groningen-Harvard)
Spektral-Durchmusterung} were taken. From \cite{Hamhist2016}, Hamburger Sternwarte,  reproduced with permission.}
\label{fig:Lippert}
\end{figure}

This survey aimed at determining spectral classes of stars in the magnitude
range 8.5 to 13. The method was to take photographic images of the sky with the
objective prism in front of one of the astrographs, so that each stellar image
was replaced with a  spectrum on the photographic plate. The other half
of the double astrograph was used to take a direct plate
of the same area at the same time so that the 
spectra could be directly related to the stars that corresponded to it. The
method was tested by taking plates in the same manner of stars with known
spectral type. The limit of the spectroscopic survey was magnitude 13.5 or so.

The fields obtained at the Lippert astrograph measured 1.5 degrees on a side, 
much larger than the 20 arcmin in the {\it Harvard(-Groningen) Durchmusterung}, 
so that it was necessary to determine apparent magnitudes
over much larger fields.
The Groningen part of the project involved the determination of accurate
positions and 
photographic magnitudes. For this, extensive new plate material was collected at
Harvard with the cooperation of director Harlow Shapley. These were measured
in Groningen using the photometer that Jan Schilt had developed, first in
Groningen as part of his PhD thesis work and improved while he worked at
Leiden Observatory.
Schilt developed this, as he noted in his thesis, following a suggestion by
Kapteyn. The practice after the start of the use of
photographic plates to determine magnitudes was to estimate the
diameter of the image. Schilt wrote
\cite{Schilt1922}:
\begmarg
At the request of Professor J.C. Kapteyn
I have made lately some experiments at the Groningen Astronomical Laboratory
concerning the determination of photographic
magnitudes by means of a thermopile and a galvanometer. The principle is this:
the image of a small circular lens is formed on
the photographic plate. The image of a star on the plate is brought to the
center of the circular bright spot projected on the plate
by the illuminated lens; the darker the image of the star, the smaller will
be the fraction of the light transmitted by the
photographic plate. This fraction is measured by means of the thermocurrent.
The reading of the galvanometer, which indicates
the intensity of this current, is thus a measure of the photographic magnitude
of the star.
\endmarg
A thermopile is a set of thermocouples operated usually (as in this case)
in series, a thermocouple being two different
materials that are electric conductors. When these have a different
temperature the energy difference gives rise to an
electric current (the Seebeck effect). The temperature difference results from
illumination through the photographic plate.
This current then is measured with a galvanometer and the strength of the
current can then be related to the amount of light transmitted by the
photographic emulsion. This requires a very good
thermopile, and Schilt had obtained an excellent one,  produced by Frits
Zernike, who after obtaining his PhD in Amsterdam had returned to Groningen in
the physics department. 

It is
illuminating to read Blaauw’s description in 1983 (\cite{Blaauw1983}, p.54 and
further; my translation):
\begmarg
The measurement program carried out [...] at the Kapteyn Laboratorium [...]
has greatly absorbed the capacity of the laboratory over a period of more than
two decades, probably much longer than van Rhijn had expected. The German
partner has discharged its task with full responsibility and German
`gr\"undlichkeit', but the cooperation with the Harvard College Observatory was a bit
more difficult. However, Shapley and his collaborators cannot actually be
blamed for this. The difficulty lay rather in the nature of the cooperation.
While in Bergedorf and Groningen the staff worked for one's own institute, it
was difficult for those at the Harvard College Observatory who were charged with the
recording of the photographic plates, always devoting themselves to it with
the necessary altruism. It went well as long as this was in the hands of pure
observers who, in the patient, careful nocturnal work with the telescope, find
full satisfaction, a mentality one finds more readily among the best of the 
astronomical amateurs than among the `professionals'. The
work at the Harvard College Observatory required more and more effort from advanced
students who were difficult to motivate and moreover, were quickly replaced
by new, inexperienced ones. No wonder, then, that the work, as I sometimes
heard afterwards, gradually got the name of `the damned van Rhijn program'.
That it turned out ultimately satisfactory, must be partly due to the
circumstance that van Rhijn's assistant Bok became a staff member of the
Harvard College Observatory and thus was able to keep an eye on things.
\endmarg

What we see here is a first crack (or one of the first) in the model by
Kapteyn of an astronomical 
laboratory relying entirely on observatories elsewhere for in this case plate
material. In part this may be the changing ‘zeitgeist’, a growing reluctance
to work predominantly for the benefit and in the 
service of others. What probably played a rol as well is that van Rhijn did
not quite have the unquestionable authority Kapteyn had had. Blaauw continued
(my translation):
\begmarg
[...] Bergedorfer Spektral Durchmusterung.
This is not a fortunate name, for the large Groningen share is not expressed
in it; in fact, the Groningen effort, counted in man-years, must have been
considerably greater than the German. But it was not in van Rhijn's
character to give
much attention to this external aspect; the main thing
was for him that the job got done.
\endmarg

In an article on the history of the Hamburger Sternwarte
\cite{Hamhist2016} it is stated (p.197):
\begmarg
In the 1920s multiple survey programmes began, among them the Bergedorfer
Spektral-Durchmusterung  by Schwassmann and
Wachmann, who used the Lippert 
Astrograph to obtain spectra of more than 160,000 stars,
\endmarg
\noindent
ignoring the positional part, the co-authorship of van Rhijn and the
collaborative nature with the Kapteyn Laboratorium and Harvard College Observatory
even though this is manifestly clear from the full title of the Durchmusterung.
Completing Blaauw’s remarks on the subject (my translation):
\begmarg
Many generations of van Rhijn's temporary
assistants we find back in his introductions to the five issues mentioned as
co-workers: in the first, of 1935, B.J. Bok, J.J. Raimond en P.P. Bruna; in
the one of 1938 P.P. Bruna; in 1947 me; in 1950 and 1953 L. Plaut. To the
measurement and calculation work participated G.H. M\"uller, W. Ebels, P.
Brouwer, P.G.J. Hazeveld, H.J. Smith, J.B. van Wijk, M. Seldenthuis, D.
Huisman, H. Schurer, M.L. Wagenaar and W.L. Gr\"ummer. During the War years,
cooperation with Bergedorf naturally became increasingly difficult.
Until it stopped completely. It must have been around 1943 that I myself, then
an assistant, was summoned to come to the office of a high German official in
Groningen to justify why Bergedorf no longer received any response from us
(van Rhijn was staying in a sanatorium with a
serious illness), but it was not difficult for me to point out to the
gentlemen that for some well-known reason the delivery of the plates from
Harvard had been interrupted. At that time in  Bergedorf they had very little
idea of the conditions prevailing here under the occupation.
\endmarg
\bigskip

Bruna is Petrus Paulus, born in Utrecht in 1906, who was assistant at
the Laboratorium from 1934 to 1939. He did publish some work in the
{\it Monthly Notices of the Royal Astronomical Society}
on the determination of the motion of the
Sun in space. The only thing in public records I could find
about him is that he got married in in 1940 and he must have left
astronomy probably to take up a job as high school teacher or so.
In the correspondence in the Oort Archives
there is one reference to him. On 15 January
1941, van Rhijn wrote to Oort (Oort Archives, website accompanying
\cite{JHObiog}, Nr. 149; my translation):
\begmarg
With Bruna it is a difficult case. He has done the routine work of the
Laboratorium as an assistant quite well, but he has no gifts for
scientific research. Added to this is the fact that he and his wife are ill
quite often and he has a busy job as a teacher. At the
moment he is busy writing down the work. But that is also going slowly because
the pieces he sends are poorly written. He has
been working on the research for about seven years. It is indeed high time
that the work is finished. I will write him again.
\endmarg
Bruna never completed the thesis.

A number of
the computers can be found in Fig.~\ref{fig:staff}. The HBS mentioned in the
caption is the 'Higher Civic School', instituted in 1863 by liberal Prime
Minister Johan Rudolph Thorbecke (1798-1872) as part of a major overhaul of the
Dutch educational system. It was designed to provide
education preparing upper or middle class boys for a career in industry or
commerce, but which as a result of the excellent training in science by
teachers with PhD's quickly became the preferred secondary school leading to
academic studies in mathematics and natural sciences.
\bigskip

 \begin{figure}[t]
\begin{center}
\includegraphics[width=0.98\textwidth]{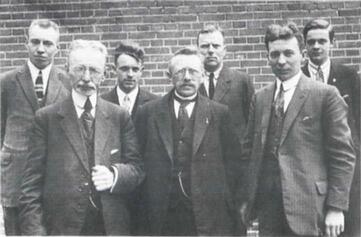}
\end{center}
\caption{\normalsize Computers and observators of the Kapteyn Laboratorium;
photograph taken in 1926 or 1927.  From left to right (caption taken from
Adriaan Blaauw in \cite{Blaauw1983}; my translation):
\textbf{J.B. van Wijk} -- appointed 1-11-1924 after HBS training and
military service, celebrated 1-11-1964
his 40th anniversary in office and in 1966 wrote his memoirs of the first
years.
\textbf{J.M. de Zoute} -- worked under Kapteyn and van Rhijn, initially
studied theology,
but had to terminate his studies to earn a living. Was a computer and wrote
letters for Kapteyn and van Rhijn in the final `neat' version
before sending. He was always
computing or writing standing at a high lectern (which in 1938 was still in the
`computer's  room'). Was released
in the crisis years due to university budget cuts.
\textbf{W. Ebels} -- appointed in the early 1920s, worked like van Wijk and
M\"uller until his retirement.
\textbf{T.W. de Vries} -- worked for 36 years under Kapteyn, then 6 years
under Van Rhijn, did in particular much measuring work.
\textbf{H.J. Smith} -- computer under van Rhijn, died in the War years.
\textbf{G.H. M\"uller} -- amanuensis, instrument maker, already worked under
Kapteyn, did measurement work and
was van Rhijn's right hand  in organizing the installation of the
telescope on top of the laboratory building.
\textbf{P. Brouwer} -- affiliated to the laboratory from Sept. 1925 to Dec.
1935, afterwards career with the police, 
helped A. Blaauw (March 1983) to identify the persons on this photograph.
  Kapteyn Astronomical Institute.}
\label{fig:staff}
\end{figure}

Eventually, the results were published in five volumes with the long title
quoted at the beginning of this section
by the Hamburger Sternwarte at Bergedorf between 1935 and 1953
\cite{Bergedorf}. Schwassmann and van Rhijn were the authors. The magnitude
range came out as between 8.5 and 13 photographic and the total number of
stars was 173,599. As Blaauw noted \cite{Blaauw1983}, the daily supervision of
the work had been the duty of the assistants; this was a full-time job, so the
preparation of PhD theses was restricted to the evenings and the weekends.
In this article I wish to honor Blaauw, van Rhijn and others in Groningen
and Harvard by
referring to it as the {\it Bergedorf(-Groningen-Harvard)
Spektral-Durchmusterung}.

It should be mentioned
here that a similar exercise was performed at the Potsdam Sternwarte near
Berlin as the {\it Spektral-Durchmusterung der Kapteyn-Eichfelder des
S\"udhimmels}, also in five volumes, published by Wilhelm Becker (1907--1996)
between 1929 and 1938,
partly in collaboration with Hermann Alexander Br\"uck (1905--2000). 

\section{Research based on the Selected Areas and the Spektral-Durchmusterung}\label{sect:PSAreserach}
One might ask whether these spectral surveys have been of much use to others 
and constituted a worthwhile investment of funds, effort and manpower. I would 
think to some extent yes, but also that in the case of Groningen it went at 
the expense of momentum in research.  The five volumes of the {\it 
Bergedorf(-Groningen-Harvard)
Spektral-Durchmusterung} together have collected in ADS of order one 
citation every two years at a rather constant rate since the 1950s. 
And since the start of the current millennium it
had and still has some ten reads per year. It certainly was not a major
advance, although it  definitely not a 
complete waste, but whether it was a wise decision in view of the future of the 
laboratory to make such a long term commitment of most of its resources is 
another matter. It is likely that the more useful part of the {\it 
Durchmusterung} was the spectral type information provided by Bergedorf rather 
than the magnitudes from Groningen. 
\bigskip

I stressed above that the {\it Plan of Selected Areas} was more than a provision
of fundamental data to the community at large. It was first and for all
designed to solve for a detailed model of the distribution of stars and their
kinematics and to study the dynamics of the Stellar System. After the `first
attempt' by Kapteyn in 1922 \cite{JCK1922} no further attempts had been made.
Van Rhijn had taken the
three next steps by refining the mean parallaxes for separate
spectral types \cite{PJvR1923}, determined a state of the art luminosity
function \cite{PJvR1925} and he had derived improved star counts in 1929 based
on the Harvard and Mount Wilson Durchmusterungs \cite{PJvR1929}, but
he stopped here and no
new model for the stellar distribution had been attempted. Undoubtedly this
had to do with the problem of how to allow for interstellar extinction.

\begin{figure}[t]
\sidecaption[t]
\includegraphics[width=0.64\textwidth]{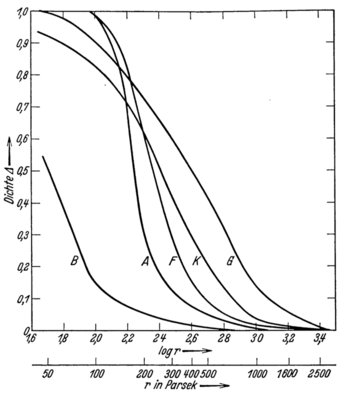}
\caption{\normalsize Density distributions (`Dichte $\Delta$')
  of stars of separate spectral class 
with distance from the Sun at latitude $+50^{\circ}$  as determined by van Rhijn 
\&\ Schwassmann  \cite{vRS35}, based on 22 \textit{Selected Areas} at roughly 
this latitude.  The values have been normalized at
the solar position. No distinction is
made between dwarfs and giants here, but this is done in tabular form
in the paper for G, K (and M)-stars. Kapteyn Astronomical Institute.}
\label{fig:RhijnSchwas}
\end{figure}

Since the aim of the {\it Plan of Selected Areas} was to elucidate the 
structure of the Stellar System, it is important to stress that van Rhijn did 
use the {\it Bergedorf(-Groningen-Harvard) Spektral-Durchmusterung} for this purpose. Probably since the 
effects of absorption could not very reliably be eliminated he and Schwassmann 
choose to study distributions of various spectral types at a Galactic latitude 
of $50^{\circ}$. Their paper \cite{vRS35} is in German, it collected only 11 
citations in ADS, of  which  9 after 1955 (the last one in 1985), and attracted 
about one read per year in the last two decades. It used determinations by van 
Rhijn of the luminosity functions per spectral type B, A, F, G, K and M, 
distinguished dwarfs and giants (and supergiants) by assuming reasonable 
ratios between these two for each apparent magnitude. The result is quite 
significant and has been  illustrated in Fig.~\ref{fig:RhijnSchwas}. They 
found that the layer of B-stars is thinnest, increasing with later spectral 
type.

This was not entirely new, earlier studies and in particular Oort’s 
well-known $K_z$ study of 1932 \cite{JHO1932} had shown this as well. 
But it is an important confirmation. In the 1930s it was not
understood at all how this came about. The explanation was
found only in the 1960s. Stars are born in the thin gas layer and have therefore initially the same distribution and mean
vertical velocity (velocity dispersion) as the gas clouds.
For any generation of stars this dispersion increases with time
due to scattering of stars by massive molecular clouds, spiral
structure or in-falling satellites and the average distance
from the symmetry plane will increase with time for each
generation. The sequence of stars corresponds to an increase
of mean average age and the effect reflects this so-called `secular evolution’.

I will elaborate on this study to show
the similarities and the differences
in approach and in scope with van Rhijn's work.
A more detailed analysis of the paper is given in my
Oort biography \cite{JHObiog}. Oort had used stars at high latitudes with
known radial velocities,  absolute magnitudes and spectral types (mostly from
the {\it Henry Draper Catalogue})  and he first determined the
dispersions of vertical velocities (the mean random motion). This showed no
significant trend with distance from the plane, especially for K- and M-giants,
on which most of his final analysis rested. He then used space distributions,
particularly from papers by van Rhijn and others, and used dynamics to find the
vertical force $K_z$ up to 0.5 kpc.  Oort then turned the problem around: if
for a group of stars (range of absolute magnitudes or of spectral types) one
knows the velocity dispersion, the force field can be used to calculate what
the vertical density distribution for that group of stars would be. By comparing
that to actual counts  he arrived at an improved form for the luminosity function.

But that was only the beginning. First he simulated star counts for fainter
stars, both apparently and intrinsically and compared
that to van Rhijn’s counts. When that was satisfactory he constructed mass
models of the Galaxy using a spherical mass around the center and a homogeneous
ellipsoid, satisfying constraints as the rotation parameters (rotation velocity and Oort
constants).
He then derived the gravitational forces, using these forces,
the corresponding star counts for
different assumptions of the velocity dispersions. He did (rather had his
computers do) this enormous amount of work first as a function of latitude
(concluding that above 15\degs\ no absorption effects were evident) and then
as a function of longitude. This resulted in two important deductions. The
first is the often quoted Oort limit, the total amount of mass in a column
perpendicular to the plane of the Galaxy at the position of the Sun,
  since he now knew the asymptotic
value for the vertical force. The second was a first approximation of the
stellar density in a vertical crosscut through the disk and the Galactic
center, at least away from the plane and extinction. This is reproduced in
Fig.~\ref{fig:crosscut1}. 
\bigskip

\begin{figure}[t]
\begin{center}
\includegraphics[width=0.98\textwidth]{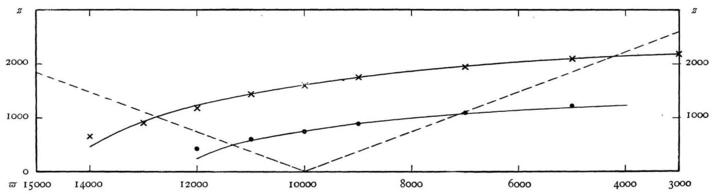}
\end{center}
\caption{\normalsize  Crosscuts through the Galactic System from Oort in 1932. 
Surfaces of equal density of 
starlight, as ellipse fits centered on the Galactic center at
1/25th (dots) and 1/100th (crosses) of that near the Sun. The Sun is at 
10000 pc, the center of the Galaxy at zero to the right. The dashed 
lines indicate latitude $20^{\circ}$. From \cite{JHO1932}. From the Oort
Archives.}
\label{fig:crosscut1}
\end{figure}

It is true that Oort in this work used many results of van Rhijn’s earlier
investigations and in fact the analysis would not have been possible without
it.  Van Rhijn \&\ Schwassmann’s conclusions
did constitute a significant strengthening of Oort’s conclusion as it
was based on many more spectral types of fainter stars. 
On the other hand, Oort’s work was far more imaginative and went much further.
He aimed at an 
understanding of the structure of the local Galaxy in terms of distributions, 
kinematics and dynamics, and the  possible
  presence of unseen matter, while van Rhijn and 
Schwassmann merely mapped distributions without addressing the deeper issue of 
understanding the structure of the Galaxy. Van Rhijn followed on the program of 
Kapteyn to collect the data, while  Oort
followed up on Kapteyn’s final analysis to use this to map the Stellar System.

\section{Interstellar extinction}\label{sect:extinction}

In addition to the distribution of stars in space, van Rhijn had a second main
line of research in studying interstellar extinction. The first half of his
thesis was devoted to this \cite{PJvR1916} on the basis of reddening of stars
with distance, and he had, as we have seen, in 1928 unsuccessfully tried
to determine this effect on apparent magnitudes on the
basis of diameters of globular clusters \cite{PJvR1928}. 

I will not review in detail the history of the discovery of interstellar
extinction. For a general listing of the chronology see \cite{Li}, and for
the argument why the discovery should be credited to Robert Trumpler, as it
usually is,  see \cite{SeelBe}.

After William Herschel had noted apparent holes
in the distribution of stars in the Milky Way, and Wilhelm Struve suggested
this to mean  that there was an apparent decline in the number of stars with
distance from the Sun, which could be due to absorption, it was Edward
Pickering’s consideration of the star ratio \cite{Pick1903},
that sparkled early in the
twentieth century the debate about extinction. The star ratio is the relative
number of stars between subsequent apparent magnitudes, which with fainter
magnitudes deviated more and more from the theoretical value of log~0.6 = 3.981
for a uniform distribution. George Cary Comstock (1855--1934)
noted in 1904 that the average proper motion decreased
much less rapidly with apparent magnitude then expected (at least at low
Galactic latitudes) and proposed that this was due to extinction
\cite{Com1904}. Kapteyn
argued from detailed modeling that the matter was more complicated and the effects
of changes in extinction and in space density were degenerate and
considerations of analysis of star counts could not resolve this.
He proposed that the observed effects were
a mix of the two. It was also Kapteyn who in the second of a set of two papers
{\it On the absorption of light in space} \cite{JCK1909}
suggested that it could be predominantly scattering and
that therefore the effects would be more severe at bluer wavelengths. He
subsequently actually found these effects of increasing reddening
with distance. So in addition to star
counts, proper motions and studies of the distribution of stars there now was a
new sign of extinction, namely interstellar reddening.

Extinction had been established beyond doubt by Trumpler using reddening
of stars in open clusters, comparing the distances of about one hundred open
star clusters determined from apparent magnitudes and spectral types with
those from angular diameters, assuming these are on average the same
\cite{Trump}. He derived an extinction of 0\magspt7
photographic magnitudes per kpc. This was
improved in several ways by astronomers, of which I take two examples without
wanting to suggest all the progress came from these two studies.  Former van Rhijn
student Peter van de Kamp -- in the mean time having moved to Leander
McCormick Observatory -- used an interesting method; he assumed it was zero at
high latitudes and then used galaxy counts as a function of latitude to
estimate average values of 0.8 to 1.4 photographic magnitudes per kpc
\cite{vdKamp1932}. Jean Jacques Raimond (for an obituary see \cite{JHO1961b})
wrote a thesis in 1933, {\it The
coefficient of differential Galactic absorption}, under van Rhijn in which he
compared colors of stars at high latitudes with those of stars of the same
spectral type in the plane. His result \cite{JJR1934} can be summarized as
follows. He used magnitudes (photographic on the Mount Wilson scale and visual
on the Harvard scale) and spectral types from various sources and estimated
distances from van Rhijn’s \cite{PJvR1923}
mean parallaxes as a function of magnitudes, proper
motions, spectral types and Galactic latitude or from spectroscopic parallaxes.
This gave him absolute magnitudes. From stars at high latitude he calibrated
the relation between colors and absolute magnitudes and then could find the
reddening for stars at low latitude. He  assumed a layer of absorbing material
with a thickness of 200 pc uniformly filled, and then found a reddening of
(converted to units used here) $0.50\pm 0.03$ magnitudes per kpc.

The problem remained what the corresponding amount of extinction would be.
This requires 
knowing the wavelength dependence of the scattering. Kapteyn had been the first 
to hypothesize that the signature of interstellar extinction would be a 
reddening of starlight with distance and he had assumed that the effect was 
similar to that in the Earth’s atmosphere giving rise to the blue color of the 
daylight sky. This is, however,
the result of scattering of light on very small particles such as {\it molecules}.
Most of this
is Rayleigh scattering first described by Lord Rayleigh (John William 
Strutt; 1842--1919) around 1870 (it is now known that Raman scattering
also plays a role,
which is inelastic while Rayleigh scattering is elastic, 
and results in exchange of energy). These processes occur 
from scattering by particles {\it smaller} than the wavelength, while it became 
clear that interstellar scattering really is from {\it dust grains} with sizes
{\it comparable} to the wavelength.
Consequently Kapteyn’s initial guess that the dependence was 
as for Rayleigh scattering, that is inversely proportional to the fourth 
power of the  wavelength, was incorrect; it actually is
in first approximation linear with the inverse wavelength. 

It was Alfred Harrison Joy (1882--1973), who cleverly derived the 
relation between reddening and extinction from Cepheids, which are bright,  
can be seen over large distances, have relatively small random motions, and 
whose absolute magnitudes follow from their period of the brightness 
variations. He used Jan Oort’s formalism in which on scales 
significantly smaller than the size of the Galaxy the radial velocity should 
increase linearly with distance as a result of differential rotation (see Sect. 
4.5 in \cite{JHObiog} or Sect. 5.4 in \cite{JHOEng}). Joy published this 
result in 1939 \cite{Joy}, but it was reported already in the 1932 annual 
report of the director of Mount Wilson Observatory.
In 1936, van Rhijn \cite{PJvR1936} 
followed this up by applying the method to open star clusters (with 
distances from the Hertzsprung-Russell diagram) and to F to M stars
(in addition 
to Galactic rotation using spectroscopic and secular parallaxes). This 
together resulted in average extinctions of 1.1 and 0.55 magnitudes 
photographic and visual respectively per kpc.

Van Rhijn, in the same paper,
used this to do a somewhat general mapping in the plane of the 
Galaxy, using the northern  Selected Areas at roughly latitude zero. 
Remarkably, beyond a kpc from the Sun there is a strong increase 
in density towards the center of the Galaxy, which is not accompanied by a 
rapid decrease in the anti-center direction. This would be consistent with the
Sun being located just outside a spiral arm, but that would have been rather
speculative and was not offered as an explanation by van Rhijn.
In fact, there is no discussion at all
on the broader view of our Galactic System or comparison to external galaxies.
The two studies by van Rhijn -- the one with Schwassmann on the distributions 
for different spectral types along a line of sight at high latitude \cite{vRS35}
and the one on 
distributions in the Galactic plane corrected with an assumed uniform 
extinction \cite{PJvR1936} --
provide no profound new insight into the structure of the 
Stellar System as a second step following and improving the 1920 analysis 
with Kapteyn \cite{KvR1920b}. It was not the case that this was impossible, or 
too early, van Rhijn simply did not address it.
\bigskip

There was one more thesis completed under van Rhijn during the 1930s and
that person would have been a better choice for the assistant position
than Bruna.
This was Broer Hiemstra (1911–??), who defended a PhD thesis in 1938
\cite{BH38}, but had not been appointed assistant.
He was born in Pingjum (a village south of Harlingen and close to where the
Afsluitdijk joins Friesland) as a son of an elementary school teacher.
He is not listed
among the staff of the Laboratorium in the ‘Jaarboeken’, the annual reports of
the University of Groningen, so he must have been supported by his parents or
been supporting himself. Hiemstra had obtained his doctoral exam in November
1936, but since 1934 the assistant position had been occupied by Bruna.
In the certificate of his marriage in Dokkum (Friesland)
in 1938 Hiemstra is listed as having the profession ‘teacher’ (according to a
newspaper clipping they were engaged to marry already in 1931!, also in Dokkum),
so a possibility is that, while working on his thesis, he supported himself as
part-time teacher, which turned into full-time after his thesis defense 
and made the marriage possible (he was 28 years by then and his bride
Arendina Maria Vlietstra, an elementary school teacher, 29). They settled in
Amsterdam where the young Mrs. Hiemstra tragically
died in 1940, one would think probably
in child birth. According to a photograph in the Den Haag Municipal Archives
(Nr. 1.21111), Hiemstra remarried and in 1976 at the age of 65 retired as
director of a large secondary school conglomerate there (the Hague).
So he pursued a
successful career in secondary education.
No more can be found on Hiemstra in official records or
newspapers, except the announcement of the death of his father in 1944.

The thesis was defended in 1938 and concerned  {\it Dark clouds in Kapteyn’s
Special Areas 2, 5, 9 and 24 and the proper motions of the stars in these
regions}, and this study is rather ingenious and
very much worthy of review here. It concerned
determinations of distances and amounts of extinction in dark clouds.
These Areas are four of the ones especially selected by Kapteyn for
this purpose

These properties had been
estimated before, usually from  counts of stars projected
onto the dark cloud compared to those in the immediate vicinity. The way to do
this is to determine from star counts and the luminosity function what the
apparent density along a line of sight is towards the cloud and one
or more just next to
it on the sky. The problem
using star counts is that this involves solving integral equations. Counts of
stars in a particular direction as a function of apparent magnitude, follow
from a summation. For each element of distance from the Sun along a line of
sight each apparent magnitude corresponds to a certain absolute magnitude. The
number of stars seen at that apparent magnitude from this element then is the
total density of stars there multiplied by the the fraction of stars of the
required absolute magnitude, which is the value of the luminosity curve at
that absolute magnitude. All such contributions from elements at other
distances along the line of sight, corresponding to different {\it absolute}
magnitudes, then have to be added up to find the total count of stars of that
particular {\it apparent} magnitude.
Mathematically this is an integral, which can
be evaluated in principle in a straightforward manner. But the solution
required is the inverse. Given the form of the luminosity curve and the star
counts at apparent magnitudes, the problem is to determine the total density
as a function of distance. That means having to
‘invert’ the integral and inversion of integral equations is notoriously
difficult. For the solution for the ‘Kapteyn Universe’
\cite{KvR1920b}, Kapteyn and van Rhijn had used Schwarzschild’s analytic method
for the special case that the luminosity curve could be approximated by a
Gaussian.
This method, which German astronomer Karl Schwarzschild (1873--1916), then from
the Sternwate G\"ottingen,
had actually described in a paper in 1912
\cite{KS1912}, would have played a role in
Kapteyn proposing Schwarzschild for an honorary doctorate from the
University of Groningen on the occasion of its tricentennial in 1914 (see
\cite{PCK2021}).

In the early 1920, Anton Pannekoek in Amsterdam had also addressed the
problem of the distance and extinction of a dark cloud  \cite{AP1921}. He used
Kapteyn’s density distribution and the luminosity function to compute the
theoretical decrease of the number of stars of different magnitudes for
different assumptions for the  distance and extinction in the cloud and
compared those to actual counts towards the cloud. In other words he solved
the inversion problem by trial and error.
Other major contributions to this field had come from Maximilian Franz Joseph
Cornelius Wolf (1863--1932) at the Landessternwarte Heidelberg-K\"onigstuhl
(\cite{Wolf1923} and subsequent papers in the {\it Astronomische Nachrichten}).
Wolf introduced what became known as Wolf diagrams, which compare star counts
for the obscured and  adjacent regions. This shows immediately at what
magnitude the curves begin to deviate, from which the distance may be
estimated, and the amount of absorption from at what magnitude the
curves become parallel again.  For an interesting review of the history of
this subject, see Pannekoek (1942) \cite{AP1942}.

All of this was based on counts of stars as a function of apparent magnitude
and solving the corresponding integral equation of statistical astronomy. The
innovative idea in Hiemstra's thesis was to use counts of proper motions and
solve the  corresponding equation, in which  the luminosity law is replaced by
the distribution law of space velocities. Hiemstra attempted to determine
distances and amounts of extinction of the dark clouds in these areas using
proper motions of stars in front of and next to these clouds. For this he 
measured proper motions from plates obtained at Radcliffe Observatory and made
available by its director Harold Knox-Shaw (1885--1970). 

The method he used is a very interesting one, so I will explain its principles.
But first I will, for those unfamiliar with this, explain the notions of velocity
ellipsoid and vertex deviation. Kapteyn used the effects of the solar velocity
on the distribution of proper motions to derive statistical distances (secular
parallaxes), which required random, isotropic motions. When he had found that
in addition
there was a pattern of two opposite Stellar Streams, Karl Schwarzschild quickly
realized that this could also be explained as an asymmetry in the velocities,
which would then be Gaussian with three unequal principal axes. Galactic
dynamics required the long axis of this velocity ellipsoid to be directed
towards the Galactic center, but Kapteyn’s Star Streams were directed some
20\degs\ away from this. This deviation of the vertex was not a fluke in
Kapteyn’s analysis, but convincingly confirmed and eventually attributed to
gravitational effects of spiral structure by Oort \cite{JHO1940}.

Hiemstra chose components of the proper motions
that line up as much as possible
with the long axis of the velocity ellipsoid as given by Oort (so not assuming
this points at the Galactic center, but taking the vertex deviation into
account) and at at least 66\degs\ from the solar motion in space, both
conditions ensuring a minimal effect of the parallactic motion. The
distribution of proper motions is a combination of the distribution of
parallaxes and that of the linear velocities and comparable integral equations
can be written down for counts as a function of proper motions as for those
of apparent magnitude. This function  follows from a study
of radial velocities and is assumed uniform with position in space.
Hiemstra inverted the relevant integral equation by trial and error
and in this way arrived at a distribution of parallaxes for the stars next to
the dark nebula. From this he determined what star counts would look like
taking only stars up to various maximum distances.
Assuming that the density of stars and their luminosities is
the same in the  obscured region and towards the dark nebula he then arrived
at an estimate of the distance of the cloud and the amount of extinction it
gives rise to. The four dark clouds were at distances ranging from 300 to
1000 parsecs and
their total extinctions at least 0\magspt5 to 2\mags in the photographic band.
This provided excellent, independent confirmation of results from other studies.

Hiemstra's paper has no citations in ADS but still averages 2 reads per year. It
seems to me the conclusions are still valid according to current insight and
constituted a very significant result, that probably did not  at the time
attract the attention it deserved. For example Pannekoek in his review in
\cite{AP1942} mentioned Hiemstra’s publication but only in passing, noting 
it as an illustration that his own
formalism can easily be adapted for proper motions. At
about the same time (1937) a study appeared by Freeman  Devold Miller
(1909--2000) of Swasey Observatory in Ohio, who used the ‘Pannekoek equation’
(the integral equation of star counts in the case of an intervening dark cloud)
\cite{Miller37}. Hiemstra compared his work to Miller’s and the results were
similar. Yet, the Miller paper currently has on average only one read per year.
The Wolf 1923-paper has become  a classic with some 1 to 2 citations and 15 or
so reads per year, while the 1921-paper by Pannekoek 
was cited for the last time in 1993. But it is not easily accessible;
it has about one read per year, probably only for the person involved
to find out that the text is not on the Web. Hiemstra’s thesis,
it seems, still is significant.

\section{Oort’s second approximation}\label{sect:Oort1938}

Actually the time was ripe for a new attempt to synthesize a model for the
distribution of the stars in space and the structure  of the Stellar
System, and it was Oort who took this step in 
\cite{JHO1938}! I quote from Oort’s historical discussion on {\it The 
development of our insight into the structure of the Galaxy between 1920 and 
1940} (\cite{JHO1972}, p.263):

\begmarg
I wanted to determine the general density distribution outside the disk as a 
function of Galactic latitude as well as of longitude, and thereby to extend 
Kapteyn and van Rhijn’s initial investigation to a second approximation. It 
was found that beyond about 500 pc on either side of the Galactic plane, the 
star density increased rather smoothly toward the center. It showed a 
logarithmic density gradient of 0.14 per kpc at levels between about 800 and 
1,600 pc  above and below the Galactic plane, and the equidensity surfaces 
made an angle of $10^{\circ}$ with this plane.
\endmarg

This paper had been summarized and discussed extensively in section 8.4 of 
\cite{JHObiog} (or 6.6 of \cite{JHOEng}),
so I will only give a brief summary here in order to 
demonstrate the ingenuity of the approach. When Oort started this investigation 
(well before the publication year of 1938) the {\it Plan of Selected Areas} 
had progressed quite well. At this point it is relevant to summarize the state 
of progress in the Plan at that time, say the middle of the 1930s. 

As for star counts, the {\it Harvard(-Groningen) Durchmusterung} down to 
magnitude 16 had been completed as well as the {\it  Mount
Wilson(-Groningen) Catalogue} 
of the northern Areas to magnitude 19.
The next step would be to determine 
spectral types and most importantly, proper motions. The spectral types were 
not fully published but mostly available from the Bergedorfer Sternwarte
for the northern fields and the Potsdam Sternwarte near Berlin for 
the south. Oort must have had access to all or most unpublished material from 
these two Durchmusterungs as far as available at any particular time, which 
meant for stars brighter than magnitude 12 or so. As mentioned above, for the 
northern Areas  Humason at Mount Wilson used the 60-inch
telescope to determine the spectral type of at least 10 stars per 
Area down to magnitude 13.5 or so. 

Another major advance was the publication in 1934 of {\it The Radcliffe 
catalogue of proper motions in the Selected Areas 1 to 115} by Harold Knox-Shaw 
(1885–1970) and his assistant
H.G. Scott Barrett. Radcliffe Observatory, which was part of 
Oxford University, had a 24-inch telescope; it was first used to try and 
measure parallaxes in the Selected Areas, but when that was unsuccessful, 
attention  was re-directed towards proper motions. This had been started in 
1909 under Arthur Alcock Rambaut (1859--1923), taking photographs in duplicate, 
one being developed immediately, the other stored for re-exposure at least ten 
years later. The second stage was started in 1924 under Knox-Shaw. The 
catalogue, published in 1934, contained proper motions of some 32,000 stars. 
The Oxford Radcliffe site was actually vacated in 1935 and the Observatory 
moved to a site near Pretoria, South Africa. Another important source for Oort 
to use was a study by Peter van de Kamp and Alexander Nikolayevich Vyssotsky 
(1888--1973) at Leander McCormick Observatory near Charlottesville, Virginia, 
which had a 26-inch refracting telescope. Van de Kamp and Vyssotsky had 
measured the proper motions of 18,000 stars with this telescope.

All these data together already constituted a substantial fraction of what
Kapteyn  must have envisioned necessary for an attack on the problem
when he conceived the {\it Plan of Selected Areas}. 

But before embarking on an analysis of the available material with the aim to 
derive a new picture of Galactic structure, Oort went through some important 
preparatory exercises improving estimates for the constant of precession
(vital for determining proper motions), 
the motion of the equinox and the Oort constants of differential Galactic
rotation\footnote{Oort of course did not call them that, but in his lectures
introduced them as `two constants of Galactic rotation’, to which he
would add `,which we will call $A$ and $B$’, hesitating a
few moments as if to make up his mind what he would decide to call them.}.
And from radial velocities 
and proper motions combined with estimated distances of spectroscopic and 
other parallaxes, he derived an improved estimate of the shape of the velocity 
ellipsoid (the distribution of velocity vectors).
This left the problem of the extinction. For this Oort used galaxy 
counts by Edwin Powell Hubble's (1889--1953)
mapping the `Zone of Avoidance', with which he was 
provided by Hubble before publication. The problem was to turn the distribution 
of galaxy counts into one of magnitudes of extinction. Oort did that by 
determining the colors of fainter stars in Selected Areas and comparing 
that to those of nearby, unreddened stars of the same spectral type, and  
found the reddening. This related galaxy counts to reddening, but he still
needed to transform reddening into the corresponding actual amount 
of absorption.  For this one would have to measure the brightness as a function 
of wavelength for the reddened and the un-reddened
stars and in this way the wavelength dependence of the 
extinction. Such a study had just been published by by Jesse 
Leonard Greenstein (1909--2002), who had calibrated spectra in terms of the 
brightness as a function of wavelength using photo-electric techniques, and 
had found the dependence of the absorption to be inversely proportional to the 
wavelength. Greenstein determined also the actual amount of absorption by 
comparing stars still visible behind dark clouds to stars just next to it, 
which also gave him the factor to turn color excess into actual magnitudes of 
extinction. 

From studies of faint stars Oort concluded that
\begmarg
\noindent
in all probability a 
considerable fraction, if not all, of the absorption derived from the counts 
of nebulae must take place within a relatively small distance from the Galactic 
plane.
\endmarg
Oort then felt justified to approximate the situation by assuming that 
the entire absorption indicated by the nebulae took place in front of the 
stars considered. He ignored the Selected Areas at too low Galactic latitude,
generally below 10\degs\ or so, but without any strict criterion.
So with that assumption and using van Rhijn’s luminosity
function, for each Kapteyn Selected Area 
he proceeded to calculate the star density along the line of sight at three 
distances from the plane. He found two important things. First a general and 
large-scale increase in stellar density towards the center of the Galaxy, 
and secondly, that there is symmetry between both sides of the Galactic plane.

\begin{figure}[t]
\sidecaption[t]
\includegraphics[width=0.64\textwidth]{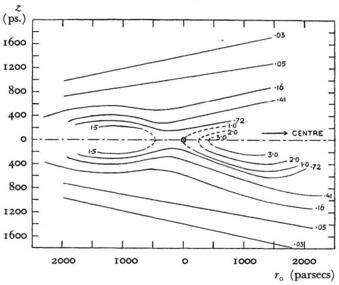}
\caption{\normalsize  Crosscut through the Galactic System from Oort in 1938. 
Surfaces of equal density of 
starlight, at various light densities. The Sun is at zero,
the Galactic center to the right at 10 kpc. From the Oort Archives.}
\label{fig:crosscut2}
\end{figure}

Oort continued by sketching from his data isodensity curves in a cross-cut 
through the Galaxy at the solar position in a plane at right angles to that of 
the Milky Way and along the direction towards the center. This has been 
reproduced in Fig.~\ref{fig:crosscut2}.
It looks very much like what would be expected if the 
Galaxy had a spiral structure like many external systems. The precise form of 
the ‘arms’ could not be determined, but Oort was relatively certain about one 
thing and that is that ‘the Sun should be between two spiral arms’. At larger 
distances from the plane the spiral structure is replaced with a general 
density gradient towards the center, amounting to a factor 1.398 in the stellar 
density per kpc. In modern terms this corresponds to a scalelength (e-folding)
of about 3 kpc; my preferred value is more like 4.5 kpc (see \cite{vdKF2011},
section 3.1.1), but this is open to different opinions.

This result is a rough, first approximation which quite well
fits current understanding. In 
Oort’s analysis, hot and bright O- and B-stars play a significant role and 
these are now known to be concentrated heavily to the spiral arms, where in a 
spiral galaxy star formation is preferentially taking place. So an analysis 
relying heavily on short-lived OB-stars probably
enhances the real contrast between arm 
and interarm regions. Modern studies have revealed the spiral structure in our 
Galaxy in much more detail. The Sun then is 
located near a local arc or 
branch between the major spiral arms, as found also in modern
studies (see e.g. \cite{Gon21}). Oort’s analysis is along the lines that 
Kapteyn must have had in mind when he defined the {\it Plan of Selected Areas} 
and Oort’s paper is therefore a major outcome of this Plan.
\bigskip

I entered into this somewhat detailed discussion to illustrate that
in the course of the 1930s the work on 
the Selected Areas actually had progressed to the point that an analysis 
as originally envisaged by Kapteyn was possible and that Oort actually had 
performed this. Of course more work needed to be completed and published, but 
I would argue that with this study by Oort the
primary goal of the {\it Plan of Selected Areas}
had been accomplished. What followed was to a large extent
provision of more catalogue data. After all,
in his presentation of the Plan
(\cite{SA}, p.1), Kapteyn wrote:
\begmarg
The aim of it is to bring together, as far as possible with such an effort,
all the elements which at the present time must seem most necessary for a
successful solution to the sidereal problem, that is: the problem of the
structure of the sidereal world.
\endmarg
It would seem that with adding more data no substantial improvement could be
achieved. A main reason for that is that the  distribution of the dust that
gives rise to Galactic extinction is {\it highly} irregular.
Van Rhijn’s 1936 study
\cite{PJvR1936} had failed to show anything fundamentally new and indeed Oort
\cite{JHO1938} does not even mention the outcome of this work. He refers to
this paper twice, in both cases related to B-stars and Cepheids being bright and
therefore more readily included
in samples selected at random. So, Oort’s paper and the
crosscut in Fig.~\ref{fig:crosscut2} is the final result of the {\it Plan of
Selected Areas} in terms of the goal set by Kapteyn. Significant progress
in the mapping of the stellar component of the Galaxy
became only feasible with the recent astrometric satellites Hipparcos and Gaia.
\bigskip

Although, as I stated, Oort’s 1938-study constituted to a large extent what
Kapteyn had defined as the aim of the {\it Plan of Selected Areas}, it was
not at all the end of the undertaking. In 1930 van Rhijn
had published the fourth
progress report in the {\it Bulletin of the Astronomical Institutes of the
Netherlands}, and after that it was taken up by the International Astronomical
Union as a dedicated Commission 32. Van Rhijn chaired this commission
this until 1958 and
edited progress reports in the tri-annual {\it Transactions of the IAU}
\cite{IAU2307}. In 1958 at a meeting of Commission 33 (Structure of the
Galactic System), chaired by Adriaan Blaauw, Nancy Grace Roman (1925--2018)
moved a proposal to incorporate it as a sub-commission to Commission 33, which
after being seconded by Bart Bok was carried. The new chair was Tord
Elvius (1915--1992); his last report to the IAU appeared in 1970. At the
corresponding General Assembly  in Brighton, Commission 33 President George
Contopoulos proposed to abandon the separate sub-commission and incorporate the
work in that of the
Commission itself. This was in fact the {\it formal} end of the {\it
Plan of Selected Areas}.

In addition to the IAU reports, two overview articles have been written on the
1960s that should be mentioned here. The first of
these is a very nice overview by Beverly Turner Lynds, discussing developments
up to 1963 \cite{L1963}, and the second a presentation in 1965 in the
famous {\it Galactic Structure} volume V of the {\it Stars and Stellar Systems}
compendium by Blaauw and Elvius \cite{ABTE1965}. 

In a sense the {\it informal} end of the Plan was marked
in 1953 at what became known as  the Vosbergen
conference, as I submitted in my Kapteyn biography \cite{JCKbiog}.
This meeting, which as number 1 marks the beginning of the IAU-Symposia
series,  was a tribute to van Rhijn and therefore was held near Groningen.
I return to this below. 

As I noted, significant progress
in the mapping of the stellar component of the Galaxy
became only feasible with the recent astrometric satellites.
Still, the {\it Plan of Selected
Areas} had a significant impact on Galactic studies, at least through the
1930s. This is not too obvious from the authoritative discussion of the
development of the field  between 1900 and 1952 by Robert Smith \cite{RHS2006}.
He mentions the {\it Plan of Selected Areas} only by its inception by Kapteyn
in 1906 and the latter’s first progress report in 1924, and does not cover
further studies of the distributions of stellar properties in its context 
during the 1920s and 1930s (such as e.g. \cite{JHO1932} or \cite{JHO1938}). 
\bigskip

\begin{figure}[t]
\begin{center}
\includegraphics[width=0.98\textwidth]{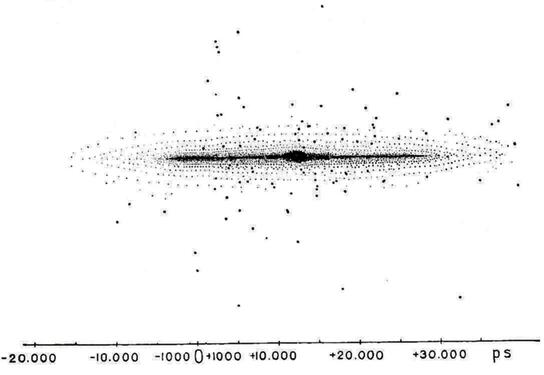}
\end{center}
\caption{\normalsize The structure of the Galaxy as it emerged towards the end
of the 1920s as summarized by Frederick Seares  \cite{FHS1928}. This drawing has
been published first by Willem de Sitter in his book \textit{Kosmos} of 1932
\cite{WdS1932}, and attributed there to `Dr. Oort'. He probably drew it earlier
than this. It was used and published by Jan Oort himself in his inaugural
lecture as Extraordinary Professor in 1936 \cite{JHO1936}. The scale is in
parsecs with the Galaxy extending from -15 to +40 kpc from the Sun.
From the Oort Archives, see \cite{JHObiog}.}
\label{fig:OortMW}
\end{figure}

\begin{figure}[t]
\sidecaption[t]
\includegraphics[width=0.64\textwidth]{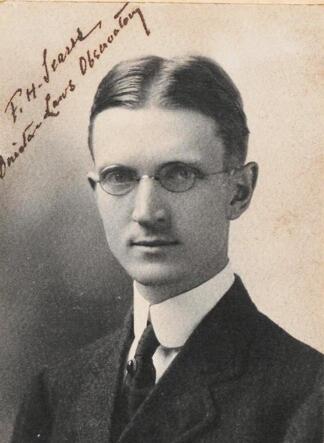}
\caption{\normalsize  Frederick Seares at the time of his involvement with
the \textit{Plan of Selected Areas} commenced. This picture comes from the
\textit{Album Amicorum}, presented to H.G. van de Sande Bakhuyzen on the
occasion of his retirement as professor of astronomy and director of the
Sterrewacht Leiden in 1908 \cite{Album}.}
\label{fig:Seares}
\end{figure}

The canonical picture prevalent in the 1930s and maybe some
years before was that the Galaxy consists of a flattened disk extending much
further then extinction allows us to see and a system of globular systems
of which the centers coincide and that it resembles extragalactic spiral nebulae.
Oort’s well-known sketch for Willem de Sitter’s book in 1932 in
Fig.~\ref{fig:OortMW} sums this up well.
This picture was preluded by various astronomers, such as Henri Norris
Russell (1877--1957) in an important review on the state of affairs at the
end of the 1910s on {\it Some problems of sideral astronomy} \cite{HNR18},
or somewhat later  Oort in his
inaugural lecture as privaat-docent in 1926 (Appendix A. in \cite{Legacy} for
an English translation). In an important paper from 1928 mostly coming out
of his involvement in the {\it Mount Wilson(-Groningen)
Catalogue} as part of the {\it Plan of
Selected Areas}, Frederick Seares \cite{FHS1928}, see Fig.~\ref{fig:Seares},
put it as follows (p.176 and
further):
\begmarg
As a directive aid to further research [...] lead to the following picture of
the probable structural relations in the stellar system: The galactic system
is a vast organization resembling Messier 33, although probably larger [...] It
includes a central condensation, scattered stars, and aggregations of stars
distributed over the galactic plane, [...] Like the spirals, the system also
includes diffuse nebulous material concentrated near the galactic plane, dark
and obscuring, or luminous if stimulated to shine by the radiation of
neighboring stars of high temperature. The central condensation, in longitude
325\degs, is hidden behind the dark nebulosity which marks the great rift in
the galaxy. Stellar aggregations, seen above and below the obscuring clouds in
the manner familiar in edge-on spirals, form the two branches of the Milky
Way, extending from Cygnus to Circinus. The diameter of the system is large --
80,000 to 90,000 parsecs, if it may be regarded as coextensive with the system
of globular clusters; [...] The sun is situated almost exactly in the galactic
plane defined by faint stars, but a little to the north of that determined by
Cepheids, faint B stars, and some other special classes of objects. Its
distance from the central condensation is perhaps 20,000 parsecs, [...]
Obscuring clouds conceal the great surface density which otherwise might be
expected in the direction of the center of the larger system.
\endmarg
Note that this precedes Trumpler’s ‘discovery’ of interstellar absorption.
Oort’s 1932 and 1938 papers significantly strengthened the understanding of
Galactic structure; van Rhijn’s work on the luminosity function was important,
but in terms of a fundamental understanding of 
the overall structure of the Galaxy his work was not groundbreaking. 

For completeness: By 1939 the distance to the Galactic center had shrunk to
10 kpc (but is still over 10 kpc in Fig.~\ref{fig:OortMW}),
e.g. on the basis of the review by John Stanley Plaskett  (1865--1941)
\cite{JSP1939}.

As Smith also noted in the article cited above \cite{RHS2006},
by the end of the 1930s the
hopes that further analysis of star counts, and the {\it Plan of Selected
Areas} would help further understanding of the structure of the Galaxy were
diminishing. Studies of light distributions and kinematics in external
galaxies, as Oort’s seminal, pioneering study of NGC 3115 \cite{JHO1940},
presented at the 1939 McDonald dedication, were promising in that he showed
the feasibility of photographic surface photometry, but the field
awaited larger telescopes and better
observing techniques to measure stellar motions in these systems. 

Star counts were not dead yet, though. Bart Bok still saw great promise in
them, especially when he and his Irish student Eric Mervyn Lindsay (1907--1974)
discovered and studied apparent windows in the obscuring layer along the
Galactic plane \cite{LB1937}. They wrote (\cite{BL1938}, p.9): 
\begmarg
Maybe the study in or near windows in the distribution of obscuring material
can help, if studies on the distribution of faint external galaxies go hand in
hand with studies on the distribution of colors and variable stars, spectral
surveys and star counts to faint magnitude limits on a definite photometric
system. For several years our ignorance concerning galactic absorption has
appeared to be the chief obstacle in the way of effective analysis of galactic
structure; this obstacle can now be removed. Our knowledge of galactic
structure can be further advanced if detailed studies of the distribution of
stars and nebulae are made for more or less obscured fields near galactic
windows as well as for the windows themselves.
\endmarg
But this program never produced much fundamental new insight.

\section{Van Rhijn’s plans for his own observatory}\label{sect:Observatory}
Already in the 1920s it had become clear that Kapteyn’s model of an
astronomical laboratory, an observatory without a telescope, had started to
show cracks. Blaauw (\cite{Blaauw1983}, p.57)
attributed this among others to the
difference in personalities between Kapteyn and van Rhijn (my translation):
\begmarg
Two character traits had made Kapteyn succeed: his exceptional gifts as a
scientific researcher and his engaging, optimistic personality. In these
respects, what may we say of van Rhijn? He was a thorough scientific researcher
and had a wide interest in the developments in astronomy, also outside his own
field of research. [...] His approach to research, however, was highly
schematic, strongly focused on further evaluation of previously determined
properties or quantities. This benefited the thoroughness of the intended
result, but came at the expense of flexibility and
reduced the opportunity for unexpected perspectives; as a result,
it did not lead to surprising discoveries.

Also, van Rhijn's personality was so different from what has been
handed down to us about Kapteyn; meetings with foreign colleagues, offering
the opportunity for casual discussion and the development of friendships
hardly occurred, because he was not very keen on travel; he liked  organizing
and conducting much of the work from behind his desk, admittedly very well and
carefully, but in an almost impersonal way that could unintentionally come
across as `pushing'. All in all, circumstances that made working with the
outside world not always run smoothly [...].
\endmarg

At the time of Kapteyn’s death and the birth of the IAU, or
to be precise the first
General Assembly, the {\it Carte du Ciel} and the {\it Plan of Selected Areas}
were integrated into the Union as dedicated Commissions. No such new
initiatives on this scale of participation were attempted after that;
in the 1950s the {\it National
Geographic Society -- Palomar Observatory Sky Survey} was performed at a
single observatory with an especially built telescope. Kapteyn had profited
from the existence of the Helsingfors Observatorion and Anders Donner, who was
interested primarily in supplying data for others: participating in the {\it
Carte du Ciel}, but in addition providing data for Kapteyn
to use for the latter's
research. No such opportunity presented itself to van Rhijn.  Major
observatories were no longer ready to act as service institutions merely to
supply large masses of observational material for others
(if they ever were). The story
of the grumbling astronomers at Harvard to take plates for the ‘damned van
Rhijn program’ illustrate this. The {\it Bergedorf(-Groningen-Harvard)
Spektral-Durchmusterung} was performed with a clear division of tasks.
Bergedorf did not supply observational material to Groningen, the observatory
obtained the spectra and classified them,
while Groningen relied on Harvard plates to
determine positions and magnitudes. 

Blaauw attributed the success of Kapteyn to both his extraordinary talent for
scientific research and his outgoing personality, a combination of
attributions van Rhijn did not
possess. This would explain both the prominence of the Laboratorium in
Kapteyn’s days as well as the decline under van Rhijn. I actually would go a
bit further: the model of an astronomical laboratory relying on others to
supply observational material, simply was not a viable one and was bound to
fail not only in the long run, but already as soon as Kapteyn had stepped down.
Kapteyn will have realized this when he presented his
{\it Plan of Selected Areas}, where in addition to the scientific goals quoted
above he put it forward as a means to secure funding, actually increased
funding, for his laboratory. As I put forward in \cite{JCKbiog} and
\cite{JCKEng}, Kapteyn
must have realized that the future of Dutch astronomy lay in
Leiden with a real observatory, which after the reorganization of 1918
consisted of three
departments lead by renowned astronomers. Groningen had no observatory, nor
the support the Sterrewacht had from its university and particularly from the
Ministry and was bound to become second rate.  It could survive only with a
long term commitment by a large group of observatories to his Plan,
concluded under Kapteyn's leadership.

Van Rhijn must have realized this also. His experience with the Groningen part
of the {\it Spektral-Durchmusterung} and the reliance on
Harvard made abundantly clear that the model of a laboratory
would fail. No wonder
that van Rhijn picked up what Kapteyn had failed to accomplish:
obtain his own telescope. 
\bigskip

I briefly recount the story of Kapteyn’s failure to obtain an observatory.
He was appointed the first professor in Groningen
with the teaching of astronomy as his
primary assignment as a result of a new law on higher education that
stipulated that the curricula at the three
nationally funded universities should be
the same. This roughly doubled the number of professors in the country and
this and  the stronger emphasis on research resulted in a comparable increase
in the funding of the universities. However, to a significant extent due to
negative advice by the directors of the observatories in Leiden and Utrecht,
Hendricus van de Sande Bakhuyzen and Jean  Oudemans, who did not wish to share
the resources for astronomy with a third partner, Kapteyn’s proposals to build
a small observatory in Groningen were rejected. This development made Kapteyn
give up hopes for an observatory and instead establish his astronomical
laboratory. In the 1890s he requested funding for a photographic telescope for
Groningen, proposing Leiden would concentrate on astrometry and Groningen on
astrophotography. However, van de Sande Bakhuyzen reversed his earlier
position, in which he questioned the usefulness of photography, and
refrained from Leiden participation in the {\it Carte du Ciel}.
He must have seen Kapteyn's proposal as a threat to Leiden losing its
dominant position in the country, and submitted a competing
proposal. This was significantly cheaper than Kapteyn’s since he only added
on to existing infrastructure and the choice was not difficult for the
Minister. 

Another possibility to secure regular access for Groningen to a telescope occurred
during the late 1910s, when rich tea-planter and amateur astronomer Karel Albert Rudolf Bosscha
(1865--1928) from Bandung, Dutch East-Indies, assisted by Rudolf Eduard
Kerkhoven (1848--1918), made plans to establish an observatory there \cite{LKBF}.
This was stimulated by
Joan George Erardus Gijsbertus Vo\^ute (1879–1963), who was
originally a civil engineer, but who had turned to astronomy and, without ever
defending an PhD thesis\footnote{\normalsize On September 21, 1922, Joan Vo\^ute acted as
leader of a joint German-Dutch solar eclipse expedition to Christmas Island to
the south of Java, for which the Friedrich-Wilhelms Universit\"at in Berlin --
now the von Humboldt-Universit\"at -- awarded him an honorary doctorate in
1923.},
worked as observer at Leiden and Cape Town observatories before joining the
meteorological office in the Dutch East-Indies.
Kapteyn felt this observatory should be managed by a national committee,
while de Sitter took the position that the Sterrewacht director (so he himself)
should be responsible for defining the research program and organizing the
interface with Dutch astronomy. De Sitter argued that the Minister’s attempts
at specialization and concentration could be diverted by agreeing that ‘Leiden
specializes in everything except the specializations of Groningen and Utrecht’
(from a letter to Kapteyn of May 24, 1920, see W.R. de Sitter in \cite{Legacy},
p.99), making the latter two in fact subsidiaries to the first.
In the end de Sitter’s coup failed and Kerkhoven  and Bosscha established the
observatory with the authority for the scientific program in the hands of 
{\it its} director, the first of whom of course was Vo\^ute.

De Sitter secured access to a telescope in South-Africa,
first through an agreement with the Unie Sterrewag in Johannesburg and later by
erecting a telescope there through funding by the Rockefeller Foundation in
order to lure Ejnar Hertzsprung to Leiden,
but this was not offered as a national
facility. Only after World War II would this situation be replaced by
establishment of national and international facilities. But until then Leiden
kept a firm grip on access to the telescope. And, as we will see, the Ministry
actively supported for many years Leiden’s dominance through its funding
decisions.

That this gradually changed after Word War II can be appreciated from the story
of the related  `Leids Kerkhoven-Bosscha Fund' 
(LKBF). In 1950, upon the death of his widow,  it became known that Rudolf
Albert Kerkhoven (1879–1940), son of Rudolf E. Kerkhoven and amateur
astronomer, had left a considerable sum of money to support astronomy in the
Dutch East-Indies. In the mean time this had become Indonesia and the result
eventually (in 1954) was the founding of the LKBF \cite{LKBF}.
A reasonable fraction of
the allocations that the Fund provided went, and still goes, to the Bosscha
Observatory. But the majority of the proceeds of the investments is spent on
grants to support Dutch astronomers; and, in spite of the ‘Leids’ in the Fund’s
name, astronomers from all Dutch institutes are eligible equally. Actually,
astronomers from Dutch universities other than Leiden, like myself, have
chaired the Board.
\bigskip

Van Rhijn revived the efforts to install a telescope in Groningen, but ran into
the same problems  Kapteyn had had before:
although having support from Curators,
being rejected funding by the government. This was not an isolated development.
The university of Groningen was systematically disadvantaged across the board
by the government.
In 1928 it induced Rector Magnificus Johannes Lindeboom (1882–1958), theologian
and historian of religion, to state in the Rector’s annual report on thr
`Lotgevallen’ (literally the report on the fates) of the university
(my translation):
\begmarg 
It is a fact that almost every year the vacancies caused by departures to other
universities have to be listed among the
misfortunes. [...] The fact that Groningen is a bit far —
actually really not that far — from the center cannot be changed. The
authorities in Groningen are doing their best to make Groningen a very pleasant
city to live in, and when the authorities in Den Haag [the Hague]
would be willing to do the same, there is nothing else [...] to worry about.
But they could, for example,
ensure that Groningen gets its fourth postal delivery back, and that it
as the third commercial city of the country and a scientific and
cultural center, no
longer is, as it is at the moment, put behind almost every national
establishment. All of this may — this in passing — be an explanation of the
fact that your Rector believed to do the right thing by, in a letter to Z.E.
[His Excellency] the  Minister of Waterways and Public Works, plead the
interests of the University in this matter. 
\endmarg

In the beginning of the 1920s significant budget cuts had been introduced when
the economy, dependent as it was on trade with Germany that suffered from the
aftermath of the World War I, stagnated, although stabilized later in the decade
\cite{KvB2017}. The situation worsened with the Wall Street crash of 1929. It
would be logical for the Minister to look for savings in the expenses on
astronomy, and with the existence of extensive infrastructure in Leiden, which
had been thriving after de Sitter’s reorganization of 1918, and that in Utrecht,
it would be understandable that any extra expenses on the small laboratory in
distant Groningen would be seen as an extravagance. However, the remoteness of
Groningen affected all of the university. In 1934/35 there was a (not for the
first time) seriously considered option to close one of the universities, and
that would then be the one in Groningen. An immediate extensive and intense
lobby prevented this \cite{KvB2017}, but it underlined the vulnerability of
the university. 

In view of all this it may come as no surprise that van Rhijn failed to receive
any support or funds for a telescope to be put on the roof of the
Kapteyn Laboratorium. I quote from Adriaan Blaauw’s article
(\cite{Blaauw1983}, p.58 and further; my translation):
\begmarg 
[…] van Rhijn wanted to have a little more control over the future of the work
himself by means of his own equipment. His wish was to install a telescope of
modest dimensions on the laboratory building.
Who would think that the fame which
the Kapteyn Laboratorium had acquired internationally — few scholars, in fact,
have contributed so much to the Dutch scientific reputation at the beginning
of this century as Kapteyn — would be rewarded by this low-cost plan is
mistaken, and can only look in astonishment at the spirit of the ministry’s
responses. After van Rhijn first had to acquire the telescope with the help of
private funds, even his attempts at ministerial level to obtain the funds
needed for the construction of the dome were rejected. In spite of the
intercession of the Curators of the university […], the ministry persisted in
its chosen position: there should be nothing from which it would appear that
Groningen astronomy could pretend to have a claim for its own equipment. […]
In 1929, in response to a  modern reader rather haughty letter from the
Ministry to the Groningen Curators, van Rhijn patiently and extensively
refuted the arguments put forward therein, but the university still got a
negative response, so he decided to look for private funds for the construction
of the dome as well.\footnote{\normalsize See the letter from Curators of the University to
van Rhijn dated April 27, 1929 (no. 617) with copy of letter from Curators to
the Minister of 10 April (no. 539) and of the reply of the Minister of 22 April
(no. 1592, department H.O.); Van Rhijn's letter of 4 May 1929 to Curators and
their message of December 16, 1929 (no. 2719) regarding the Minister's
dismissive answer; Van Rhijn's letter of 3 March 1930 to the Curators and their
reply of 17 July 1930 with copy of the Minister's letter of 14 July (no. 31621
dept. H.G.); all this correspondence in the van Rhijn Archives at the Kapteyn
Laboratorium.}

In March 1930 he could report to the Curators that he had succeeded. In 1931
the telescope, a reflecting telescope of 55 cm diameter, was installed.

The state of affairs described above must have been a bitter experience for
van Rhijn, and it was not the only one: also in attempts around that time to
bring the remuneration of his computers and observers and their future
prospects to the same level as elsewhere, especially in Leiden, he also
encountered this unreasonable rejection by the higher
authorities. What, for example, to think of the following introductory
remarks in a letter from the Minister to the Curators of 2 October
1929\footnote{\normalsize No.1498 Department of H.O.}:
‘With reference to the letter mentioned, I have the
honor to inform your Board that I am prepared to consider a further proposal
from your Board, on the understanding that, in my opinion, in applying the
salary regulation to the personnel of the Groningen Laboratorium, the
University should take into account the difference between the other
institutions and the Sterrewacht Leiden, also with regard to the grades to
which the personnel can be promoted.'\footnote{\normalsize From a copy attached to the letter from
the Curators to van Rhijn of October 24, 1929, no. 2429.} But he was too much
of a man of character to speak out about these matters to his co-workers at
the laboratory.
\endmarg
Blaauw’s reference to letters in the footnotes
in the van Rhijn archives no longer
is a valid one. As explained in the Introduction, these 
archives (at the initiative of Blaauw himself) have been transferred to the
central archival depot of the university and from 
there, after inventories and removal of duplicates and irrelevant parts, to
the municipal Groningen Archives for permanent 
keeping. In this process these letters have been lost.

\section{Realization of van Rhijn’s telescope}\label{sect:telescope}
So who funded the telescope? All I can find is the remark ‘private sources’.
There is no correspondence on the acquisition of the telescope at all in the
van Rhijn Archives.  Whether
private sources were individuals or funds is not known, so it could
in fact have been a kind
of crowd-funding {\it avant la lettre}. The only remark on the subject in the Oort
correspondence is a remark van Rhijn made in a letter to Oort, dated November 24,
1952 (more of this letter is cited in section 20; from the Oort Archives Nr. 149,
my translation):
\begmarg
Kapteyn has tried to establish a proper possibility for scientific research in
Groningen, which he has been refused. The only
possibility, which I saw to change this, was the purchase of a second-hand mirror,
which happened to be offered to me at the
time.
\endmarg
How and by whom this offer was made is not mentioned.
What is known is that in the end the {\it Kapteyn Fund} and the {\it Groningen
University Fund} bore most of the cost of the construction of the dome.
These funds might have contributed to the purcvhase of the telescope as well.

I digress a bit to describe very briefly the histories of these funds.  The
Kapteyn Fund was formally founded on February 21, 1925, by the signing of a
deed at the office of  a notary in Groningen. The first board encompassed the
directors of the astronomical institutes in the Netherlands: Pieter van Rhijn,
Willem de Sitter, Albert Nijland and Anton Pannekoek, but also a person called
W. Dekking. This was Rotterdam merchant Willem Dekking (1873--1923), who was a
friend of Kapteyn’s elder daughter Jacoba Cornelia (1880--1960). The latter  had
studied medicine and married a fellow-student in medicine Willem Cornelis
Noordenbos (1875--1954) and lived for a short period in Rotterdam. Decking had
been very much impressed by Kapteyn and offered his help managing the Fund and
soliciting donations. The deed mentioned a starting capital of ‘eleven hundred
sixty eight guilders and sixty-eight cents, brought together by Professor
Kapteyn during his life’, which corresponds to a current purchasing power of
18,650\textgreek{\euro}\footnote{\normalsize A useful currency conversion site is provided
by the International Institute of Social History \cite{IISG}.}.
This sentence survived in the latest update, of 2010, which I signed in my
capacity of the chairman
of the Board at that time. The current capital is of
order 70,000\textgreek{\euro}. The Fund is called the {\it
Sterrenkundig Studiefonds J.C.
Kapteyn} and aims to support studies of students or research by astronomers
from all over the Netherlands. It was and still
is not a particularly large fund and the
contribution to the cost of the dome must have been modest.

\begin{figure}[t]
\begin{center}
\includegraphics[width=0.468\textwidth]{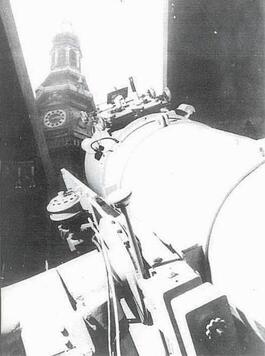}
\includegraphics[width=0.512\textwidth]{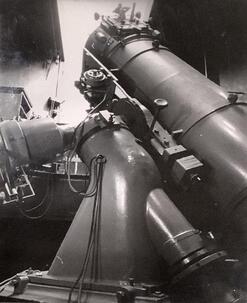}
\end{center}
\caption{{\normalsize Left: The 55-cm
telescope on top of the Kapteyn Astronomical 
Institute, pointing at the tower of the central Academy Building  of the
University of Groningen. The Laboratorium was located in the center of
Groningen just behind this main university building with its 66 meter high
tower, which is positioned to the southeast of the telescope at only some
20 meter distance (see Fig.~\ref{fig:air})}.
{\normalsize Right: View of the telescope and mount. Note that 
the spectrograph and plate holder at the Newtonian focus.   Kapteyn Astronomical Institute.}}
\label{fig:tel}
\end{figure}

\begin{figure}[t]
\begin{center}
\includegraphics[width=0.98\textwidth]{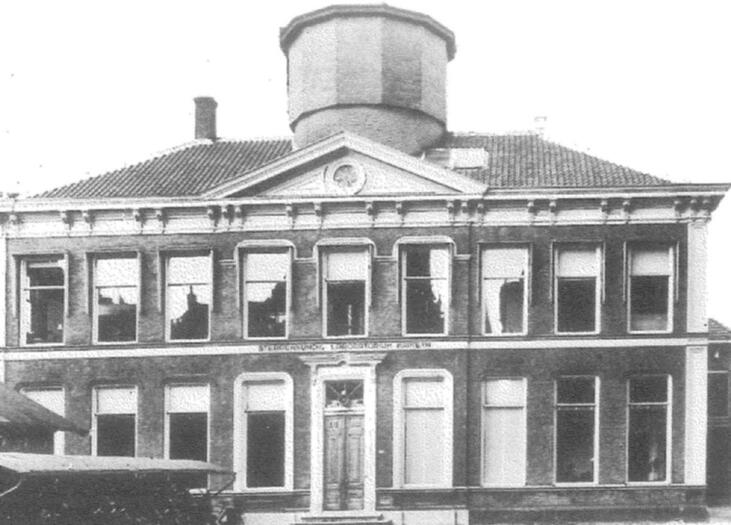}
\end{center}
\caption{\normalsize  The Kapteyn Astronomical Institute with the telescope
`dome' on the roof.   Kapteyn Astronomical Institute.}
\label{fig:labtel2}
\end{figure}

\begin{figure}[t]
\begin{center}
\includegraphics[width=0.98\textwidth]{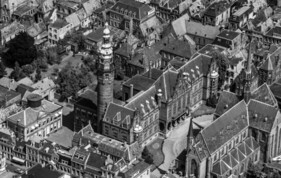}
\end{center}
\caption{\normalsize The Academy Building of the University of Groningen
(center) with the Kapteyn Laboratorium to the left with the
  cylindrical telescope dome on the roof. The photograph dates from 1952
  and has been taken from the southwest.
The telescope has been installed in 1931 and removed in 1959 (but the dome left
intact) and the building was demolished in 1988. The church on the right
opposite the
Academy  is the Roman Catholic Sint-Martinus Church, not to be
confused with the Martini Church and Tower near the Grote Markt.
It was demolished in 1982 and replaced with the Central University Library.
  Aviodrome Lelystad.}
\label{fig:air}
\end{figure}

The {\it Groningen University Fund} has a history \cite{GUF},
which is relevant to discuss here in the context of the  support
it gave to
van Rhijn where the Minister had failed to do so. The ‘GUF’ was founded on
March 4, 1893 by the professors of Groningen University, following examples
at the other universities, where in chronological order at Utrecht,
Am\-sterdam and Leiden such funds had established.
It grew out of a much older fund,
the ‘Professor’s Fund’, providing grants for promising students that had no
other means to enroll in  universities. It had been established in 1843 and in
addition to voluntary contributions from professors and parts of the tuitions
they received,
levied fines on professors that for no good reason had failed to turn up at
meetings of the Senate. A direct reason for replacing this fund with the GUF
was that the law on higher education of 1876, the same that was behind the
creation of Kapteyn’s professorship, had severely reduced the freedom to choose
where to spend the finances provided by the Minister. The GUF would, among
other things, provide a buffer to offset this. 

The new Fund hoped to increase its capital from private donations by professors
and other supporters of Groningen university. The original statutes state as
the aim (my translation):
\begmarg
The University of Groningen will use the funds for the promotion of study at
the University and of the flourishing of the University in the broadest sense.
To this end, the following will be considered: assisting with teaching materials
and financial support for talented students with limited means; expanding the
University library and the academic collections; promoting scientific research
and trips, to be undertaken by professors, lecturers, private tutors or students
at the Groningen University; temporarily employing private tutors to the
University in scientific subjects, which are not taught at a national
university, and promoting actions and measures, which may increase the prestige
of the University at home and abroad.
\endmarg 

The starting capital was 4000 Guilders, equivalent to nowadays about
55,000\textgreek{\euro}. In 1930 the capital had grown to about 123,000
Guilders (in current purchasing power a little over a million
\textgreek{\euro}). In more recent years (I was on the board between 1997 and
2015, seven years as treasurer and eleven as chairman) the capital varied due
to economic ups and downs and new donations between some 7 and 9 million
\textgreek{\euro},
while the returns on the capital were used mostly to co-fund foreign travel
related to studies of several hundred students per year and scientific
symposia, etc. in Groningen. The need to be dependent as little as possible
upon funding for research from the government was well justified, as the story
of the financing of van Rhijn’s telescope and dome
clearly shows. I continue now with the story of the telescope.
\bigskip

The ‘Lotgevallen’ for 1931 mention (my translation):
\begmarg
The Kapteyn Sterrenkundig Laboratorium has undergone a metamorphosis with the
construction of an observatory dome. The scientific work in this renowned
laboratory is based on determining the position of stars with the help of
photographs provided by domestic and foreign observatories. The director,
colleague van Rhijn, has purchased an astronomical reflector in order not to be
depending on others, being able to take  photographs
of the sky himself. The observer,
Mr. G.H. M\"uller, has shown himself capable not only of observation, but
also of designing what is required for that purpose. The costs of the dome
have been covered by private funds, mainly from the J.C. Kapteyn Fund and
the University Fund.
\endmarg

The telescope (Figs.~\ref{fig:tel}, \ref{fig:labtel2} and \ref{fig:air})
was a reflecting one
with an aperture (mirror diameter) of 55 cm and a focal length of 2.75 m,
and was originally planned apparently to be used for astrometric work.
Later it had a spectrograph in the Newtonian focus, that means
near the focus of the primary mirror the light was
reflected at a right angle towards the
telescope tube near its top, where at that location a spectrograph was mounted.
The van Rhijn Archives contain extensive correspondence with the Karl Zeiss
Company in Germany about auxiliary equipment such as a slitless spectrograph
(1936), and an ‘Astro-spectrograph for one or three prisms’ (1952), but the
latter file also contains a letter from Dr. J.H. Bannier, the director of the
Netherlands Organization for Pure Research, stating that it will not honor a
request for funding. There are also quotes for other extensions from  Zeiss.
The idea to use it spectroscopically seems to stem from the 1930s.

The only formal description of the telescope is in a few sentences in
a paper in 1953 \cite{PJvR1953}, not even mentioning the constructor,
but it is very likely to
be the Karl Zeiss Company as well. 
For this paper  \cite{PJvR1953}, the first one
based on material obtained with the telescope, spectra had been taken with a
slitless spectrograph.

\begin{figure}[t]
\begin{center}
\includegraphics[width=0.57\textwidth]{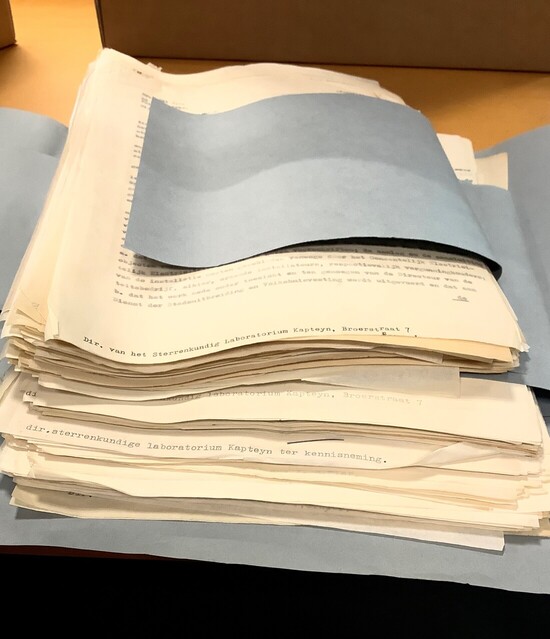}
\includegraphics[width=0.41\textwidth]{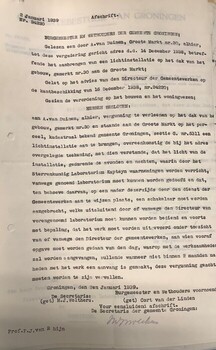}
\end{center}
\caption{\normalsize Left:
The stack of copies of the allocation of permits issued
by the Municipality of Groningen for outdoor lighting for advertising.
Right: One example of a permit. From the Archives of the Kapteyn
Astronomical Institute.}
\label{fig:permits}
\end{figure}

The location in the center of the city of  Groningen posed a major problem of
disturbing city lights, and when van Rhijn was intending to work
spectroscopically in particular those of advertising installations on the
outside of buildings, would be a serious problem.  Such observations
would be seriously affected by strong neon spectral lines. Van Rhijn 
arranged that special conditions were attached to the granting of permits by
the municipal authorities for such installations. The van Rhijn Archives
contain a large folder of copies of letters from the City of Groningen
concerning such permits (Fig.~\ref{fig:permits}).
These run from the early 1930s to well into the 1950s.
I estimate that this concerns some two hundred such permits. They
invariably contain the following passage (with sometimes minor variations on
this) (my translation):
\begmarg
... to grant permission [...]  to install a lighting installation in accordance
with the attached drawing on the understanding that the light of the
installation can be extinguished during those evenings and nights when
observations are made by Kapteyn Sterrenkundig Laboratorium, and that for this
purpose, at a location to be determined by the Municipal Works Department, a
switch shall be installed which can be operated either by or  on behalf of the
Director of the aforementioned laboratory and ...
\endmarg
According to Blaauw very little use has been made of this. It would be
something to see on a clear winter night van Rhijn or his assistant
taking their bikes and going
around the city to switch off all advertising before opening up their dome.

Maarten Schmidt, who was an undergraduate student in Groningen in the
1940s after the war, in his interview for the American Institute of Physics
Oral History Interviews \cite{Schmidt75} recollected:
\begmarg
It was a reflector of about fifty centimeters in diameter and it stood on the
roof of the observatory building which was next to the University in the
center of the city. I took spectra of blue stars, spectra of stars to study
the effects of dust in between the stars and us, [...]

That observatory, being in the middle of the city, they
had a funny system that was hardly ever used, I think, but they had gotten the
municipality to require that every building that had roof top advertisements
in lights would take care that there was a switch at ground level that could
be handled with a particular key so that if any observer found it necessary
to observe low in the sky in the direction, he would get a bike, go to that
part of the city, go to that particular part and turn off that darn light.
[laugh] I think we once used it or never did it. I have the inkling that we
once did it and I felt very guilty, you know, these advertising lights.
\endmarg

There even was the following addition to the formal Regulations:
\begmarg
\noindent
23 August 1949:
Regulation amending the Police Regulations for the Municipality of Groningen
[...] After article 113, a new article is inserted, numbered 114, and reading
as follows:\\
‘It is forbidden to have a illuminated advertisement, facade illumination or
any other light source on or in a building or in a yard, the light of which
radiating upwards could be a hindrance to astronomical observation by the
University or one of its institutions. This prohibition does not apply
to light sources for which a permit has been granted by the Mayor and Aldermen,
as long as the conditions attached to this permit are complied with.’
\endmarg
\bigskip

The question is what van Rhijn had in mind using the telescope for. 
Let me first quote Adriaan Blaauw \cite{Blaauw1983} (p.58 \&\ 59;
my translation)
and then fill in the details in the next section. 
\begmarg
Van Rhijn imagined to attack with the telescope especially the problem of the
reddening of the stars by the interstellar medium, and collected the necessary
spectrophotometric equipment in the course of the following years. However,
the work got off to a slow start. There was little technical assistance among
the staff, which had been reduced in the crisis years, and the scientific
assistant had to give attention to  the main measurement programs.
After the war, the work got somewhat under way and results of this color
dependence research were published by van Rhijn himself and by Borgman.
However, it never came to a substantial contribution, in the first place
because new photoelectric  techniques quickly superseded the photographic
ones. Moreover, the situation of the telescope, located in the middle of the
city, where city lights and smoke obstructed the sky, became an impossible
one. When, during my directorate, around 1959, the new wing of the university
building was being built and the walls of the laboratory that carried the heavy
telescope began to show cracks, the presence of a tall construction crane was
taken advantage of to lift the instrument out of the dome.
\endmarg
As Adriaan Blaauw told those interested on a few occasions (among
which myself)
he bribed the crane operator by presenting him with a box of cigars.
My colleague Jan Willem Pel, who worked at the Kapteyn Sterrenwacht
(which was established in
the 1960s as part of the Kapteyn Laboratorium) in Roden near Groningen informs
me: ‘... this mirror lay in a chest under the dome in Roden; after that it
ended up in the optical lab in Dwingeloo, where I
think it still is.’ The staff and workshop of the Kapteyn Sterrenwacht has
been moved to the Radio-sterrenwacht in Dwingeloo together
with a similar workshop and staff from the Sterrewacht Leiden to
form the ‘Optical IR Instrumentation group’ there. We will
meet Jan Borgman, mentioned here, extensively below.

\section{Adriaan Blaauw and Lukas Plaut}\label{sect:ABLP}

Adriaan Blaauw and Lukas Plaut joined the Kapteyn Laboratorium in 1938 and
1940 respectively;
Blaauw when a position as assistant became available, Plaut as a result of the
antisemitism brought along by the Nazi party and the German occupation of the
Netherlands. Blaauw was appointed assistant, which was normally
(and indeed in this case) associated
with the preparation of a PhD thesis. In order to put things in context I
have to backtrack a little.

In the years up to 1940 much of the resources and manpower was used to carry
the work on the {\it Spektral-Durchmusterung} with the Hamburg and Harvard
collaborators along.
As documented above, some other very significant and notable work was performed
as well, although no breakthroughs or major new insight had resulted from this.
That is not to say that nothing of significance happened.  I have already
mentioned the PhD  theses by Bart Bok and Jean-Jacques Raimond in 1932 and 1934
and Broer Hiemstra in 1938.
The position of assistant, occupied by Bok and Raimond when they had completed
their theses was filled by P.P. Bruna, who, as mentioned above, left without
completing a PhD thesis, and this opened up a position for Blaauw.
\bigskip

On October 1, 1938 Adriaan Blaauw (see Fig.~\ref{fig:NAC1941})
joined the Kapteyn Laboratorium as the new
assistant. Blaauw has himself summarized his biographical
details in his prefatory, autobiographical
chapter in {\it Annual Reviews of Astronomy \&\
Astrophysics} \cite{Blaauw2004}. Blaauw was born in Amsterdam, the son of a
bank auditor and a governess.  He entered Leiden university in 1932 to study
astronomy and had obtained in 1938 his ‘candidate’ degree (now called
Bachelor), but when he was appointed in Groningen, not yet the ‘doctorandus’
degree (equivalent to the current  Master). In
his AIP interview \cite{Blaauw1978}, Blaauw stated that he obtained the degree
in 1941 in Leiden  before the university closed. The University of Leiden was
closed in the fall of 1940 after the famous lecture by Rudolph Pabus
Cleveringa (1894--1980), protesting the dismissal of Jewish
professor of Law Eduard Maurits Meijers (1880--1954) and others.
However, exams were allowed again after April 30,
1941, but forbidden again after November 30, 1941.
Blaauw must have obtained his degree in this period.
For more on Leiden University during World War II,
see \cite{Ottersp}. The Germans gave Leiden students
permission to continue their studies at another university
and do their exams there.

Blaauw, as we saw, at first was involved extensively in supervising the ongoing
Groningen contribution to the {\it Bergedorf(-Groningen-Harvard)
  Spektral-Durchmusterung}, but
this slowed down after the outbreak of World War II and stopped completely
when the USA entered the war,  because of course no more
new plate material arrived
from Harvard. In 1942 he married Alida Henderika van Muijlwijk (1924--2007).
Because the work at the Laboratorium had essentially come to a standstill he had
ample time to work on  research for a PhD thesis, which would concern the
structure of the Scorpius-Centaurus Association. This is a star cluster 
which  contains many hot, young stars (of spectral types O and B) and 
in fact is a very young cluster at about
130 parsec distance. Blaauw used improved proper motions of the B-stars in
order to derive a better parallax for the association. He found that the group
was expanding and dissolving, but, as he stated later, he failed to see the
obvious fact that the association had to be have formed {\it very}
recently. I will return to Blaauw further on. 
\bigskip

\begin{figure}[t]
\sidecaption[t]
\includegraphics[width=0.64\textwidth]{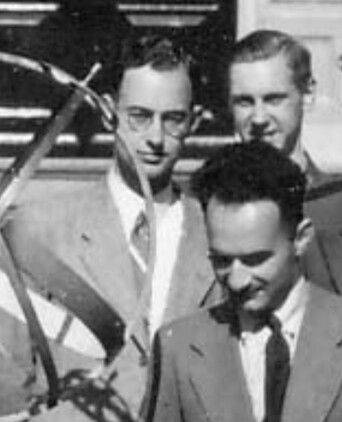}
\caption{\normalsize  Detail of the group photograph of the Nederlandse
Astronomen Conferentie in Doorn in 1941. The person on the
left is Adriaan Blaauw. In the front we see Lukas Plaut. From the
Website of the Koninklijke Nederlandse Astronomen Club, the Dutch Society
of professional astronomers \cite{KNAC}.}
\label{fig:NAC1941}
\end{figure}

Lukas Plaut (1910--1984) joined the  Kapteyn Laboratorium in 1940 after the
outbreak of the war. Since no autobiographical  notes from him exist I will
spend some more space here on Plaut than on Blaauw. A biographical article,
written by Barbara Henkes recently in 2020, focusing on his suffering from
antisemitism and based on extensive interviews
with and material provided by his widow in the 1990s, is very much worth
reading \cite{Henkes}. 
Adriaan Blaauw has written a short obituary of Plaut in {\it Zenit},
the successor
of the amateur periodical {\it Hemel \&\ Dampkring} \cite{Blaauw1978}. His
contribution to the {\it Biographical Encyclopedia of Astronomers} \cite{Bea}
is modeled on this. He wrote:
\begmarg
Occupation of the Netherlands by the German army in 1940 forced Plaut to move
first to Groningen (the Kapteyn Laboratorium, followed by a modest teaching
job), then to a labor camp, and finally to the concentration camp in
F\"urstenau, Germany.
\endmarg
This is a bit misleading, at least it  gives the impression
of  having been imprisoned for some extensive time in an extermination
camp.\footnote{\normalsize In an interview remininiscing on Plaut in 2010, a few months before his death,
Blaauw gives a much more balanced, but in details not fully accurate,
account (vimeo.com/16073067).}
I will detail the story here somewhat  extensively. Plaut's experience certainly
has been traumatic and this is not intended to take away from
this aspect.

Lukas Plaut (see Fig.~\ref{fig:NAC1941}) and his identical twin brother
Ulrich Hermann were born in Japan from Jewish, German parents. Their
father was a teacher of arts
and an art dealer;  the parents gave the children (there was
also a younger sister Eva) a liberal upbringing, hardly following Jewish
rituals and only attending the synagogue exceptionally.
In fact Lukas was not circumcised, which
may have saved his life later. When the children were teenagers, the parents
sent them to their native Germany.  Lukas went on to study mathematics and
physics at the Friedrich-Wilhelm Universit\"at in Berlin, after which in 1931 he
obtained a position at the Neubabelsberg Sternwarte in nearby Potsdam. The
boys must have continued their lives separately at some stage; the brother
became an architect, married in Shanghai in 1935 and moved to Los Angeles.
Lukas had in the mean time denounced Judaism, but this would not save him
from measures against Jews in Germany.
In April 1933 he was expelled from the observatory in the wake of the
growing antisemitism, resulting in a boycott by the Nazi’s of Jewish businesses
and suppression of Jews. With the
help of his mother, who, feeling he should move to the Netherlands,
came from Japan to visit the Sterrewacht Leiden with him,
as a result of which he took up a study in astronomy in Leiden,
financially supported by his parents. He studied with Jan Oort and Ejnar
Hertzsprung, eventually earning a PhD in 1939 under the latter on {\it
Photogra\-phi\-sche photometrie der veranderlijke sterren CV Carinae en WW
Draconis} (Photographic photometry of the variables CV Carinae and WW Draconis). In
1938 he had married a (non-Jewish) Dutch woman, Stien Witte (1911--2007).
This was not straightforward, as
it was forbidden by German law for a (German) Jew to marry an Aryan without
permission. Dutch authorities had adopted this measure (to avoid tensions
with Germany), but with the support and intervention of Jan Oort and a friendly
and helpful layer this was performed with a exceptional permit, but as
quietly (and quickly) as possible. When the permit came they were called away
from their work to marry within hours lest it
would be repealed, which robbed Stien Witte, much to her chagrin, the
opportunity to ever wear a wedding dress.

After the outbreak of WWII Plaut remained at first  in Leiden, where
Hertzsprung had arranged some modest allowance when transfer of funds from his
parents in Japan became impossible.  But increasingly measures against Jews
were introduced, resulting in Plaut being ordered in September 1940 to leave
Leiden and the neighboring Leiderdorp where they lived. On the long list of
places that were also forbidden for him, Groningen was lacking, so that is where
he went. His wife and the daughter they had by then came a few days later. This
whole affair turned Lukas Plaut from being a supporting husband in to one being
protected by his wife, as she now, being `Aryan',  had to take initiatives and
deal with authorities. They rang the doorbell of the van Rhijn residence, who
declined to provide them shelter (probably apprehensive of the possible
consequences), but showed them a list of places to rent modest rooms.
But van Rhijn gave Plaut the opportunity to enter the Laboratorium.
From Blaauw’s biographical notes it would follow he did this with some
regularity, even being paid for a little bit of  (illegal?) teaching.

But things got worse and worse. They apparently kept a stipend Hertzsprung
had arranged, which was however
barely sufficient to survive. At the insistence of his
brother in Los Angeles the Plauts filed for emigration to the USA at the
American Consulate in November 1940,
providing written guarantees from the brother and positive
recommendations from Oort and Hertzsprung. Yet the application was turned down
(the German authorities would still have had to agree as well, which also
happened rarely). In 1941 Plaut had to register as non-Dutch Jew, which could
mean deportation. Being married to an Aryan wife saved him from this however.
On the other hand,
because of their marriage Lukas lost his German citizenship and his wife
her Dutch, so they had become stateless! Of course as of May  1942 Lukas had
to wear the yellow star, identifying him as a Jew; in principle he could
be picked up any time in the street for deportation. So he and his wife  lived
as inconspicuously as possible, always terrified he might be
deported. In 1943 they had a second daughter.

However, the inevitable happened. In March of 1944 he was called upon to enter a
forced labor camp to work on the construction of  an new airport near the
village of Havelte, not far from Meppel, some 60 km to the southwest of
Groningen. The Germans had decided to build a new airport, `Fliegerhorst
Havelte',  to relieve Amsterdam Airport  Schiphol, which was the major airfield
in the country, and have a second airfield further to the east
for military purposes. The construction had
started in 1942. On satellite photographs one can still easily discern the
outline of the more than one kilometer long, east-west runway. He was allowed
to write  home regularly, reporting that the food was reasonable and sufficient
but complaining of fatigue and discomfort like flu, and actually was paid a
small salary. His wife was allowed to send him food, clothes and books, etc.
In fact the treatment of mixed Aryan-Jew marriages offered some protection.
But the strenuous physical effort took its toll, just like the lack of privacy
and the strain of the sharing of the sleeping quarters with
twentyfive noisy men which
he found difficult to endure. On September 5, 1944,  ‘dolle dinsdag’ (mad
Tuesday), when incorrect news spread throughout the country that the
Allied forces were about to liberate the Netherlands, guards flew and Plaut
returned to Groningen. Although the emerging airfield was well camouflaged,
the Allies of course knew about it pretty well and when
it was almost completed the airport
was destroyed on March 25,  1945 in a massive bombing attack, leaving  over
2000 bomb craters.

Plaut went into hiding in Groningen in the home of Johannes Gualtherus van
der Corput (1890--1975), a professor of mathematics. Van de Corput has been
active in the resistance movement throughout the War.
He survived and after the War he first
moved to Amsterdam and later the USA. Plaut was given  a new non-Jewish,
Aryan identity; he was supposed to be an agricultural engineer from the
School of Higher Education in Agriculture,
the predecessor of the current Wageningen
University \&\ Research. However, he no longer could send
letters to his wife, who even was not allowed to know where he was, although a
courier picked up his dirty laundry, brought it to her and after cleaning
returned it, and she provided the
necessary financial support. On February 23, 1945, a mere eleven weeks before
the end of the war, Plaut, van de Corput and a  third person 
were arrested following a tip the Germans had received
that there were persons in hiding at the address where he lived and
taken to the prisoner camp F\"urstenau.  Not as a Jew, but as an Aryan male
hiding to avoid the {\it Arbeitseinsatz} of Dutch men in Germany, the forced
labor to keep German (war) industry and economy going. His liberal Jewish
upbringing, resulting in him not being circumcised, supported his forged
identity and very likely saved him from being transported to an extermination
camp. As I noted above, Blaauw, in his contribution to the {\it
Biographical Encyclopedia of Astronomers} \cite{Bea} described F\"urstenau as
a concentration camp, and those citing him have copied that, but it really was
a prisoner camp and not an
extermination facility for Jews as this word would suggest. 

F\"urstenau lies some distance into Germany  from the Netherlands at the
latitude of Zwolle, at about 130 km from Groningen. After the liberation of
the camp some time in April 1945,
Plaut was released and he walked the distance to Groningen, appearing
unannounced at his wife‘s doorstep. She had not heard from
him since his arrest. In a letter of Plaut to Oort on May 19 (see
Section~\ref{sect:recover} below) he mentioned that he worked on a farm for a
few weeks and returned back in Groningen ‘three weeks ago,’
so this must have been before the German
surrender. Throughout the war
he had not been in immediate extermination threat in a concentration camp but
the danger of arrest and deportation had been hovering over him and his wife
the full five years of the war.
He survived this ordeal but never was able to free
himself from the trauma of the injustice, humiliation and indignation of being
subjected to all of this only because of his Jewish descent. I remember him as
a kind, soft-spoken man with an obvious all-consuming, but untold and
unresolved grief, suffering from the post-traumatic stres syndrome that
held him in its grip. His mental well-being deteriorated more and more when
he grew older. He died in 1984 from an heart attack likely not unrelated to
the terror and distress he had been subjected to. His twin brother had died 
already in 1971; his parents survived the War also, they moved the the USA,
probably before Japan entered the war. 

As a sideline:
when Lukas Plaut retired in 1975 I was offered (and I accepted) the vacancy at
the staff he left at the Laboratorium. I took over from him the introductory
astronomy lectures he had been giving for many years to first-year students
in physics, mathematics and astronomy.
He had obviously enjoyed doing that; the book he used to accompany his course,
{\it Abri\ss\ der Astronomie} by Hans–Heinrich Voigt, he translated with Nancy
Houk into an English edition in two paperback volumes \cite{Abriss},
of which he presented me with a copy with obvious pride.

\section{Van Rhijn and the Laboratorium during the War}\label{sect:WWII}
When the war
started, van Rhijn had been in the middle of his one-year term as
Rector Magnificus of Groningen University (see Fig.~\ref{fig:RM}). He 
completed his term before all the troubles with Jewish staff started.
In September 1940, when he was scheduled to present the Rector’s departing
lecture, he was expressly ordered
by the German authorities to avoid any reference to the political situation.
This lecture  concerned {\it The continuity in the development
of research in the natural sciences} (September 16, 1940). He mentioned that
this subject was chosen `in connection with a lecture course [...], given by
Professor Zernike on ‘Principle, direction and purpose of physics’.’ He asked
the following questions (my translation): 
\begmarg
But why should I believe what science teaches today when I know with a fair
degree of certainty that after only a few decades natural scientists will
reject our present views and perhaps teach the opposite of what we now believe
to be true? What value can one place on a science that has repeatedly
contradicted itself in the course of time and continues to do so to this very
day?
\endmarg
\noindent
His answer is continuity. Van Rhijn sums the characteristics of this up as
\begmarg
First, the present solution of a problem is built on the previous one
and forms the basis of the next solution.

Secondly, the older conception is understood in the newer theory as a special
case, so that for this special case the older conceptions are also considered
admissible according to the newer theory. 
\endmarg
He then proceeded to illustrate this with a description of the development of
our understanding of the motions of Sun and planets, from Ptolemy via
Copernicus, Brahe, Kepler, Galilei, Newton to Einstein.
\bigskip

\begin{figure}[t]
\sidecaption[t]
\includegraphics[width=0.64\textwidth]{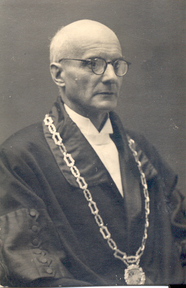}
\caption{\normalsize Pieter Johannes van Rhijn as Rector Magnificus
(1939--1940) of the University of Groningen.  He wears the chain of office,
decorated with the university’s coat of arms, which includes (in abbreviation)
the motto  \textit{ Verbum Domini Lucerna Pedibus Nostris} (the
word of the Lord is a light for our feet). From the
Website of the Stichting `Het geslacht van
Rhijn' \cite{PJgenea}, with permission.}
\label{fig:RM}
\end{figure}

As successor of van Rhijn  as Rector Magnificus, the Germans appointed a
sympathizer of the Nazi regime by the name of Johannes Marie Neele Kapteyn
(1879--1949) -- no known relative of Jacobus Cornelius. He had been professor
of German language and literature in Groningen since 1924 (later also Frisian).
In his traditional report on the `Lotgevallen’ of the University in the past
year, van Rhijn actually did refer -- in spite of the instructions -- to
the political situation. He apparently got away with it. It started as follows
(my translation):
\begmarg
\noindent
Ladies and Gentlemen,\\
Looking back on the past Academic year, one single coherent series of
events dominates our thoughts: mobilization of our armed forces to protect our
neutrality, war with Germany, capitulation of the Dutch army and the
occupation of the Netherlands by the German Wehrmacht. We have all suffered
bitterly under these battles and are suffering indiscriminately from the
consequences these events have had for our country and our people.
\endmarg

And he finished before transferring the office of Rector Magnificus to his
successor  with:
\begmarg
As Dutchmen, we have a reputation to sustain in the field of science
and the practice of science according to our best traditions is of the utmost
importance for the preservation of our independent national existence, which
we so ardently wish to maintain.

And finally, ladies and gentlemen students, if the courage to continue your
studies fails you because the future is too dark and uncertain, let me remind
you of the motto of our University: an open Bible with the words: {\it Verbum
domini lucerna pedibus nostris} [The word of the Lord is a light for our feet].
It is this Word that has been the support and strength of our leaders in very
difficult times.

{\it Verbum domini}. The Word that leaves us in no doubt about the situation
of this world that has turned away from God, but also opens a view far beyond
this world to the Kingdom of Heaven that has been and is to come, the Word
that calls us to work in the service of that Kingdom.

And what remains, with all the events of the world in which we find ourselves
today, is the classic phrase: {\it ora et labora} [pray and work].

My rectorate has come to an end and I now hand over the office of Rector
Magnificus of this University to my successor, Dr. J.M.N. Kapteyn. 

Dear Kapteyn. You are assuming the rectorate of the University in an eminently
difficult period of its existence. I express the wish that you may be granted
the wisdom to lead the University in such a way that it will fulfill its
mission also in these turbulent times. I ask you to come forward and while I
decorate you with the sign of your dignity, I greet you with the eternal wish
of salvation.

\noindent
{\it Salve, Rector Magnifice, iterumque salve!} [Hail, Rector Magnificus, and
hail again!].
\endmarg

After van Rhijn’s speech mentioned above and the transfer of the rectorate
van Rhijn and all, or at least most, of the
professors stayed away from the usual reception to congratulate
the new Rector and wish him well (see \cite{Berkel05}).

In his
correspondence with Oort, van Rhijn wrote (my translation):
\begmarg
\noindent
18 November 1940.\\
After the transfer of the rectorate Rein and I spent two weeks in Zuidlaren
and we very much enjoyed that. Beautiful weather and the surrounding
countryside is also very nice. It is easier to forget all the misery outside
than inside the city. Furthermore we were a week at the academic summer
conference in ter Apel. Approximately 200 students and professors were 
gathered there. It was a very successful attempt to bring more unity among the
students and also between the students and the professors. It is good that
professors and students do not only meet in the lecture room but also go on a
fox hunt together like in ter Apel.
\endmarg
This summer conference, which actually took place in July 1940 was intended
to be a, what in our current
parlance would be, team building exercise. As Rector van Rhijn had spent the
full week there, underlining the
need for good relations and unity among the academic community.
Vossenjacht or fox hunt is a type of scavenger hunt in which participants
must search for
foxes within a certain area.
The foxes are game leaders who walk around dressed in
distinctive clothing, for example of a certain profession. In this case the
foxes might have been the professors and students the hunters.

During the War the work at the Kapteyn Laboratorium came to an almost complete
standstill. Arrival of new plate material from
Harvard ceased, and the staff, mostly male computers, went
into hiding to escape internment as prisoner of war when having served in the
army or to avoid transportation to Germany in the context of the
{\it Arbeitseinsatz} (labor deployment), etc. 

In a letter of November 24, 1941,  Pieter van Rhijn wrote to Jan Oort in Leiden
(my translation):
\begmarg
There is so much misery and so much injustice that it is almost impossible to
bear. But we will not lose courage. One can tyrannize us and destroy all that
is valuable in our country. But if we remain courageous, they cannot change
our disposition and thus be  victorious over us. The strong language and the
compassion and the trust in God and in the Wilhelmus van Nassouwe can renew our
strength in these days. 
\endmarg
The ‘Wilhelmus’ is the Dutch national anthem. It would have been forbidden
even to mention this, but obviously both Oort and van Rhijn trusted each other
enough to do so anyway. Van Rhijn referred to God; he was religious
and a church-goer but Oort was not.

At first teaching at the universities in the Netherlands went on as usual. In
November 1940 the German occupation ruled that  Jewish professors had to
resign, which resulted in a well-known, strong protest in Leiden
mentioned above by Rudolph Cleveringa, at that time Dean of the Faculty of Law,
supporting his mentor, Jewish professor also of law, Eduard Meijers.
Students distributed transcripts of this lecture. As a result of
this Leiden University was closed and the possession of such a transcript was
declared illegal. Many students went to other universities, mostly Amsterdam.
Unlike Leiden University, the University of Groningen did not close during
the war.

In a letter of November 7, 1941, van Rhijn asked Jan Oort whether there were
any students in Leiden that would
want to do their exams in Groningen. A week later he wrote that a student van
Tulder has visited him. Johannes Jacobus Maria van Tulder (1917--1992) wanted
to obtain his doctoral (Masters) degree with van Rhijn in Groningen.  There is
some more correspondence from the next few weeks on this in the Oort Archives,
in which Oort recommends him for it. However, van Rhijn is worried that van
Tulder sympathized with the Nazi’s, since he had been seen by the personnel of
the Laboratorium with a well-known police officer that was a member of the NSB,
the Dutch Nazi party (NSB is the Dutch acronym of  the National Socialist
Movement). Van Rhijn wrote to Oort on this because this person (identified
only by his initial K) had betrayed various persons that had been transported
to Germany, and cautioned Oort to be careful in not mentioning this to van
Tulder. Fortunately, van Tulder is eventually cleared when it turned out that
the person showing him to the Laboratorium was another person with the same
surname (van Rhijn now identifying these persons as K$_1$ and K$_2$) living in
the same street, but this K$_2$  definitely was not a ‘colborateur’ of the
Germans. He had been identified incorrectly by the suspicious personnel of the
Laboratorium.

The {\it Yearbooks} \cite{Yearbooks}
of Groningen University lists van Tulder as having passed
this doctoral
exam in April of 1942. Oort and van Tulder published three papers in 1942
in the {\it Bulletin of the Astronomical Institutes of the Netherlands}, but
the latter dropped out of astronomy and eventually obtained a PhD in sociology
and became a professor in that discipline in Leiden. In Groningen no
other exams in astronomy were taken during the War. 

The Germans initially left the university alone. But in the fall of 1940
-- as already alluded to -- Jewish
professors were fired. Contrary to Leiden this was accepted resignedly. Next,
Jewish students were told to leave the university, which again was accepted
without major resistance. 

Teaching continued until early 1943, when students were rounded up for
transport to Germany as workforce to keep the German war economy going. The
students protested this time and stayed away from lectures. To turn he tide,
the Germans promised to leave them alone if they would return attending
lectures, but only after signing a loyalty statement promising no further
resistance to the occupying authorities. Only a small fraction (of order 10\%)
signed. On April 18, 1943, van Rhijn wrote to Oort about this:
\begmarg
The situation at the universities is uncertain and changes with the day,
sometimes with the hour of the day. In Groningen not even 10 percent of the
students have signed, elsewhere a little more. But in any case the declaration
is a failure. The Germans don't want to withdraw the declaration. The case is
hopelessly deadlocked and nobody can see a way out. It will be the end of
higher education in the Netherlands for the time being.
\endmarg
\noindent
And on May 8:
\begmarg
A few hundred students were deported to Germany. Only the NSB-ers, the girls
and the good Dutchmen who went into hiding, remain. I have no idea what will
happen to the universities now. 
\endmarg

The university remained officially open until the liberation in April 1945.
But teaching practically came to a standstill. There were no astronomy students
in the first place, but it should be remembered that for van Rhijn teaching
involved mostly general
astronomy lectures for physics and mathematics majors. And in the last years
of the War he was away being nursed for tuberculosis and formal replacement
was not really necessary. 
\bigskip

\begin{figure}[t]
\sidecaption[t]
\includegraphics[width=0.64\textwidth]{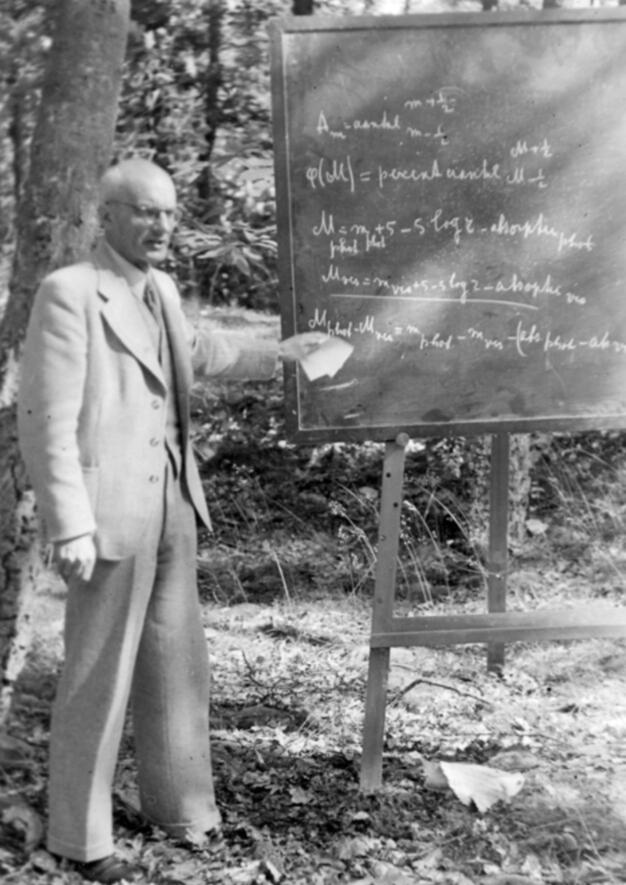}
\caption{\normalsize
  Pieter van Rhijn lecturing in the open air during the
second Netherlands Astronomers Conference in 1942. From the Website of the
Royal Netherlands Astronomical Society \cite{KNAC}.}
\label{fig:NAC}
\end{figure}

In the first years of the War there were still some national activities in
astronomy, such as one-day meetings of the society of professional
astronomers, the Netherlands Astronomers Club NAC. These were
supplemented by a new kind of meeting, in the beginning independent from the
NAC. The first Netherlands Astronomers Conference had been organized
in Doorn, a village not too far from Utrecht, in June
1941 by Marcel Minnaert, see Fig.~\ref{fig:NAC1941}.
These conferences were meant to bring together the Dutch astronomical
community, including students, for longer than the single half-day of meetings
of the Netherlands Astronomers Club (see my more
extensive discussion in chapter 9.3 of \cite{JHObiog} or ch. 7.4 in
\cite{JHOEng}).\footnote{\normalsize On the occasion of its centennial the predicate Royal
has been awarded. Contrary to my optimistic and hopeful expectation in
\cite{JHObiog} the change in the name from a ‘Club’ to the more professional
‘Society’ unfortunately did not happen, at least not in the Dutch name.
However the abbreviation NAC was changed into KNA, at
least removing the confusion with the Astronomers Conference.\label{note:KNA}}
A second meeting
in the same place was organized in July 1942. Adriaan Blaauw attended both
meetings, Lukas Plaut only the first (see Fig.~\ref{fig:NAC1941}),
and Pieter van Rhijn only
the second one (Fig.~\ref{fig:NAC}). Some more information and photographs
are available on the Website of the NAC \cite{KNAC}. After 1942 even such
conferences ceased to be possible,
but that was picked up again after the War and they are
still being organized annually, but now under the auspices of the KNA
(see footnote~\ref{note:KNA}).
During the war, meetings of the Club were being held every now and then,
sometimes preceded by special colloquia such as  the famous one in 1944
where the possibility of using the 21-cm line of neutral hydrogen in astronomy
was announced by Henk van de Hulst.
\bigskip

During the War, in November of 1943 according to his letter to his
sister-in-law in 1945 referred to in section~\ref{sect:Early},
van Rhijn started suffering from tuberculosis for which he was
hospitalized for a long time in a sanatorium. Well after the War, in 1948 or
1949, he returned
to his job (part-time before that), but his health remained fragile, and this
probably played a role throughout
the rest of his career. In relation to van Rhijn’s attitude after the War to
start radio astronomy in the Netherlands, Adriaan Blaauw has said in an
unpublished interview with Jet Katgert-Merkelijn (conducted in preparation
for what came to be my Oort biography \cite{JHObiog}; my translation):
\begmarg
So the last year of the War he [van Rhijn] was in a sanatorium. He was in
Hellendoorn and maybe in another place too. But when
he was admitted to that sanatorium, he forced himself to continue to provide some
leadership at the institute. I think
he actually could have retired quite well, because he had had a severe form of
tuberculosis after all.
But I always felt that -- because he had a very young family, he
married very late, so those children were younger even than the children of
Oort. In any case, that he forced himself to continue anyway, to put it
crudely, to keep his salary for that family. Well, so it was a man who really
couldn't handle much more than he did at the time anyway. And starting such an
adventure in radio astronomy, which was something very new, I think his mental
and physical condition prevented him from saying, ‘I'm going to participate
in that enthusiastically.’ 
\endmarg
Hellendoorn was one of the larger sanatoria in the Netherlands; it now is a
nursing home for persons suffering from dementia and comparable ailments.
It is the name of a nearby village 90 km directly to the south from Groningen,
15 km to the west of the city of Almelo, in a forest area with conditions
favorable for treatment of tbc. Van Rhijn was nursed there from December 1943
to December 1944, when he was transported back to Groningen (see below).
\bigskip

Through Gert Jan van Rhijn, keeper o the van Rhijn genealogy,
I obtained some more information on this period from
van Rhijn’s daughter. She mentioned that in the first months he was
much too ill to read, just was lying in bed in a small room. This
must have been very difficult. His wife was allowed to visit him
twice per month, but later during the War this had become
much too dangerous. He did have a few other people visiting,
like his brother Maarten, but this was exceptional. When this
became possible he wrote letters to his wife, about four per week on size A5
paper. By the time he was able to read he asked his wife repeatedly for two
books: {\it Geloof en openbaring} (Faith and
revelation) by G.J. Heering, who was a remonstrant clergyman, and
{\it Physik der Sternatmosph\"aren mit besonderer
  Ber\"ucksichtigung der Sonne} by A. Uns\"old. These could not
be sent for danger of getting lost, so only a visitor could
deliver them and visiting was difficult and rare. It shows
that he divided his interest between religion and science.

The Oort Archives contain a number of letters from van Rhijn to Oort during
the first’s treatment, that contain some more information on
his illness (Oort Archives, website accompanying \cite{JHObiog}, Nr. 149). On
March 18, 1944 (last letter before this is dated July 1943 before he fell ill)
he wrote from the sanatorium to Hulshorst, where Oort and his family were in
hiding (see \cite{JHObiog} or \cite{JHOEng} for more background to this;
my translation):
\begmarg
I am also doing well, according to the circumstances. My temperature has been
normal for weeks, the coughing is much better, and the sick spot in my lungs
has shrunk a bit. I can write letters and read very
simple reading material (Ivans for example) only a few hours a day.
But that is progress already.
Some months ago I was too tired to write or read anything.
\endmarg
Van Rhijn refers here to Jakob van Schevichaven (1866--1935), author of
detective stories, acclaimed as the first professional writer in this genre in
the Netherlands. His pseudonym, Ivans, was derived from
abbreviating his name to J. van S.

And on  18 October 1944 (same source; my translation):
\begmarg
Our expectations have since been disappointed and we have to be patient. I
would very much appreciate it if you could visit me sometime. But there can be
no question of that for the time being. Of course Rein cannot come either. It
is quiet lying in bed day after day. The sanatarium has no shortage yet. The
electricity is produced here. There still appears to be some supply of fuel.
Recently my blood was tested, the result was positive. I also feel better than
I did in the summer and less tired. Sometimes I even have the urge to get up!
The last few weeks I have been reading through Blaauw's dissertation. Every
now and then patients who lie here ask me astronomical questions, such as: How
do you weigh the Earth?
\endmarg
On December 18, 1944 van Rhijn wrote from one of the  three Groningen
hospitals, called the Diaconessenhuis (Deaconess Hospital). In addition to
this hospital for Protestants there was a Roman-Catholic and an academic
hospital in Groningen. Van Rhijn would choose for a hospital
for people of his religion if at all possible. Hellendoorn on
the other hand was a `volkssanatorium', `volks' (meaning  people's)
indicating open to all
religions. Part of the letter reads (same source again; my translation):
\begmarg
\noindent
Dear Jan,\\
Thank you for your letter of 3 December, which I received in Hellendoorn. I am
now back in Groningen in the Diaconessenhuis. The matter is that in the last
four weeks V2 flying bombs have been launched near the sanatorium. Several of
these failed; one of the failed bombs landed a few kilometers from the
sanatorium and killed 18 workers who were working in the fields. It became too
unsafe in Hellendoorn. With great difficulty Rein rented a car that
transported me to Groningen. It gives me peace that I am here now. It is a
wonderful feeling to be so close to Rein and the children. It was difficult in
Hellendoorn, especially the last few months when no more visitors came.., [...]

My illness is healing. At the end of November I was examined with favorable
results.  The blood is completely normal again and the sick spots in my lungs
have become smaller. The doctor who treats me here is also satisfied. When I
traveled in the car from Hellendoorn to here I had to get dressed for the first
time in 1944. You can understand my dismay when it turned out that my top
pants could not be closed at all anymore! There was a gap of 4 to 5 centimeters!
You won't recognize me when you see me again. I have become a fat puffy guy
with a double chin. This is a punishment from heaven because I used to look
down on fat people. It gives something smug about the whole person. And now I
am becoming such a fat person myself.
\endmarg
Traveling by car was rather dangerous as very often they were fired
upon from Allied airplanes, but
van Rhijn fortunately made it. It was only during the last few months
that his children were allowed
to visit him, but -- as his daughter recollects --
sitting in the window sill with the windows opened.
\bigskip

During the last two or three years of the War in effect only Blaauw and
observer M\"uller (and whenever possible Plaut)
kept the Kapteyn Laboratorium going. Blaauw
spent all his time on his thesis subject, but very little other research was
being done. In 1946 van Rhijn wrote a short article summarizing the
astronomical work performed in the Netherlands during the War \cite{PJvR1946}.
He mentioned his own work on using interstellar H
and K line intensities (from singly ionized calcium) as a distance criterion
and derived the density and luminosity functions of O and early B stars.
This was published after the War in the {\it Groningen Publications} and I will
return to it. Then there was Blaauw’s work already noted and a paragraph
on Lucas Plaut:
\begmarg
Dr. L. Plaut has worked at the Kapteyn Laboratorium since 1940. He has been
imprisoned for some time in a camp in Germany but returned safely after the
liberation. He has investigated a field of 10\degs x 10\degs\  for variable
stars with the blink comparator of the Kapteyn Laboratorium. The plates have
been taken by Dr. van Gent with the Franklin-Adams camera at Johannesburg.
\endmarg
So, Plaut was regarded as working at the Laboratorium since his arrival in
Groningen. 
\bigskip

In his autobiographical article in 2004 \cite{Blaauw2004}, Adriaan Blaauw wrote:
\begmarg
A curious incident occurred toward the end of the war. One day in February
1945, my doorbell rang and two lads of high school age, amateur astronomers,
asked to see me. With all the bad things the War brought us, the forced
blackouts at least provided amateur astronomers with a unique opportunity to
watch the sky. These lads now also wanted to observe the sky and speak to a real
astronomer, and of course they were welcome to enter. The lads were Maarten
Schmidt and Jan Borgman; both later had excellent careers in astronomy. As one
of them reported years later, my wife met them with the words, ‘You are lucky,
he has just been released from prison!’ Why had I been in prison? I had been
accused of listening to and spreading news broadcast from London -- something
strictly forbidden. Luckily, the War was approaching its conclusion, and
although nobody could tell how long it would continue, some lower-ranking
German officials were ripe for some gentle bribery. My relatives and friends
collected a sufficiently persuasive sum and succeeded in getting the prison
doors opened for me.
Two things I recall of that memorable release on February 8,
1945 -- it was my little son’s second birthday, and I could at last satisfy my
hunger, as far as circumstances allowed.
\endmarg

\section{Recovering from the War}\label{sect:recover}
The Second World War ended in the Netherlands on May 5, 1945. Soon after that
astronomers started looking forward, and in renewing contacts they did relay
to each other their experiences and circumstances.
The following comes from the
Oort Archives. Van Rhijn wrote Oort immediately on May, 5 (Oort Archives,
website accompanying \cite{JHObiog}, Nr. 149; my translation).
For clarity I remind the reader, that Joos is his elder sister, who lived with
her four children in Groningen (see section`\ref{sect:Early}).
\begmarg
Last night the news that the German army in
[the Netherlands] has capitulated; it is almost unbelievable
that for us the mistreatment by that execrable people has come to an end. There
was quite some fighting  here from April 13 to 16. Relatively few casualties
(about 100). But many houses burned, [...] Rein and the children and Joos and
her family came out of the fight unharmed; the houses only have glass damage.
[...] Blaauw and his family got off well. Plaut has miraculously returned from
Germany on foot, where he was imprisoned for a few weeks; he only had to walk.\\
\noindent
[long list of family and friends that did not survive]
    
I am doing well, the diet has been now so good that I have to eat a little less
to keep from getting too heavy! At the end of March there were no more bacilli
in my sputum, indicating recovery.

The laboratory has not suffered any damage, [...]. 

There was hard fighting in the vicinity of the Diaconessenhuis. The Krauts
shot from the garden of the house. It was dangerous for us though. We were in
the basement of the building.
\endmarg

\noindent
Adriaan Blaauw wrote Oort on May 15 (Oort Archives, website accompanying
\cite{JHObiog}, Nr. 161; my translation):
\begmarg
First, that Plaut did not fall prey to persecution, but resumed his work at
the Laboratory. It was a close call, however: [details the hiding and capture
of Plaut and his return and his own short stay in prison without mentioning
why he was arrested].

We went through very nervous days during the fighting around Groningen. Much
has been destroyed in the city, and with more unfavorable winds the Laboratory
would also have been affected by the fires. Now we got away with a few bullet
holes. The staff that had been in hiding is all but one present again, so that
the programs have been resumed. I myself am busy preparing my article on the
Scorpio-Centaurus cluster for the printer. Unfortunately Hoitsema's entire
printing facility has burned down, so printing is suspended waiting for attempts
to get help. Nevertheless, I hope that the work can be started within a few
days. Also burned in the fire are the 150 pages of the Special Areas
and the two publications of Prof. van Rhijn, completely corrected.
\endmarg

\noindent
And Lukas Plaut on May, 19 (Oort Archives, website accompanying \cite{JHObiog},
Nr. 163; my translation):
\begmarg
\noindent
Dear Professor Oort,\\
Now that the postal connection seems to be getting a little better I would
like to write to you in a little more detail. Since last year
March it has not been
possible for me to do anything for astronomy. I had to go to Havelte to
work on the airfield. In September I was discharged and went home again. Many
were rounded up at that time and taken to Westerbork, moreover, there was
compulsory digging for all men, so I preferred to go into hiding. However, in
February I was arrested and after a few days taken to Germany. Compared to
many I had a good time there, mainly because it was not
discovered that my identity card was false. When the Allied armies arrived, I was released and
worked on a farm for a few weeks before returning home. That was three weeks
ago now. My wife and the children have fortunately come through the last few
months well. The house did not suffer from the fighting either. 

Since September my wife has received financial support from an illegal fund.
The fund has now been absorbed into the `Nationaal Steunfonds' [National
Support Fund], of which I do not
dare knock at the door, as so many need to be supported who have lost
everything. Now I would like to ask you about the possibilities of getting a
job as an assistant or another scholarship. If at all possible, I would like
to continue working at an observatory. I do not believe there are any other
possibilities for me.
\endmarg
The `illegal' fund was in fact illegal in the sense that the underground
resistance movement had arranged this to support persons that had no income.
The National Support Fund had been set up by the Dutch government upon its
return to the Netherlands.
\bigskip

When the War ended in 1945 slowly business was being  picked up again.
In 1956 van Rhijn would turn seventy, so he still had a bit more than a
decade to go. But he had not yet recovered from
tuberculosis and did not fully return on his job. The other personnel did
get back to the Laboratorium. For many years there were in addition to the
professor and assistant four persons on the staff, of which one
was amanuensis (to copy letters or manuscripts  from drafts), one observer and
two computers. Three of these were already at the Laboratorium before the
partial closure and returned. Already before the liberation Oort had offered
Adriaan Blaauw a position in Leiden when things would be back to normal.
Blaauw accepted and started his appointment in October 1945. He finished his
thesis {\it  A study of the Scorpio-Centaurus cluster} and defended it in
Groningen in July 1946. 

What about Lukas Plaut? Henkes \cite{Henkes} noted:
\begmarg
The much-praised ‘modesty’, which Lukas Plaut demonstrated immediately after
his flight to the Netherlands, had not diminished in the years of his
persecution. Making himself invisible had become second nature to him,
reinforced by the advice to the few Jews who had survived not to draw
attention, ‘to show modesty, and to be grateful’.
\endmarg

Lukas Plaut had asked Oort in his letter in May for support with finding some
kind of employment. Oort had taken this op by recommending him to van Rhijn as
assistant, now that the position had been vacated by Blaauw.
In the Oort Archives there is
an envelope on the back of which Oort has made a note for himself that he had
replied on June 14, 1945 to the personal letter from van Rhijn of May 5 (see
above) and after some personal matters he wrote (Oort Archives, website
accompanying \cite{JHObiog}, Nr. 149):
\begmarg
As quickly as possible proposal to appoint Blaauw. Appointment will take
some
time and he can stay in Groningen for a while if necessary. Need him to help
build up the new spirit.

What does he think of Plaut as his successor? Recommended him highly. Also
mentioned van de Hulst, Objection: mainly theoretical.
\endmarg
Oort had taken the initiative to see if Plaut could be appointed as assistant
in Groningen. Van Rhijn will
have hesitated, since Plaut worked on variable stars, and with himself still
suffering from tuberculosis he would have preferred someone more suited to
take the supervision of work related to the {\it Plan
of Selected Areas} upon himself.
Variable stars had been a major line of research in Leiden under Hertzsprung.
The latter had stayed on as director after reaching the age of seventy in 1943
and now was moving back to Denmark. It would seem reasonable to expect that
Pieter Theodorus Oosterhoff (1904--1978) would continue this (see
Fig.~\ref{fig:NAC1952}).  The formal appointment of Oort as professor at Leiden
University and director of the Sterrewacht was made on November 8, 1945. And
indeed on the same date, Pieter Oosterhoff was appointed lecturer and
adjunct-director (Oosterhoff would be appointed extra-ordinary (honorary)
professor in 1948; for an obituary see \cite{JHO1978} ).
So, why would van Rhijn use his single staff position for
someone working in a field that was already a major line of research in Leiden?
Furthermore, the assistant had for many years been a person supervising work
on the {\it Plan of Selected Areas} and the {\it 
Spektral-Durchmusterung}, preparing at the same time a PhD theses  and then
vacating the position. No wonder van Rhijn hesitated to propose Plaut for the
assistantship. 

This left Plaut uncertain about his future and his wife decided to write Oort
without telling her husband. She did so on August 1, because, she explained, 
she no longer could handle the situation. She did not directly ask
him to approach van Rhijn about the Groningen assistantship, but asked for
support in a more indirect manner (Oort Archives, website accompanying
\cite{JHObiog}, Nr. 163; my translation):
\begmarg
My husband is seriously ill (jaundice). This is otherwise a nasty disease, but
harmless. However, my husband is mentally totally exhausted and no longer
believes that he can get better. For a long time he has been pondering about
his future, but saw no light. Now he is so exhausted physically and mentally
that he is no longer able to think about it. I would like to ask you, isn’t
there anyone in the whole astronomical world who can help this man?
\endmarg
She proposed that Oort may write to Plaut to cheer him up. At the top of the
letter Oort has scribbled that he had written a note to Plaut on August 15
and had ‘tried to give him some
encouragement’. And also that he hoped Plaut and Blaauw could visit Leiden after
his trip to England. However, it took a while for Oort could make the trip.

Jan Oort was General Secretary of the International Astronomical Union. The
president Arthur Eddington had died in 1944, and through vice-president Walter
Sydney Adams (1876--1956) of Mount Wilson Observatory it had been arranged that
Harold Spencer Jones (1890--1960), director of the Royal Greenwich Observatory
and Astronomer Royal, would be interim president. Oort was planning to fly to
England to discuss the course of action with Spencer Jones. The trip was
arranged by Gerrit Pieter (later Gerard Peter) Kuiper (1905–1973), a graduate
from Leiden, but by then naturalized American, who had come to Europe as part
of the ALSOS Mission, which investigated and collected intelligence on the
German atomic, biological and chemical weapons research. He has been reported
to have staged a daring rescue mission into eastern parts of Germany, during
which he took famous physicist Max Planck into Allied care hours before the
Soviets arrived. Due to visa and other formalities Oort's trip kept being
postponed but eventually took place from September 23 to October 7. Obviously
Oort had many things on his mind, busy as he was  with putting the Sterrewacht
and the IAU back on track.

In the meantime Lukas Plaut had writen
to  Oort again on August 30 (Oort Archives, website
accompanying \cite{JHObiog}, Nr. 163; my translation):
\begmarg
Yesterday morning I had a meeting with Prof. van Rhijn about the possibility
of becoming an assistant here. Prof. van Rhijn, however, would have liked to
have some certainty about my future after the assistantship ended and only
wanted to take further steps with the trustees or the Board of Rehabilitation
when, for example, I could show that there was a possibility for me to go to
America. So I wrote to my brother in Los Angeles and asked him to take care of
the official papers again for obtaining a visa after a few years. It is
unfortunate that the decision has been delayed, so I am still in limbo for
some time.
\endmarg
So, van Rhijn was still hesitant to allocate his assistant position to Plaut
and wanted some assurance that the position could become available again after
a few years and that Plaut would have a future when it had ended.
There is no further
mention of this situation in the Oort Archives. In the end van Rhijn did
arrange for Plaut to be appointed assistant, filling the vacancy created by
Blaauw’s departure. 

The appointment started on October 1, 1945, the date on which Blaauw’s had
ended, so van Rhijn’s hesitations would not have lasted long after August 30.
Plaut had to take over part of van Rhijn’s teaching because of the latter’s
illness and for this he was given (by Ministerial decree as required) a
teaching assignment in 1946 in prodaedeutic (first year introductory) astronomy.
He eventually remained on the staff until his retirement. Later Plaut’s
presence in Groningen played a major role in defining the {\it
Palomar-Groningen Survey}, which concerned mapping of a particular type of
stars, RR Lyrae variables, combining the study of variable stars, on which
Plaut was an expert, with Galactic structure, the research field of van Rhijn
and the Laboratorium. But that was only in 1953 and cannot
possibly have played a role in Plaut’s appointment in 1945.

\section{Staff and workers at the Laboratorium}\label{sect:staff}

At this point it may be appropriate to spend some time on the development of the
actual manpower (no women yet) of the Laboratorium over the years. Table 1
shows the personnel between 1898 and 1960.
This is compiled from the lists published annually in the
{\it Yearbook} \cite{Yearbooks}
of the university of Groningen and since 1946 partly in a supplement to
this, the  {\it University Guide} \cite{UGuide}.

From 1900 onward Kapteyn had had an assistant. The first appointment
had been done at the beginning of the academic year
1899-1900, in the ‘Lotgevallen’ in the {\it Yearbook 1898-1899}
\cite{Yearbooks} it is mentioned (my translation): 
\begmarg
An increase in personnel is to be recorded at the `Astronomische Laboratorium',
where for the first time an assistant was appointed, namely Mr. L.S. Veenstra. 
\endmarg

\begin{table*}[t]
\centering
\begin{threeparttable}[b]
\caption{\normalsize Table of personnel at the Kapteyn Laboratorium from 1898
to 1960 according to the {\it Yearbook} \cite{Yearbooks}
(Jaarboek der Rijks-Universiteit
Groningen) and since 1946 in part to the {\it University Guide}
\cite{UGuide} (Groninger
Universiteitsgids). Years in which the personnel is the same as in
the preceding one, haven been omitted.}\label{table:first}
\begin{tabular}{rlll}
\toprule 
Year\tnote{a} \ \ \ &  \ \ \ Professor & \ \ \ Assistant & \ \ \ \ \ \ \ \ \ \ \ \ \ \ \ \ \ \ \ \ \ \ \ \ \ \ Other personnel\tnote{b} \\
\midrule
1898\ \ \ & \ J.C. Kapteyn\tnote{c} & \ \ \ \ \ \ \ \   -- & \ \ \ \ \ \ \ \ \ \ \ \  \ \ \ \ \ \ \ \ \ \ \ \ \ \ \ \ \ \ \ -- \\
1900\ \ \ & \ J.C. Kapteyn &  L.S. Veenstra & \ \ \ \ \ \ \ \ \ \ \ \ \ \ \ \ \ \ \ \ \ \ \ \ \ \ \ \ \ \ \ -- \\
1901\ \ \ & \ J.C. Kapteyn &  W. de Sitter\tnote{d} &  \ \ \ \ \ \ \ \ \ \ \ \ \ \ \ \ \ \ \ \ \ \ \ \ \ \ \ \ \ \ \ -- \\
1904\ \ \ & \ J.C. Kapteyn &  W. de Sitter &  T.W. de Vries\\
1908\ \ \ & \ J.C. Kapteyn &  H.A. Weersma\tnote{e} &  T.W. de Vries\\
1910\ \ \ & \ J.C. Kapteyn &  H.A. Weersma &  T.W. de Vries, J. Jansen \\
1913\ \ \ & \ J.C. Kapteyn & F. Zernike &     T.W. de Vries, J. Jansen\\
1915\ \ \ & \ J.C. Kapteyn & P.J. van Rhijn\tnote{f} &  T.W. de Vries, J. Jansen \\
1918\ \ \ & \ J.C. Kapteyn & P.J. van Rhijn\tnote{g} & T.W. de Vries, J. Jansen \\
1919\ \ \ & \ J.C. Kapteyn & P.J. van Rhijn &  T.W. de Vries, G.H. M\"uller, J.M. de Zoute, \\
 &&& D. van der Laan, J. Jansen\\
1921\ \ \ & \ P.J. van Rhijn\tnote{i} & J.H. Oort &   T.W. de Vries, G.H. M\"uller, J.M. de Zoute, \\
 &&&  D. van der Laan\\
1922\ \ \ & \ P.J. van Rhijn\ \ \ \ \  & P. van de Kamp\ \  &   T.W. de Vries, G.H. M\"uller, J.M. de Zoute,\\
&&&  D. van der Laan\\
1923\ \ \  & \ P.J. van Rhijn  & W.J. Klein Wassink &  T.W. de Vries, G.H. M\"uller, J.M. de Zoute,\\
 &&&  D. van der Laan, W. Ebels \\
1927\ \ \  & \ P.J. van Rhijn  & W.J. Klein Wassink &  W. Ebels, G.H. M\"uller, J.M. de Zoute,\\
 &&& D. van der Laan\\
\bottomrule
\end{tabular}
\begin{tablenotes}
  \normalsize
\item[a]{Situation at the start of the academic year.}
\item[b]{Functions were described as amanuensis, observator, computer or clerk.}
\item[c]{Professor of astronomy, probability theory and mechanics since 1878.}
\item[d]{Also privaat docent (additional appointment for teaching) in astronomical perturbation theory and related subjects.}
\item[e]{Also privaat docent in orbital determination and perturbation theory of celestial bodies.}
\item[f]{Also privaat docent in theoretical and stellar astronomy, later astronomy, probability theory.}
\item[g]{In addition between 1918 and 1922 A.E. Kreiken was listed as assistant b.b. (‘without burden to the country's treasury’).}
\item[i]{Professor of  astronomy and probability theory.}
\end{tablenotes}
\end{threeparttable}
\end{table*}

\addtocounter{table}{-1}

\begin{table*}[t]
\centering
\begin{threeparttable}[b]
\caption{\normalsize {\it (continued from previous page)} The last line of the first part has been repeated.}
\begin{tabular}{rlll}
\toprule
Year \ \ \ &  \ \ \ Professor & \ \ \ Assistant\tnote{a}   & \ \ \ \ \ \ \ \ \ \ \ \ \ \ \ \ \ \ \ \ Other personnel \\
\midrule
1927\ \ \  & P.J. van Rhijn  & W.J. Klein Wassink &  W. Ebels, G.H. M\"uller, J.M. de Zoute,\\
 &&& D. van der Laan\\
1928\ \ \ & P.J. van Rhijn &  B.J. Bok &  W. Ebels, G.H. M\"uller, J.M. de Zoute,\\
 &&&  D. van der Laan,  H.J. Smith \\
1929\ \ \ & P.J. van Rhijn & J.J. Raimond  &  W. Ebels, G.H. M\"uller, J.M. de Zoute, \\
&&&  D. van der Laan,  H.J. Smith\\
1934\ \ \ & P.J. van Rhijn  & P.P. Bruna &   W. Ebels, G.H. M\"uller, J.M. de Zoute, \\
&&&    D. van der Laan, H.J. Smith\\
1939\ \ \ & P.J. van Rhijn  & A. Blaauw &  W. Ebels, G.H. M\"uller, D. van der Laan,  H.J. Smith\\
1941\ \ \ & P.J. van Rhijn  & A. Blaauw &  W. Ebels, G.H. M\"uller, J.B van Wijk, H.J. Smith\\
1945\ \ \ & P.J. van Rhijn  & A. Blaauw  & D. Huisman, G.H. M\"uller, W. Ebels, J.B. van Wijk\\
1946\ \ \ & P.J. van Rhijn  &  L. Plaut\tnote{b} &   D. Huisman, G.H. M\"uller, W. Ebels,  J.B. van Wijk\\
1956\ \ \ & P.J. van Rhijn  &  L. Plaut, J. Borgman &    D. Huisman, G.H. M\"uller, W. Ebels,  J.B. van Wijk \\
1957\ \ \ & A. Blaauw\tnote{c}  &  L. Plaut, J. Borgman &   D. Huisman, W. Ebels, J.B. van Wijk,  H.R. Deen,  \\
&&&  E.R.R. Timmerman \\
1958\ \ \ & A. Blaauw   &  L. Plaut,  J. Borgman & \ \ \ \ \ \ \ \ \ \ \ \ \ \ \ \ \ \ \ \ \ \ \ \ \  --\tnote{d} \\
&& H. van Woerden & \\
1960\ \ \ & A. Blaauw  &   L. Plaut, J. Borgman,  & \ \ \ \ \ \ \ \ \ \ \ \ \ \ \ \ \ \ \ \ \ \ \ \ \ -- \\
 && H. van Woerden, & \\
 && A.B. Muller & \\
\bottomrule
\end{tabular}
\begin{tablenotes}
\item[a]{In 1947 this changed into conservator, and from 1948 onward scientific staff member.}
\item[b] Also a teaching assignment between 1946-1949 in propaedeutic (first year introductory) astronomy.
\item[c] Professor of astronomy.
\item[d] After 1957 no longer available in \textit{Universiteitsgids}.
\end{tablenotes}
\end{threeparttable}
\end{table*}

But in the same {\it Yearbook} \cite{Yearbooks}
Veenstra is not listed as the personnel of the
Laboratorium (Table 1 is based on  these listings and presents this such that
for example
the line 1900 refers to the situation at the start of the academic year
1900-1901).  Schelto Leonard Veenstra (1874--1951) is listed in the 1898-1899
{\it Yearbook}
as a fifth-year student in mathematics and natural sciences, but I
have not found him in other {\it Yearbooks}
around that time in lists of students or
persons that passed an exam.  So, who was this Mr. Veenstra?
The information I have been able to uncover in municipal
archives is the following. In March 1901 Veenstra (26 years of age) married in
Groningen 19-year old  Doetje Heidstra and on the certificate he is described
as having the position observator. They seem to have moved to Utrecht in 1908.
The `Canon Autisme Nederland', an organization concerning persons suffering from
autism, notes in the description of a {\it Cannon on the history of autism},
that in 1910 (my translation)
\begmarg
by the Ministry [of Law] an Inspector of Probation was appointed to supervise
the government support of probation.  According to this organization the
first inspector was `the energetic
Schelto Leonard Veenstra (1874–1951), former captain of the Salvation Army,
characterized [...] as `deeply religious, broad-minded'.
\endmarg
Veenstra probably terminated his studies and academia. It is likely he never
contemplated a career in astronomy in the first place, but took the position
for a short period to earn a living.

Willem de Sitter came next (see \cite{Guich} for an informative biography).
He had passed his Doctorandus (Masters) degree in 1900, after having his
studies interrupted 1897–1899 for a period of observational experience with
David Gill at the Cape Observatory. While he was away Kapteyn had arranged
the assistantship for him when he returned, but on a temporary basis
had appointed
Veenstra on it. In 1901 de Sitter
defended  a thesis on the orbits of Jupiter’s Galilean satellites -- a
subject that remained his principal interest throughout his career --,
and his appointment continued until he moved to Leiden in 1908. Like some of
his successors, de Sitter also held an appointment as privaatdocent
to give lectures. His assignment (and that of his successors) is given
in footnotes to Table 1. When de Sitter left for Leiden,  Herman Weersma took
over, but he left astronomy after a few years. Frits Zernike came next, but he
left to do a PhD in Amsterdam in physics.
Van Rhijn was given the position in 1915 (like de Sitter
and Weersma, but unlike Zernike, this included also
one as privaat docent, but only after he had defended his PhD thesis). 

The position of assistant apparently could be filled either as a temporary
appointment of a student working on a PhD thesis, or on a longer or indefinite
basis by a person holding a PhD. De Sitter and Weersma
resigned rather than having come to the end of some term.
So from Oort up to and including Blaauw, who resigned in 1945 to accept an
appointment in Leiden, van Rhijn used the assistant position for a person,
who had obtained a Doctorandus (Master) degree and was now preparing a PhD
thesis.

Starting with Kapteyn’s share in the {\it Cape Photograph Durchmusterung} in
the 1880s the measurement of photographic plates and the associated reductions
was performed by persons employed from grants obtained especially for that
purpose. Kapteyn started with T.W. de Vries paid from grants from a number of
funds, and more personnel followed. This soft money did not provide any
long-term security to the employees and Kapteyn over many years sought ways to
provide that to de Vries (see \cite{JHObiog}). As table 1 shows this finally
succeeded only in 1904. When the {\it Plan of Selected Areas} had
been  started, George Hale had adopted it as the primary project for Mount
Wilson and his 60-inch telescope, and Kapteyn had been appointed Research
Associate to the Carnegie Institution of Washington. But that was not all; the
Carnegie Institution also provided funding to hire personnel in Groningen
to measure up the 60-inch plates. The situation did change in
1919, when the University of Groningen appointed three more computers on the
university staff in addition to the amanuensis de Vries and the clerk Janssen.
This was still the situation at the time van Rhijn took over from Kapteyn.

So at the start of his term as director van Rhijn had one assistant and four
positions for technical support. Various astronomers were
subsequently employed on the
assistant position. Jan Oort filled it for a short period.
Then followed Peter van de Kamp, Willem Klein Wassink, Bart Bok,
Jean-Jacques Raimond, Petrus Bruna and Adriaan Blaauw. Except Bruna all
these assistants finished and defended a PhD thesis under van Rhijn
(see Fig.~\ref{fig:vRhijn3}). 

Not long after van Rhijn took office a fifth position was added to the
Laboratorium, but at the end of the thirties during the a Great Depression
this disappeared again, not to come back before van Rhijn’s retirement. So,
the net result was that as far positions on the university pay-roll are
concerned,  the number of staff had remained the same as it was at the start
of van Rhijn’s directorship. 
\bigskip

So, how does this compare to the university as a whole? For this I refer to
Table~\ref{table:Univstaff},
which covers the period 1920 to 1960 in steps of a decade, so roughly
encompassing van Rhijn’s  tenure of the directorship of the Kapteyn
Laboratorium. The growth of the university is understandably modest
during the Great Depression and the
Second World War, but picks up after this,
especially during the 1950s.

The number of institutes is somewhat more than half the number of professors.
The column professors, however,  includes extraordinary professors
(appointments on an individual basis in addition to the normal contingent),
honorary
professors and so-called ecclesiastical professors (who in the Theological
Faculty teach particular subjects on behalf of churches or religious
organization), so ignoring these it follows that as a rule the majority of
ordinary professors were also directors of an institute.
The number of professors
increased by a factor 2.4 and of institutes by 2.1, so this situation did not
change much. Not surprisingly, astronomy remained having one institute with
one professor.  Roughly speaking one professor out of four was supported by a
lecturer, a senior scientist involved in research and teaching. Although the
total number of lecturers
increased, it remained relatively constant at only three or  four
for the natural sciences and -- maybe a not unreasonably --
none for the Kapteyn
Laboratorium. The growth university-wide was in assistants, as we have seen
junior scientists, in Kapteyn’s day as long term employment (de Sitter to van
Rhijn), but later as temporary appointments and usually to prepare a PhD thesis
(under van Rhijn from Oort to Blaauw). From the end of the War to the end of
van Rhijn’s term this was a junior permanent position occupied by Plaut. There
were no new positions for junior staff in astronomy,
while this increased by a factor three
in the university and Faculty of Sciences. The increase to four assistants
occurred after Blaauw took office and was part of the conditions he bargained
for as bonus for his acceptance of the directorate. Whereas de Sitter, Weersma
and van Rhijn were appointed privaat docent in addition, this was not the case
of the ‘graduate students’ under van Rhijn. Plaut was given a teaching
assignment, but  only temporarily during van Rhijn’s illness. While the number
of such appointments increased, none was available structurally to astronomy.
In this respect van Rhijn was worse off than his predecessor and contemporaries

\begin{table*}[t]
  \caption{\normalsize University personnel employed at the University of Groningen}
\centering
\begin{threeparttable}[b]
\begin{tabular}{r|rrrr|rrrr|rrrr|rrrr|rrrr|rrrr}
\toprule 
Year\tnote{a}\ \  &  \multicolumn{4}{|c|}{Institutes\tnote{b}}
& \multicolumn{4}{|c|}{Professors} & \multicolumn{4}{|c|}{Lecturers} & \multicolumn{4}{|c|}{p.d./l.o.\tnote{c}} & \multicolumn{4}{|c|}{Assistants, etc.} & \multicolumn{4}{|c}{Further personnel} \\
  & u & n & k & \ & u & n & k &\ & u & n & k &\ & u & n & k &\ & u & n & k &\ & u & n & k &\ \\
\hline
1920\ \ &  \ \ 25 &  \ \ \ \ 9 & \ \ \ \ 1 &\ &  \ \  46 & \ \ \ \ 11 & \ \ \ \ 1 &\ &  \ \ 9 &  \ \ \ \ 2 &  \ \ \ \ 0&\ & \ \ 7 &  \ \ \ \ 4 &  \ \ \ \ 1 &\ &  \ \ \ 52  & \ \ \ \ 15 & \ \ \ \ 1 &\  & \ \ \ 83 & \ \ \ \ 35 & \ \ \ \ 5&\  \\
1930\ \ &  31 & 11  & 1 &\ & 54 & 12 & 1 &\ &  9 & 1 & 0 &\ & 13 &  5 & 0 &\ &   81 & 32 & 1 &\ & 101 & 45 & 5 &\  \\
1940\ \ &  36 & 15 & 1 &\  & 57 &  14 & 1 &\ & 13 & 3  & 0 &\ &  21 &  9 & 0 &\ & 116 & 39 & 1 &\ & 107 & 52 &  4 &\ \\
1950\ \ &  53 & 14 & 1 &\ & 76 & 18 & 1 &\ & 18 & 4 & 0 &\ &  35 & 12 & 0 &\ & 171 &  47 & 1 &\ & 135 & 60 & 4 &\ \\
1960\ \ &  60 & 16 & 1 &\ &  97 & 25 & 1 &\ & 28 & 4 & 0 &\ & 21 & 8 &  0 &\ & 172 & 41 & 4 &\ & --\tnote{d} & -- & -- &\ \\
\bottomrule
\end{tabular}
\begin{tablenotes}
\item{a} Situation at the start of the academic year.
\item{b} Here and in the further columns: u = University, n = Faculty of Natural Science, k = Kapteyn Laboratorium.
\item{c} p.d. = privaat docent (additional appointment for teaching), l.o. = teaching assignment.
\item{d} No longer listed in the Universiteitsgids.
\end{tablenotes}
\end{threeparttable}
\label{table:Univstaff}
\end{table*}

In terms of scientific staff astronomy did not profit at all from the
national and local growth of university budgets,
student numbers and staff during van Rhijn’s
directorship. In further personnel the situation also showed a negative trend;
while there was a significant increase (by a factor 1.6 university-wide and 1.7
in the Faculty) for astronomy it went went from 5 to 4 and stayed at that level
for a long time. In this category
there was a modest increase as well under Blaauw
as part of the conditions under which he accepted his appointment. 

\begin{figure}[t]
\sidecaption[t]
\includegraphics[width=0.64\textwidth]{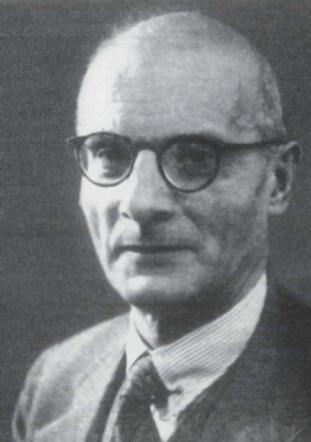}
\caption{\normalsize Pieter Johannes van Rhijn, probably in the late 1940s.
 From the collection of
Adriaan Blaauw and Kapteyn Astronomical Institute.}
\label{fig:vRhijn3}
\end{figure}

The Kapteyn Laboratorium remained a marginal effort in Groningen; while
the university grew by a factor of about three or so between 1921 and 1957, it
remained at the same level, probably below critical mass in the long run. On
a national level this falling behind of Groningen astronomy was even more
evident. For  nation-wide comparison the
available literature in English is limited.
David Baneke’s book {\it De ontdekkers van de
hemel} on the history of Dutch astronomy in the twentieth century \cite{DB2015}
is not available in English, as is also the case for a number of his other
informative papers on related issues except for a few, e.g. \cite{DB2010}.
From all of this I compile the following numbers. In
the 1930s there were eleven staff positions (professors, assistants, etc.) in
astronomy at Dutch universities: six in Leiden, two in Utrecht and Groningen
and one in Amsterdam.  Compared with the Sterrewacht Leiden the Groningen
Laboratorium was insignificant. In 1933, Willem de Sitter wrote a history of
the Sterrewacht on the occasion of its tricentennial \cite{WdS1933} and listed
the staff at that time: two professors (of which one extraordinary), two
conservators (senior staff comparable to lecturer), two observators
(scientific staff), four assistants (plus six voluntary, so not paid from the
regular budget), ten computers, two technical staff, a typist and a carpenter.
The observatory itself had various telescopes and a station in Johannesburg,
South-Africa, which included one telescope and access to at least two more.
The teaching
load (astronomy staff taught astronomy courses to physics and mathematics
majors, and some times physics or mathematics courses) was divided between at
least four persons. This is no comparison to the other universities, where all,
or at best almost all, the teaching had to be done by a single person. 
Utrecht had an observatory as well, concentrating on solar research
\cite{Utrecht370}, but expanding into astrophysics in particular hot, massive
stars. The Amsterdam Institute of Pannekoek was not significantly
different from the Groningen establishment \cite{Pannekoek}. Anton Pannekoek
had modeled his institute along Kapteyn’s lines in the sense of relying on
observational material from elsewhere. Pannekoek concentrated furthermore on
theoretical work, pioneering the field of stellar atmospheres. The institute
remained small under his leadership, and also under his successor Herman
Zanstra (1894-–1972), who took over after Pannekoek’s retirement in 1946. 

Not only did Leiden dominate Dutch astronomy and therefore making it more
difficult for other universities to strengthen their astronomy
departments, more importantly for the Kapteyn Laboratorium is that Groningen
astronomy had under van Rhijn lost ground by failing to increase its size in
step with the rest of the university. 
\bigskip

As far as students contemplating a possible career in astronomy are concerned,
the situation in Groningen was also minimal. Maarten Schmidt had  enrolled as
a student with a strong interest in astronomy in 1946, and obtained his
Candidaats (Bachelor) in 1949. But on invitation of Jan Oort he moved to
Leiden at that stage. The single promising student left, Jan Borgman, although
of almost exactly Schmidt’s age, entered Groningen university only in 1950. He
progressed in a satisfactory way, obtaining his ‘Doctoraal’ (Masters) in 1955
and a PhD in 1956. However the PhD degree was granted with physics professor
Hendrik Brinkman (1909--1994) as supervisor, and concerned electronic variable
star recognition on plate pairs.
No further PhD students came forward during the rest of van Rhijn’s
professorship.

Up until the first years of WWII (the period
1921--1942) the number of PhD theses under
van Rhijn had been significant compared to the other universities, namely eight (Groningen)
versus fourteen (Leiden), eleven (Utrecht) and  two (Amsterdam). But between 1945 and 1957
van Rhijn had only one, compared to eight (Leiden), four (Utrecht) and four (Amsterdam).
This one PhD was Adriaan Blaauw in 1946, but after him there were none for
almost two decades, unless one counts the physics thesis of Jan Borgman. Of the
total of nine Groningen PhDs under van Rhijn, however, 5 had been recruited as PhD students
from elsewhere: Jan Schilt and  Peter van de Kamp  from Utrecht, and Jean
Jacques Raimond, Bart Bok and Adriaan Blaauw from Leiden. Egbert Kreiken,
Jan Oort, Willem Klein Wassink and Broer Hiemstra had obtained their Doctoraal
(Masters) in Groningen, but the first two were in a sense students of Kapteyn.
Students in physics and mathematics do attend astronomy lecture courses in
their first years and inspiring astronomy professors might ignite an interest
in their field. Van Rhijn did not recruit many new astronomy students, although
it must be said that after the War his tuberculosis would have made this
difficult. Klein Wassink and Hiemstra started as Groningen undergraduates
and went on to a PhD; they are the only ones that can be seen as having been
pure students of van Rhijn. The small number of students would not help
persuading Curators or the Minister to create or assign new staff
positions at the astronomy department.

\section{The late forties and early fifties}\label{sect:forfif}

During the first years after the War work at the Kapteyn Laboratorium was slow
in picking up.
The work on variables was painstakingly slow, but very competently executed
by Plaut
and van Rhijn was out of action a large part of the time. Work on the telescope
was performed but no results appeared. Van Rhijn and Plaut made up the full
scientific staff, which had never grown compared to Kapteyn’s days
(see Fig.~\ref{fig:NAC1952}). Teaching went
on as normal, lectures to a large extent being given by Plaut. There was one
astronomy major student, Maarten Schmidt, but as already mentioned,
he left for Leiden halfway through his studies.
\bigskip

During the War the work on the {\it Bergedorf(-Groningen-Harvard)
Spektral-Durchmus\-te\-rung}
had first slowed down and then essentially halted.
Van Rhijn had been able to do some research, part
of it probably while being nursed for tuberculosis. In any case he did publish
two papers in the {\it
Publications of the Astronomical Laboratory at Groningen}.
In his review of Dutch astronomy during the Second World War \cite{PJvR1946}
he had noted: 
\begmarg
The H and K lines are absorption lines due to calcium and can only be
distinguished from stellar lines of
these have not such lines in their spectra, which means they have to be
spectral types O or B.

During the fighting in Groningen in April 1945 two publications, written during
the War by the director, have been burned, but a copy of the manuscripts is
undamaged.

Dr. P.J. van Rhijn investigated the accuracy of the interstellar line
intensities as a criterion of distance and reduced the known equivalent widths
of the H and K lines, which appeared to be most suited to the purpose, to
distances. The density and luminosity functions of the O and early B stars were
derived; contrary to former results it appears that the density does not
decrease with increasing distance from the Sun.
\endmarg  

After the War van Rhijn had not sit
still in spite of his tuberculosis, from which he recoverd only very slowly.
A large research project that he completed in 1949, was a mapping of
extinction on the basis
of color excesses measured at various places all over the sky \cite{PJvR1949}.
According to the Introduction:
\begmarg
It is the purpose of the present paper to derive the mean differential
absorption and the distribution of the amount of the differential absorption
for stars at a specified distance and Galactic latitude. If, as seems probable,
the photographic and differential absorption are in a constant ratio, the mean
photographic absorption and the distribution of this absorption for stars at a
specified distance and latitude can be found. These functions can be used
advantageously in the solution of the equation of stellar statistics, [...]
\endmarg
For clarity I remind the reader that differential absorption is what we
would now call reddening, in those days
usually between photographic (blue) and photovisual (yellow) wavelength bands.
The distribution of actual values in a particular direction was
found to be approximately Gaussian and could  be
explained by a chance distribution of absorbing clouds in regions near the
Milky Way. The number of clouds on a line of sight near the Milky Way was
6 per kpc on average, and the
photographic absorption of a single cloud  van Rhijn estimated as 0\magspt27.
The differential absorption of such a cloud turned out to be
0\magspt035. The
distribution of cloud sizes was not addressed. Interesting as this may be the
use of this to solve the equation of stellar statistics is limited, because of
the severe irregularities in the distribution of extinction on the sky and
along sight lines. As far as I am aware these results have never been used for
such a purpose. In ADS van Rhijn's paper
has a single citation in 1963 and only an occasional
read (the peak in 2021 is me preparing this paper).

It should be recorded that van Rhijn after the war
faithfully fulfilled his task as
coordinator of the {\it Plan of Selected Areas}; he wrote a comprehensive
report for Commission 32 for the IAU General Assembly in Z\"urich, Switzerland,
in 1948. But he did not attend and during the meeting Jan Oort took the lead as
acting President of that Commission together with Adriaan Blaauw as Secretary.
\bigskip

Plaut did vigorously pursue his research in variable stars. Some of it during
the War was probably completing work he had  started in Leiden before he
moved to Groningen. This means extremely labor intensive and time consuming
work of which the yield was small and progress slow. But is shows the kind of
work Plaut did and how very patient a person he was. In 1946 he published
a paper on an eclipsing binary from 341 plates taken between February 1930 and
August 1936 \cite{Plaut1946}. The plates had been obtained  at the Leiden
southern station using the 10-inch Franklin-Adams Telescope of the Unie
Sterrewag in Johannesburg, which had  been donated by John Franklin-Adams
(1843--1912), the maker of the first all-sky photographic survey. In the deal
de Sitter had negotiated, Leiden astronomers had access to this telescope.
Hertzsprung and collaborators used it primarily to study variable stars from
repeated exposures. Plaut had measured these plates with the Schilt photometer
(see above) to construct a detailed light curve, the form of which gave among
others information on the eccentricity of the orbit. 

This is not all.
As noted above, van a Rhijn \cite{PJvR1946} had noted in his article on Dutch
astronomy during the war, that Plaut
had been searching for variable stars on plates
taken at the Leiden southern station with the Franklin-Adams telescope.
Obviously, this involved  plates taken
by the Leiden representative there, which in this case was Hendrik van
Gent (1900--1947), who had worked on variable stars with Hertzsprung
and obtained a PhD under him in 1932.
Plaut used the blink comparator of the Kapteyn Laboratorium
(rapidly switching the view between two plates of the same area)
and published a
paper in 1948 [56], in which he reported 91 variables by comparing
12 pairs of plates, subsequently estimating the magnitudes on a full set of 300
plates. These estimates were performed by eye. This was a matter of patience
and extreme care. And in
1950 Plaut published a review collecting the available data on all 117 known
eclipsing binaries brighter than magnitude 8.5 at maximum and determining their
orbital parameters and updated that in 1953 (\cite{Plaut1953}
and earlier references therein). Plaut had established himself as an authority
on variable stars.

\begin{figure}[t]
\sidecaption[t]
\includegraphics[width=0.64\textwidth]{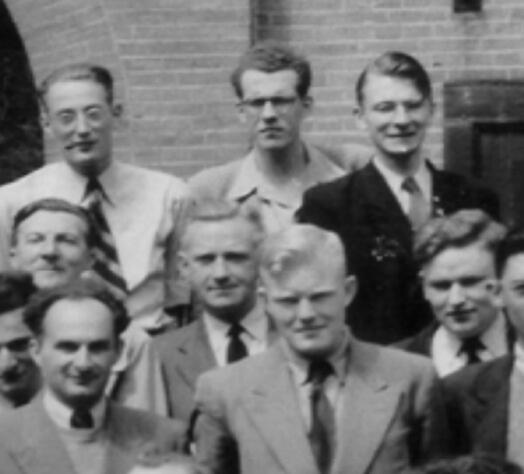}
\caption{\normalsize Detail of the group photograph of the Nederlandse
Astronomen Conferentie in 1952. The full delegation from Groningen consisted
of only two persons: on the lower-left  Lukas Plaut, on the lower right
Jan Borgman. The latter was a second year student. In the 
top we see Maarten Schmidt, who by then had moved to Leiden.
Other notable persons are Cees Zwaan from Utrecht top-right and Pieter
Oosterhoff from Leiden between Schmidt and Plaut.
From the Website of the Koninklijke Nederlandse Astronomen Club \cite{KNAC}.}
\label{fig:NAC1952}
\end{figure}

As we have seen, van Rhijn was out of action for most of the rest of the 1940s
after the War and the Laboratorium relied mostly on Lukas Plaut for teaching.
In the ‘Lotgevallen’ for 1949, covering the previous academic year, Rector
Magnificus Pieter-Jan Enk (1885--1963), Professor of Latin, noted (my translation):
\begmarg
It fills us with great joy that Prof. Dr. P.J. van Rhijn, after an illness that
lasted more than four years, has now completely recovered and is again going
to devote himself to his work with full vigor. We wish our colleague all the
best in this respect.
\endmarg

\noindent
And
\begmarg
The teaching assignment given to Dr. L. Plaut, in connection with Prof. van
Rhijn’s illness, to teach Propadeutic Astronomy was withdrawn by Ministerial
Decree of November 18, 1948.
\endmarg
\bigskip

We have seen in section~\ref{sect:extinction}
that van Rhijn had been very much involved in research
into interstellar extinction, in fact it was part of his PhD thesis. 
Summarizing the background  very briefly,
following Kapteyn it had become clear that it was due to
some kind of scattering which caused reddening and part of the problem was to
turn amount of reddening into amount of extinction. This is given by what is
usually referred to as the ratio total-to-selective absorption. This is a
dimensionless property (in fact magnitudes per magnitude) and once known can be
applied without any other information: measure the reddening and with the ratio
gives the extinction. The ratio depends of course on the wavelength
dependence
of the scattering and has a different numerical value for different photometric
bands. The total-to-selective absorption depends not only on the physics of the
scattering,
but also on the size distribution of dust particles and cannot simply
be calculated from first principles of physics.

Van Rhijn’s telescope was originally intended for stellar photometry,
but it soon became clear that the
sky conditions in the center of Groningen were too poor for this.
He therefore resorted to spectroscopy
for which condition are less critical. We will seen that he
corresponded in 1936 with the Zeiss
Company about the purchase of a slitless spectrograph,
which he probably obtained in that year or
shortly after that. At that time the selective-to-total absorption
was still very poorly known and he wished to
improve this with his telescope taking spectra of obscured stars and determine
the energy distribution over wavelength and do the same for unreddened stars of
the same spectral type. Eventually he was aiming at using stellar spectra to
more fully determine the distribution of dust in the Galaxy, and address
questions on what the size distribution of the particles was and whether or not
this changed substantially with position in space.
Work on relatively bright stars in the sky was the only kind of
observational astronomy that could
conceivably be done as relevant research from the center of a city like
Groningen. The reason is that a good
fraction of the bright background in a city is concentrated
in small wavelength regions of bright spectral lines. Outside these
wavelength regions the city sky is more amenable to astronomical research.

This project took a very long time getting started. In the 1930s getting the
telescope ready for operation was a tedious and slow process, to a large extent
due to the economic depression, which prevented the hiring of additional staff,
but also as a result of the enormous effort required for the {\it
Bergedorf(-Groningen-Harvard) Spektral-Durchmusterung}.
The telescope required a
spectrograph, but the – what looks like final – quote in the correspondence
with the Zeiss Company in the van Rhijn Archives on a ‘Spaltlose
Einprismen-Spectrograph’ (slitless single prism spectrograph) stems from 1936.
Where the required 4100 RM (ReichsMark), corresponding to an estimated 70000
\textgreek{\euro} at present, came from
is not explained. It may have taken a few years
to raise that amount. Some observing did take place during the war, but
whether useful material for this project was obtained then is doubtful. The
collecting of spectra only proceeded successfully apparently after the war, as
noted by Blaauw in the quote at the end of section~\ref{sect:telescope} and
Maarten Schmidt described his experience with learning observing from Lukas
Plaut as a student in the late forties in the citation preceding that of
Blaauw. The first results on the wavelengths dependence of
interstellar extinction were published by van Rhijn in 1953 in a single
authored paper, \cite{PJvR1953}, in which he acknowledged Plaut for taking
most of the plates. His technical and computational staff, M\"uller and Huisman,
are noted for help with the reduction (and M\"uller also with the observing).
It seems that the work at the telescope was mainly done by Lukas Plaut.
This would have been a major effort, as it also involved obtaining experience
with the equipment and the problem of the poor observing conditions in the
center of the city of Groningen. Plaut has not published any paper on data
obtained with the Groningen telescope and did not use the telescope for studies
of variable stars; for useful research purposes it has been used only with
the spectrograph.

Since the design and installation of the telescope astronomy had changed
considerably. Would the idea of measuring brightness distributions of reddened
stars as a function of wavelength using photographic recordings of spectra have
been timely in the late 1920s or the 1930s, it was definitely not state of the
art after the war. The field of stellar photometry had been revolutionized by
the introduction of photoelectric photometry (and spectophotometry),
and indeed applied to the problem
of the wavelength dependence of interstellar extinction by astronomers as
Jesse Greenstein, Joel Stebbins (1878--1966), Albert Edward Whitford
(1905--2002) and others, and these studies had established that it was
approximately inversely proportional to the wavelength. In fact, van Rhijn did
acknowledge in the first line of his paper [98] that the detailed nature of
the wavelength dependence had been addressed by Stebbins and Whitford. These
had published a series of six papers entitled {\it Six-color photometry of
stars}, starting in 1943 with \cite{SW1943}. They used the Mount Wilson 60-
and 100-inch telescopes to photoelectrically measure the intensity distribution
over the wide wavelength range of 3530 to 10,300 \AA\ (they included
measurements on galaxies) and had found relatively small, but
significant deviations from a simple inverse-$\lambda$ law.

In \cite{PJvR1953} van Rhijn described how the photographic spectra
were measured. For this he used ‘the thermoelectric photometer of the Kapteyn
Laboratory’, which must be still the ‘Schilt photometer’ built by Jan
Schilt in his thesis research and among others used for the
{\it Spektral-Durchmusterung}
(see section~\ref{sect:Spektral-D} and
\cite{Schilt1922}). The method van Rhijn had designed was to use a pair of stars
of the same spectral type, one significantly reddened and one unreddened, so
respectively at low and high Galactic latitude.
The stars used are of spectral type late O or
early B, since these are stars with the weakest spectral features, the
spectral lines present predominantly come from hydrogen and helium but
with no significant lines of other elements (early O-type stars are unsuitable
because of strong helium lines, and in later B-stars the hydrogen lines become
stronger). Their apparent magnitudes were between 5.5 and 8.0. These were
exposed various (usually four) times on a single plate together with exposures
of a late-B or early-A star. The latter are stars with spectra with strong and
sharp hydrogen lines that could be used for wavelength calibration. The spectra
of the two program stars were measured at a set of seven wavelengths ranging
from about 3900 to 6300 \AA, or a factor of about 1.6. The research reported
concerned 14 pairs and it confirmed the Stebbins and Whitford result.

This was followed up soon by Jan Borgman.
He had spent some years working at the Philips Company,
developing his technical skills and gaining
experience in building electronic instrumentation. He obtained his doctorandus
degree (Masters) in 1955.
His PhD thesis work (defended in 1956) will be described below, but as a student
he published in 1954 and 1956 two more studies (with the same title as van
Rhijn’s paper), extending van Rhijn’s work with another 10 and 27 late O- and
early B-stars, \cite{JB1954}\cite{JB1956}.
Again the agreement with the Stebbins and Whitford curve was very good.

The work suffered from all problems associated with quantitative photographic
work in that is was tedious and had to be carried out with extreme care and
perseverance. The photographic emulsion is a very a-linear detector: it suffers
from under- and over-exposure and the relation between the photographic density
-- how opaque the emulsion is, zero when the transmission is 100\%, unity when
this is 10\%, etc. --, known as the characteristic curve, varies
from plate to plate and sometimes even in different parts of the
same plate. On top of this there is
low-intensity reciprocity failure -- at faint light levels the exposure of,
say twice, fainter light source takes more than twice the exposure time to
reach the same photographic density. Van Rhijn, Borgman and Plaut did overcome
all these problems and produced reliable results.

The program was conceived by van Rhijn twenty
or so years earlier and indeed worked
out excellently, and could in the 1930s in all likelyhood have resulted in
papers that would have made a major impact, rivaling others worked on this
subject at that time (see e.g. the 1934 review by Trumpler for more details),
\cite{Trump}. But due to very long delays van Rhijn was not breaking any new
ground, merely confirming what had been determined in the mean time in
particular by photoelectric techniques that held more promise for the future.
But the fact that the photographic work in Groningen actually worked and could
produce results that were comparable in accuracy to the photoelectric data is
a tribute to the quality of the work of van Rhijn and Borgman.

\section{The Vosbergen conference}\label{sect:Vosberen}
In June 1953, the International Astronomical Union held its Symposium Nr. 1 on
{\it Co-ordination of Galactic Research}. The initiative for this meeting had
been taken by Jan Oort, who was President of Commission 33 of the IAU on
‘Stellar Statistics’ (at the IAU General Assembly of 1958 in Dublin this was
changed into ‘Structure and Dynamics of the Galactic System’). The purpose and
character of the conference, quoted from a circular letter sent to all
participants -- and obviously written by Oort -- was described as follows:
\begmarg
During the first third of this century an important concentration of work on
Galactic structure and motions has been promoted by the Plan of the Selected
Areas, initiated in 1906 by Kapteyn. Although the scheme outlined in 1906
has not lost its significance, it is widely felt that further research into
structure
and dynamics of the Galaxy should be extended beyond the original Plan.
At the same time it is felt by many that some kind of coordination of effort
remains highly desirable, if only because the value of many observations is
greatly enhanced if the data can be combined with other data for the same
stars.

Several observatories have asked for suggestions with regard to future work
on Galactic structure. The recent advent of several large Schmidt telescopes
furnished with objective prisms, red-sensitive plates, etc., brings forward with
some urgency for each of the observatories concerned the problem of how to
organize and restrict the work with telescopes that can produce much more
than can be measured and discussed by the existing observatories. The question
arises, whether it is desirable to formulate a new plan of attack, and whether
such a plan should also include recommendations for concentrated work on
particular regions. If the answer to the latter is affirmative, which regions
should be selected and what data would be the most urgent.
\endmarg

The meeting was hosted by Pieter van Rhijn and the Kapteyn Laboratorium; the
venue was the estate Vosbergen, which was used by he University of Groningen as
a conference center. It is located near the village of Eelde some 5 km south of
Groningen. Lukas Plaut had done the local organizing. Adriaan Blaauw edited
the proceedings, which was unusual according to current standards, in that
it did not consist of papers or articles of the presentations made,
but instead Blaauw
had summarized all presentations and the long discussions (Blaauw would
move to Yerkes before finishing the proceedings). The format
of the meeting was that
there were introductory presentations (some of which have been published
separately in full), followed by discussion. The Scientific Organizing Committee
(SOC), of which Oort was chairman and Blaauw secretary, consisted of in total
eight persons. Van Rhijn was not part of this (nor Plaut); it was dominated by
Oort and Blaauw from Leiden  and Whilhelm Heinrich Walter Baade (1893--1960)
(see \cite{OstBaade} or \cite{JHO1961a}) with whom Oort had been
conducting extensive correspondence and who visited Leiden in the months
preceding the meeting. This meeting also marked the first discussions on the
European Southern Observatory, taking place first in Leiden the Sunday before
the meeting. Oort had invited a few leading astronomers from
European countries to travel to the Netherlands a day earlier for this purpose.
More disucssions took place during a boat trip on the IJsselmeer (the dyked off
former Zuiderzee) as part of the symposium. 

If the meeting was intended to be at Groningen as a tribute to van Rhijn’s
dedication to the {\it Plan of Selected Areas} and guardian of the Kapteyn
legacy, it would seem a bit of an affront not to invite him on the SOC. Even
as host this would have been appropriate. After all, van Rhijn was only three
years away from retirement by now and this would be a excellent occasion to
honor him on an international stage. It is somewhat reminiscent of
the story of the Board of the Foundation Radio Radiation of Sun and Milky Way
that Oort had founded after the war
to realize the construction of radio telescopes (see \cite{JHObiog}
or \cite{JHOEng}), where
for a national effort it would have seemed appropriate that the directors
of the institutes interested in using the facilities would be board members.
But van Rhijn was not invited.
Although this was later corrected (upon protests by van Rhijn), it remained a
remarkable move. In addition to the eight SOC members of the Vosbergen meeting,
twenty more astronomers were invited to attend
\begmarg
because they represented institutions which might take part in
future galactic research, or because of the character of their research.
\endmarg
In
total ten introductory papers were presented in seven sections. The members of
the SOC all held introductory reviews (and a few of the other persons invited
to attend). Van Rhijn was not invited for this.
From the proceedings the impression is that van Rhijn
remained somewhat in the background. He is mentioned in the proceedings only
when the need for understanding the wavelength dependence of interstellar
extinction was discussed as an example of one having addressed that issue and
for his stressing (with Oort), that
\begmarg
\noindent
a very important task for the
observatories equipped with Astrographic Catalogue refractors will be the
determination of proper motions in regions centered on the Kapteyn Selected
Areas, but much larger than those given in the Pulkovo and Radcliffe
Catalogues.
\endmarg
\bigskip

The desiderata arrived at during the Vosbergen symposium showed a change in
emphasis on collecting as much data of various kinds as possible on stars in
the Selected Areas to large programs concerning each a particular set of
objects to elucidate well-defined issues. For example the inventory of stars
in the areas defined by Kapteyn
was replaced by especially designed surveys to determine
the distribution of stars in the halo or the disk using tracer objects, an
inventory of the distributions of stars in the local volume as a function of
age, surveys for proper motions and radial velocities of various types of
objects, or programs designed to address special problems. For example the
formation and evolution of the disk could be traced by studying the
distribution of stars  of different spectral types on the Main Sequence or
late-type giants. B-stars could be used to determine from the distribution of
their radial velocities a better estimate of Oort constant $A$ of
Galactic rotation.

And two more actions would be made effective.
First the establishment of two IAU sub-committees, one  within IAU Commission
33 on the coordination of observational programs and one within  Commission 27
(Variable Stars) to organize further work on variable-star surveys. Memberships
were rather limited; for the first it was Baade, Blaauw, Lindblad, Oort,
Parenago, and van Rhijn, for the second Baade, Blaauw, Kukarkin, Oosterhoff.
These persons have been identified in Fig.~\ref{fig:Vosbergen}. Note that the total
attendance was only twenty five or so. I will go into
this list of members in some detail since it is remarkable,

\begin{figure}[t]
\begin{center}
\includegraphics[width=0.98\textwidth]{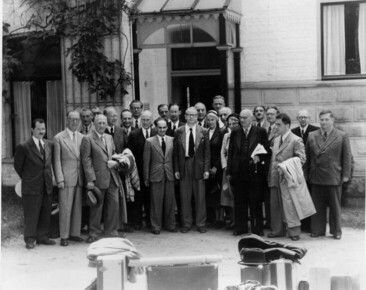}
\end{center}
\caption{\normalsize Participants at the 1953 Vosbergen Conference,
\textit{Coordination of Galactic Research}, with some of their coats, hats,
suitcases, etc. stored not really out of sight in the front.
Persons referred to in the
text can be identified as follows: 
Second from the left Bertil Lindblad, next to him Walter Baade, four persons to
the right Jan Oort, in front of Oort Lukas Plaut and
standing next to him Bill Morgan, between and behind them Adriaan Blaauw,
and with papers in his hand Pieter van Rhijn, to the left of him, partly
hidden by the lady directly to his left, Jason Nassau, and behind Nassau 
Pieter Oosterhoff. The 
two persons on the right in the second row (numbers  2 and 4 from the right)
are respectively P.P. Parenago and B.V. Kukarkin.   Kapteyn Astronomical
Institute.}
\label{fig:Vosbergen}
\end{figure}

One may argue the membership of these sub-commissions within the IAU was
chosen to optimize the interface with the IAU: Lindblad had just finished as IAU
President and was adviser to the Executive Committee, Oort and Boris
Vasilyevich Kukarkin (1909–1977) were at the time Presidents of Commissions 27
and 33, while Baade  was President of Commission 28 (Galaxies),
Pieter van Rhijn of Commission 32 (Selected Areas) and Pieter Oosterhoff was
IAU General Secretary.  Pavel Petrovich Parenago (1906--1960) and
Kukarkin were both from Moscow (the first specializing
in stellar structure of the Galaxy and the latter in variable stars) and were
to cover the Eastern countries; both were very influential in the Soviet Union.

This makes sense, but the rest of the membership involved persons
that were all special to
Oort and the Dutch dominance is both obvious and out of balance. Oosterhoff,
Blaauw and van Rhijn were all  Dutch (Blaauw was appointed Secretary of both
sub-committees). Bertil Lindblad (1895–1965) had for a long time
been a leading figure in the field of structure of the Galaxy, 
Oort had special relationships with him and with Baade, and although
they were clearly
leaders in the field, for example some American prominent astronomers that
were in the audience would have been fitting to include, such as William
Wilson (Bill) Morgan (1906--1994) or  Jason John Nassau (1893--1965), who both have
contributed very important studies involving extensive surveys of respectively
early and late type stars. I have also identified Morgan and  Nassau in
Fig.~\ref{fig:Vosbergen}. As a sideline: Donald Edward Osterbrock (1924--2007),
in his biography of Walter Baade \cite{OstBaade}, noted about Baade and Nassau:
\begmarg
The two men had the same general build, height, hair style and color, prominent
hawk nose, and liked to dress alike at scientific meetings which they both
attended. Frequently in group pictures they stood at opposite ends of the front
row, often striking very similar poses, which they must have rehearsed or at
least discussed in advance.
\endmarg
This picture would be a notable exception to this.

The second desideratum was to hold another meeting soon. It was
thought to possibly take place during the 1955 General Assembly in Dublin, but
was eventually deferred to 1957 as IAU Symposium  Nr. {\bf 7}, {\it Second
Conference on Co-ordination of Galactic Research}, held near Stockholm. There
was reasonable, but no considerable overlap with the first meeting among the
participants. Van Rhijn was absent. By now he was beyond retirement age, but had
taken the directorship upon himself as long as no successor had arrived.
The format was the same, but Adriaan Blaauw this time coordinated a
small group of editors, who shared the burden of this work. It is beyond the
scope of this article to discuss its proceedings in more detail.
\bigskip

The change from emphasis on completion of Kapteyn’s {\it Plan of Selected Areas}
that had been the focus of van Rhijn’s efforts was replaced by directing the
efforts to specific surveys related to specific questions concerning aspects
of the quest to determine the structure, kinematic, dynamics and eventually
formation and evolution of the Milky Way Galaxy.  In this sense it was the
unofficial end of the {\it Plan of Selected Areas}, since -- as noted above --
IAU Commission 32 remained operational until 1958, continued as a sub-commission
within Commission 33 and was formally disbanded only at the General
Assembly in Brighton in 1970. This last action then
should be seen as  the official end. In the end 28 observatories from 11
different countries had contributed to the Kapteyn's {\it Plan of Selected
Areas}.

For the Kapteyn Laboratorium the outcome of the first meeting was profound in
that it provided a focus on an extensive, long-term undertaking that would
occupy Lukas Plaut and a large part of the Laboratorium’s manpower
for may years. I will turn to this next.

\section{Groningen-Palomar Variable-Star Survey}

One of the major desiderata of the Vosbergen meeting was to undertake what
became known as the  {\it Palomar-Groningen Variable-Star Survey}, which
concerned variable stars in the Galactic bulge and halo. It was proposed at
Vosbergen by Walter Baade as a means of finding out what the stellar
distribution in the bulge and halo of the Galaxy was and to investigate the
differences and similarities between the Milky Way Galaxy and the Andromeda
Nebula. The halo was defined primarily by the globular clusters as tracers of
Baade’s Population II, introduced in 1944 upon resolving stars in the Andromeda
Nebula and companion dwarf galaxies. Baade had noted that the brightest stars
in the halo Population II were red (giant) stars, while in the disk Population
I these were blue OB-stars. This resulted in the realization that there was a
fundamental difference between stars in the disk and halo. At the time of the
Vosbergen meeting the situation was (see \cite{Vosbergen}, p.4-5):
\begmarg
From all the evidence available at present it seems that the distribution of
the population II objects is symmetrical with respect to the axis of rotation
of the Galaxy. (Distribution of globular clusters in the Galaxy, distribution
of population II objects in extra-galactic nebulae.) The first task of future
work, therefore, will be to determine the way in which the density varies with
the distance from the Galactic Centre, […] The following proposal was made by
Baade.

For the determination of the surfaces of equal density in the halo, we can
restrict ourselves to determining the density distribution in a cross-section,
perpendicular to the Galactic plane and going through the Sun and the axis of
rotation. […] It is proposed that, to begin with, three fields be chosen,
centered at approximately 
$l$=327\degs,$b$=about 20\degs (latitude as close to the nucleus as
interstellar absorption permits);
$l$=327\degs,$b$=45\degs;
$l$=147\degs, $b$=somewhere about 10\degs.\\
The ideal instrument for such a survey would be the 48-inch Palomar Schmidt,
which may become free in the next few years after the sky survey is concluded.
It gives fields of 7\degs$\times$7\degs, free from vignetting and can
easily reach the 20th photographic magnitude in 10-min. exposures. This
provides an ample margin to make sure that the survey will be complete up to
17.5 median magnitude corrected for absorption, i.e. distances up to 30
kiloparsecs for the RR Lyrae variables — even if the absorption in one of the
fields should amount to one or one and a half magnitudes. In the case of the
two latter fields it is very probable that regions of heavy or irregular
absorption can be avoided. As to the first, it is a matter of searching the
sky survey plates in this general direction before deciding on the most
suitable co-ordinates.
\endmarg

These coordinates are of course ‘old’ Galactic coordinates with the zero-point
of longitude is at the crossing of the Galactic and celestial equators.
Longitude 327\degs\ then
is that of the center of the  Galaxy. The fields were as close as possible
above the Galactic center, 45\degs\ up
from the center and towards the anticenter close to the Galactic plane.
The sky survey referred to is
the {\it National Geographic Society – Palomar Observatory Sky
Survey (NGS-POSS)}, which was substantially completed in the middle of 1956.

\begin{figure}[t]
\sidecaption[t]
\includegraphics[width=0.64\textwidth]{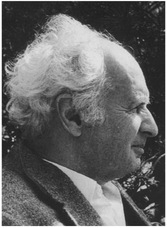}
\caption{\normalsize Lukas Plaut at later age. From the collection of
Adriaan Blaauw, Kapteyn Astronomical Institute.}
\label{fig:LPlaut}
\end{figure}

At the Vosbergen meeting at had become clear that Lukas Plaut
(Fig.~\ref{fig:LPlaut}) and the Kapteyn
Laboratorium would be prepared to take this project, the
{\it Palomar-Groningen Variable-Star Survey}, upon them. For the precise choice
of the fields Plaut relied on Baade, who had surveyed the region around the
center extensively both with the Palomar 18-inch Schmidt and the Mount Wilson
100-inch telescopes. The 18-inch Schmidt had been an initiative of Fritz Zwicky
(1898--1974).  He and Walter Baade had proposed in 1934 that supernovae were
the source of cosmic rays, and Zwicky wanted a small, wide-angle telescope to
search external galaxies for supernovae. The choice for a Schmidt telescope is
undoubtedly due to Baade. The design of a wide-angle telescope with
a spherical mirror and objective corrector plate to correct for spherical
aberration was first made by Bernhard Woldemar Schmidt (1879--1935)
at Hamburg Observatory where Baade worked
before joining Mount Wilson Observatory in 1931.
The 18-inch was the first telescope on
Palomar Mountain, becoming operational in 1936, well before the 48-inch (1948)
and 200-inch (1949) telescopes. Zwicky was a stubborn and ill-tempered person,
and regarded the 18-inch as his telescope. In the beginning he and Baade were
on good terms and published together, but later they avoided each other. In
the late 1930s Baade used the 18-inch, which recorded images of the sky on
photographic film over a circular area of almost 9 degrees diameter, to map a
large region around the center of the Galaxy. He did this in red light (with a
panchromatic film and a red filter), identifying areas of little extinction.

At the Vosbergen meeting, Plaut estimated the time required to perform the
blinking of plates and determination of light curves of RR Lyrae stars and
other variables as `something of the order of 15,000 hours or six man-years'.
At the start of the project four different fields were selected, three near
the center in new coordinates ($l$,$b$) at (0\degs,+29\degs),
(4\degs,+12\degs), and (0\degs,-10\degs) plus a field in a direction
more or less
perpendicular to this at (82\degs,+11\degs). Most observing was done for the
first three fields during two extended periods in California by Plaut in
1956 and 1959, and some plates were taken in between by other observers. The
three-year time-span allowed discovery of long-period variables. The fourth
field was never completed.

The organization of the survey took much time and also irritation on the part
of van Rhijn and Plaut. Much depended on Walter Baade, who was responsible for
choosing the exact location of the fields and
organizing the access to the Palomar 48-inch Telescope with Ira Sprague Bowen
(1898--1973), the director
of the Mount Wilson and Palomar Observatories.
In  a letter to Jan Oort of May 4, 1954 (in the Oort
Archives, Nr. 149), Pieter van Rhijn complained (my translation):
\begmarg
The correspondence between Baade and me has been as follows: Baade wrote me
Oct. 13, 1953, that he himself was in favor of the plan to search for the
variable stars in some fields in the plane perpendicular to the Milky Way
with the Schmidt Palomar, but that he still had to consult with Bowen about
the execution of the plan, Also the possibility, that Plaut would cooperate
in the execution of the plan at Mount Palomar, would be considered. In a short
letter of  Jan. 8, 1954, I asked Baade if he had already spoken to Bowen about
the matter and if there was a reasonable chance, that the matter would go
ahead. I repeated the same question on Feb. 9, `54. To date I have received
no reply from Baade. The way Baade has handled this case has annoyed me.
Surely the least he could have done was to answer me to my reasonable question
of whether there was a good chance that the plan would be realized. Instead,
he makes me wait six months and then asks you to tell me that he cannot give
accurate dates for the work. Dates that I, you should note, had not even asked
for. The sentence in which he says he was 'much amused by my writing' I will
leave unmentioned. Such remarks will not promote good cooperation.
\endmarg

\noindent
Oort wrote back two days later (same source, again my translation):
\begmarg
\noindent
Dear Piet,\\
You should not take such statements from Baade too seriously, nor his
non-answers to letters. I do nor know at all what he was referring to in the
phrase in question, but I find it hard to imagine that it would be worthwhile
to get angry or annoyed about. After all, people often have strange manners of
correspondence. If I wanted to be annoyed with Baade I would have jumped out
of my skin last year before his visit here. He did not let me hear anything
about his arrival for months, despite my repeated questions, and then finally
wrote, after the date that had been firmly agreed upon had already passed,
that he had come into conflict with other appointments and promises that he
had made before and had forgotten again. And so on and so forth. We also have
had more experiences like this. When he was here, it was most pleasant. We
benefited enormously from his stay, also the students did, with whom he was
very kind.

This is to help you get over your annoyance, which is never pleasant or helpful.
\endmarg
\bigskip

It must have been quite difficult also to find finances for Plaut's necessary
travel to Pasadena and Palomar Observatory. According to Plaut’s acknowledgments
in the publications resulting from the survey it involved  originally the
United States Educational Foundation in the Netherlands (Fulbright Travel
Grant), Commission 38 of the International Astronomical Union, and the
Netherlands Organization for the Advancement of Pure Research (ZWO).
Later the Leids Kerkhoven-Bosscha Fund was added. The Kapteyn Laboratorium would
have contributed also, but that is (understandably)
not mentioned in the acknowledgments. 

Almost 400 plates were taken in the photographic wavelength region plus about
90 in the photovisual to measure colors. 
In the course of the project
a field called {\it Baade’s Window} was added as a final part of the
{\it Variable-Star Survey}. This field had been discovered
also by Baade as part of his survey of the region around the Galactic center
and is a field very close (only four degrees) from the actual direction of the
center, ($l$,$b$) = (1\degs,-4\degs). It is in the brightest part of the Milky
Way in this general direction
and very uniform, suggesting very little extinction. It is centered on the
globular cluster NGC 6522, which is at about 7.5 kpc along the line of sight, so
in space close the center of the
Galaxy; the extinction is not zero, but about ‘only’ 2.5 magnitudes in
photographic magnitudes according to Baade \cite{Baade1946}.
Baade had studied this window extensively with the
Mount Wilson 100-inch telescope and discovered some 159 RR Lyraes (also known
as cluster-type variables, because they occur in large numbers in globular
clusters). Plaut reanalyzed these data, redetermining their
periods and the photometric zero-points, using the 1-meter telescope by then
available at the La Silla site of the European Southern Observatory. The
results of the {\it Palomar-Groningen Variable-Star Survey} were published by
Plaut in six papers, the last one appearing in 1973 (\cite{Plaut1973}, which
has references to the earlier papers). It was concluded in 1975 by Jan Oort and
Lukas Plaut in a discussion  of the whole project in terms of the distribution
of  RR Lyrae variables in the inner region and halo of the Galaxy and its
implication for the distance to its center and the rotation constants,
\cite{OP1975}.
\bigskip

The main effort in the program was to find the variable stars. The problem was
the fact that blinking plates was tedious, tiring, slow and difficult to do
consistently for anything more than a few hours at most. And therefore time
consuming and requiring an experienced astronomer. It was known very well that
completeness levels were poor except for the largest amplitudes. In general
three methods had been in use that provided useful results, always involving
two photographic plates recorded at different epochs with the same telescope
and the same exposure time. These were the following:
The first was direct blinking in a comparator or
microscope, where the  two plates were examined by rapidly switching the view
between the two. The second method involved superimposing of a
negative of one of the plates on a
positive of the other, so that the images of the non-variable stars were
canceling. A third option was to examine
the plates in a stereo-comparator, which means
viewing two plates at the same time each with one eye. All these methods
produced rather low completenesses, as could be judged from repeating the
process later by the same or a different person. New variables then
turned up, but
also some were not rediscovered. Of course the chance of discovery depended
also on the average magnitude of the star and magnitude difference between
the two observations. Whichever method was used, searching for variable stars
remained a tedious business.

The classical study in this issue is that by Hendrik van Gent in 1933,
who worked for a
long time as the Leiden observer at the Unie Sterrewag in Johannesburg,
\cite{vGent33}.
Using a blink microscope, he found the chance of discovery on a single pair
of Franklin-Adams plates around magnitude 13 or 14 and variations of 1 to 2
magnitudes to be usually below 10\%. The actual value depended
of course on the amplitude of the variation and the phase in the light curve.
Astronomers used often ten or more pairs of plates, but even then the
incompleteness can be still quite significant. Plaut in the famous volume V of
the compendium {\it Stars and Stellar Systems}, \cite{Plaut1965},
his Appendix I, summarized
the completeness issue, and provided further details such as the dependence
on type of variable. Obviously completeness in detecting variables was a
major concern. 

This had worried Plaut for a long time. At the suggestion of a young
physicist in Groningen (Herman de Lang) he  tried a new method by
observing one plate in the blink 
comparator through blue and one through a red filter, such that
non-variable stars were black while variable stars because of their differing
dimension on the plates had a blue or red ring around them. He involved student
Jan Borgman in this, which resulted in a short note \cite{PB1954}, reporting
that the method worked well, but also announcing they (this really was Borgman)
were developing an electronic device for the purpose of identifying variable
objects more completely and conveniently. Indeed Jan Borgman designed and
built an instrument for electronic discovery of variables in the framework of
his PhD research. The thesis that resulted from this was defended in 1956
under Hendrik Brinkman  and was entitled {\it Electronic scanning
for variable stars}, \cite{Borgthesis}. Brinkman was a professor of
experimental physics in Groningen. He worked in many areas of physics, and is
particularly noted for the founding of the nuclear physics accelerator
institute in Groningen.

\begin{figure}[t]
\sidecaption[t]
\includegraphics[width=0.64\textwidth]{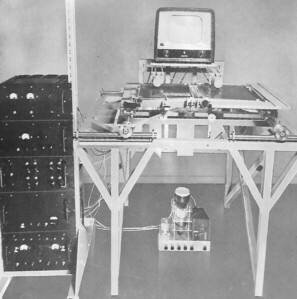}
\caption{\normalsize  The electronic instrument designed and constructed by
Jan Borgman to identify variable stars
by comparing a pair of photographic plates. See
text for explanation. This work constituted his PhD
thesis under physics professor
Hendrik Brinkman. From Borgman’s PhD thesis \cite{Borgthesis}.}
\label{fig:Borgmanapp}
\end{figure}

The design was in principle straightforward. It made use of the principle of
so-called flying-spot scanning developed for television. On a cathode ray tube,
sitting on the floor in Fig.~\ref{fig:Borgmanapp}, a rectangular grid or raster
is formed on the screen, and a bright spot is formed moving across that grid.
This spot is focused by an optical system from below onto the
two photographic plates lying on the table. The optical system divides the light
in two by a semi-transparent/semi-reflecting mirror. This system is positioned
such that the same portions of the plates are covered by this grid. The light
spot on the screen on the cathode ray tube results in two beams of light on
the photographic emulsions, at any time illuminating the same position. As the
spot moves across the raster on the screen of the cathode ray tube the
corresponding spots on the photographic plates are illuminated simultaneously.
Two photomultiplier tubes above the plates convert the light transmitted by the
photographic emulsions into electrical impulses (a video signal). These signals
are then subtracted electronically and the resulting signal is amplified and
transferred to  the grid of another cathode ray tube, whose dot is synchronous
with the dot on the first tube. This then is displayed on the television screen
on top. The image shows noise (resulting chiefly from the grains in the
photographic emulsions) as a gray background level in positions where the two
plates are the same, but show a white or black dot where a variable star is
present. Fig.~\ref{fig:JB}  shows the results.

\begin{figure}[t]
\begin{center}
\includegraphics[width=0.88\textwidth]{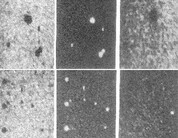}
\end{center}
\caption{\normalsize The left panels show images of parts of plates obtained
with the Palomar 48-inch Schmidt telescope. The middle panels show the same
areas from a different plate but now with the video signal’s phase reversed.
The right panels show the difference signals. The variable star in the top
panels  changed from magnitude 12.1 to 13.5, and at the bottom from 17.9 to
below the detection limit of 19. Adapted from Borgman’s contribution to
\cite{Voetspoor}.   Kapteyn Astronomical Institute.}
\label{fig:JB}
\end{figure}

Plaut used the Borgman machine to measure up the plates for the {\it
Palomar-Groningen Variable-Star Survey}. The Borgman machine was essential
for the project because it speeded up the blinking
process substantially; with classical methods it would have been
impossible within any
reasonable timescale. This was the first time the device
was used extensively. He reported that it took 40 hours to search a plate pair
(square of 14-inch sides or about 6\degspt5). The work was still strenuous and
could not be done for more than 4 hours a day, so searching each pair had
to be spread over at least 10 days. The detection probability for
amplitudes in excess of 1\magspt5 was over 80\%, and still of order 50\%\ for
1 magnitude amplitude, at least for the first field.  For the more crowded field
it was worse. As mentioned, Plaut blinked ten pairs of plates for each of the
three fields. Altogether the three fields yielded about 1100 RR Lyraes,
with 110 more from the reanalysis of Baade’s Window,  and on top of that
determined magnitudes for all objects found on the total set of almost
500 plates. In addition the survey
yielded similar numbers of long-period and other types of variable stars,
including eclipsing binaries. 

The estimation of brightnesses was done visually using the technique introduced
by Friedrich Wilhelm August Argelander (1799--1875) and designated the
`Argelander step estimation method', in which a brighter and a fainter
comparison star are used and it is estimated where the variable’s brightness
is positioned as a step value of the fraction 
of the magnitude difference between these stars.
The fascinating history of these magnitude estimations from the Herschels to
Argelander and beyond is summarized by John B. Hearshaw (\cite{JBH}, chapter 2).

The {\it  Palomar-Groningen Variable-Star Survey} was a major, long term
undertaking, extending well beyond van Rhijn’s directorate, actually even
beyond that of his successor Adriaan Blaauw. It dominated Plaut’s research
efforts for the rest of his academic career. The  most important
results announced in the paper by Oort and Plaut in 1975 is that the
distance to the Galactic center is 8.7 kpc, with an uncertainty of 0.6 kpc,
and that the density distribution in the bulge and halo
to about 5 kpc is almost spherical and
varies approximately as $R^{-3}$ \cite{OP1975}. This constitutes fundamental,
very important results of a very long-term investment of 
time and effort of Lukas Plaut, but of course does not generate a stream of
press releases announcing breakthroughs.

\section{Van Rhijn’s final research and seventieth birthday}
The final installment of the {\it Bergedorf(-Groningen-Harvard)
Spektral-Durchmusterung} appeared at the time of the Vosbergen Symposium
(in the same year 1953), as did van Rhijn's paper with the first results of the
spectrographic work with his
telescope addressing the issue of the wavelength dependence of interstellar
extinction. In the year before, 1952, finally the measurement of the Harvard
plates for positions and magnitudes of stars in the {\it Selected Areas Special
Plan} had been completed and published. It was presented as a separate
publication by the Kapteyn’s Laboratorium with van Rhijn as first author and
Kapteyn posthumously as the only
other author \cite{vRK1952}.  But other work was going on in Groningen -- other
than the programs by Lukas Plaut on variable stars and by Jan Borgman on
spectral energy distributions -- in the framework of the {\it Plan of Selected
Areas}, as can be seen from the IAU Commission 32 reports.

In the 1955-report there is a paragraph on collaborations between `Alger and
Groningen'. This involved proper motions near the equator. The measurements of
these had been performed at the  Kapteyn Laboratorium and the plates had been
taken at Alger. This institution had been established by l’Observatoire de
Paris as l'Observatoire astronomique de Bouzareah, so named after the suburb
where it had been located.
The observatory had been involved in the {\it Carte du Ciel}
project, in which it had been assigned the equatorial zone. These plates served
as first epoch plates. Their useful extent covered 30-80\%\ of the 
3\degspt5$\times$3\degspt5 area of those of the {\it Spektral Durchmusterung},
which served as second epoch plates. At Alger some special plates were taken
to investigate possible systematic errors.

The project has been described in some detail by Plaut in \cite{Voetspoor}.
First relative proper motions were derived for 10,960 stars.
These were made absolute using various fundamental
catalogues for which there was overlap of 850 stars. The results were published
in 1955 and 1959 in two volumes of the {\it Publications and
of the Kapteyn Astronomical Laboratory at Groningen}, the first part  
on the relative proper motions by van Rhijn and Plaut \cite{vRP1955} and the
part on the absolute motions by Plaut by himself \cite{Plaut1959}. This enormous
volume of proper motions constituted another major investment of time. It is 
not clear what use has been made of it, since until recently these publications
had not  been included in ADS (see below).

The second project concerned also proper motions and was a collaboration with
the Yale-Columbia Southern station. It was aimed at measuring proper motions of
the stars in the zone of declination -15\degs.  The first epoch plates date
from around 1927 and the  interval was twenty-four years, which meant that the
probable error of the proper-motion components was $\pm$0\secspt0022. The
telescope was the 26-inch refractor originally erected in 1924 in Johannesburg
under Frank Schlesinger  (1871--1943), director of Yale Observatory. Actually,
when he was at Yale in 1922-1924, Jan Oort was  briefly considered to be
appointed the astronomer responsible for the setting up of the telescope. 
Oort went to Leiden instead, but Schlesinger’s intervention had meant
for Oort an
earlier and shorter military service (see \cite{JHObiog}).
In 1952 the telescope was
moved to Mount Stromlo near Canberra, Australia, where it was destroyed
in a bush fire in 2003. The proper motion survey was set up by Dirk Brouwer
and Jan Schilt after they had as directors of respectively Yale and Columbia
initiated the collaboration and orchestrated the move from Johannesburg to
Canberra. The plates were measured in Groningen, but as far as I am aware
the results were never published. 

These two programs did constitute the final contribution from Groningen to
the {\it Plan of Selected Areas}.
\bigskip

As indicated above, Borgman took over the program of determining the wavelength
dependence of interstellar extinction with the Groningen telescope. As already
described above he published two more papers on this subject \cite{JB1954} and
\cite{JB1956}. As he stated in the
Introduction of the second of these two papers :
\begmarg
The publication of these results concludes the present investigation; the
material available for our instrument and method is almost exhausted and,
moreover, it is considered that the present state of photo-electric techniques
does no longer justify a photographic investigation of the absorption law.
\endmarg

So the paper by Borgman in 1956 marked the end of research with the  Groningen
telescope. The total harvest was three papers on wavelength dependence of
extinction, when the technique was actually no longer state of
the art.  Most likely the telescope
had also been used for student training (as far as there were students).

I will briefly summarize Borgman’s further career.
Borgman after completing the work with the Groningen telescope first developed
the instrument to identify variable stars for Plaut and the {\it
Palomar-Groningen Survey} and then moved on to photoelectric photometry. He was
introduced to the Yerkes and McDonald Observatories, undoubtedly through
Adriaan Blaauw, who was adjunct-director at Yerkes under Gerard Kuiper as
director, and in 1957 the new director in Groningen.  Borgman collaborated with
Bill Morgan and Bengt Georg Daniel Str\"omgren
(1908--1987) at Yerkes and McDonald Observatories and with  Harold Lester
Johnson (1921--1980) of Lowell Observatory and observed extensively at McDonald
and Lowell. Borgman  developed a seven color narrow-band (width 5 to 10\%\ of
the wavelength)   photometric system and designed a new d.c. amplifier (which
is used to  amplify the output photocurrent of the 1p21 photomultiplier used
in photometers). With this seven color system he analyzed the spectra of
OB-stars and attacked again the problem of the wavelength dependence
of extinction.
Eventually he became director of the new
Kapteyn Observatory in Roden (15 km SW  from Groningen) and his career
moved to infrared, ultraviolet, and balloon and space astronomy and
in the 1970s on to
science administration. He became Dean of the Faculty of
Mathematics and Natural Sciences, Rector Magnificus and Chairman of the
Governing Board, all of Groningen University, before moving to Den Haag
(the Hague) as
chairman of the Netherlands Organization for Scientific Research NWO and
finally Brussels as first chairman of the  European Science and Technology
Assembly of the European Commission.
\bigskip

In 1956 van Rhijn turned seventy, the age of retirement. There was a
small event  celebrating this milestone, honoring him for his accomplishments
and major contributions. A special booklet was produced entitled
in English translation {\it In
Kapteyn’s footsteps: Sketches of the work program of Dutch astronomers,
offered by friends and colleagues to Prof. Dr P.J. van Rhijn on the occasion of
his seventieth birthday}, published by The Dutch Astronomers Club and dated
Groningen, April 7, 1956. Van Rhijn’s birthday was March 24. The date April 7,
1956 was a Saturday, so it was offered at a
special meeting of the NAC on that date held in Groningen. 
The contributions to the booklet do not look like written accounts of oral
presentations, but rather short contributions written for the occasion.

\begin{figure}[t]
\begin{center}
\includegraphics[width=0.465\textwidth]{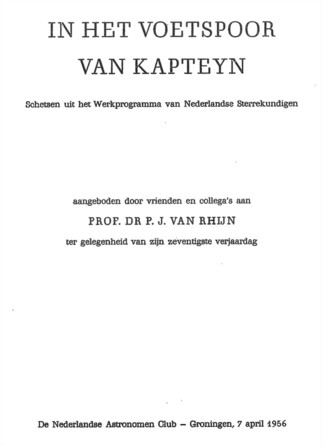}
\includegraphics[width=0.525\textwidth]{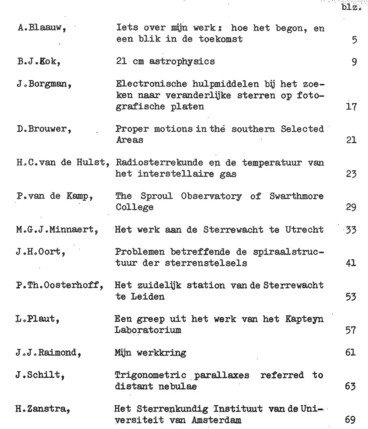}
\end{center}
\caption{\normalsize Title page and table of contents of the booklet
 \textit{In het voetspoor van Kapteyn} presented to van Rhijn at his 70th
birthday.   Kapteyn Astronomical Institute.}
\label{fig:voetspoor}
\end{figure}

\begin{figure}[t]
\sidecaption[t]
\includegraphics[width=0.64\textwidth]{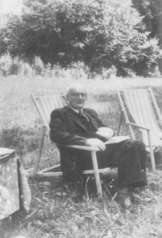}
\caption{\normalsize  Photograph of van Rhijn as frontispiece of the
`Voetspoor' booklet of Fig.~\ref{fig:voetspoor}. There is no caption or
accompanying text.   Kapteyn Astronomical Institute.}
\label{fig:frontispiece}
\end{figure}

It gives the impression of a sloppy, hastily prepared document (see
Fig.~\ref{fig:voetspoor}). It is produced using offset printing, bound
cheaply with a soft cover without any illustration. On the Contents page the
entry has Bart Bok’s name typed as  ‘B.J. Kok’, and then corrected by
over-typing the ‘K’ with a ‘B’. There is no introduction, and the editor
or coordinator is not identified . There is a picture of
van Rhijn in the front of the booklet (Fig.~\ref{fig:frontispiece}),
but without a caption.
Lukas Plaut has, as follows from the correspondence below, organized
the  whole and collected the contributions into a single booklet.

The title refers to an expression quite similar in Dutch and in English,
meaning ‘following someone's example’ or ‘going to do something similar to what
someone else has done before’. Maybe unintended this  points at a lack of
originality; after all in a literal sense putting your feet on the same spot
as another person means you are not breaking any new ground. This choice for the
title of what could have been a festschrift is a rather minimal token of
appreciation, just like the fact noted in the Introduction of the lack of any
prominent obituary in international journals after van Rhijn’s death.

There is a little bit more background in the Oort-Blaauw correspondence.
Jan Oort mentioned on February 21, 1955, that he had received a letter from
Lukas Plaut (not present in the Oort Archives), and noted he is
uncertain what to do about this. It is a typed letter in Dutch, so what follows
is my translation.
\begmarg
From Plaut I received a letter these days, a copy of which you will
find enclosed. I do not really know what to make of it. I think it is very
difficult for the people he mentions to write something in 2 or 3 pages that
would be of use to van Rhijn. That is why I think it would be more appropriate
to offer his complete work in beautifully bound volumes, and perhaps also an
album with photos and signatures of all the people who have just collaborated
with him. Let me or Plaut hear what you think about this.
\endmarg
On 27 February 1955  Blaauw wrote to Oort from Yerkes
(handwritten in Dutch, my translation).
\begmarg
I am as unenthusiastic as you are about Plaut's proposal concerning the tribute
for van Rhijn. I seriously doubt that van Rhijn would care much for a
collection of essays as proposed by Plaut, especially if we apparently do not
think it is worthwhile to print them. I also doubt that I could get people
like Ambartsumian and Baade to write such an unpublishable article.

I believe that the tribute should in any case have a rather intimate character
and should be limited to those who have worked with van Rhijn. I would feel
very much for an album with signatures and photographs and perhaps letters
addressed to van Rhijn from these fellow workers. One could then indeed expand
it with a nicely bound collection of van Rhijn's works, but I do not find that
solution ideal. The collection cannot be complete, for how can the work on the
Bergedorfer Spektral Durchmusterung be represented in it. And would van Rhijn
really care enough? He will probably rarely look in it, since he has his notes
in his much-used hand copies anyway. What do you think of the idea of having
all those who received their doctorates from him write a short article about
the current view of the subject of their dissertation? Such articles, which at
the same time give a picture of the development of astronomy, might be of
interest to a wider public and could be printed somewhere in Dutch or English.
Writing this would not be any problem for most participants. Would the articles
not be suitable material for Hemel en Dampkring, e.g. a special issue. That
would reduce the costs.
\endmarg
Interestingly they refer to him as ‘van Rhijn’ and not  ‘Piet’ or ‘Pieter’,
which if talking about a mutual good friend would be more logical
(however, they refer to ‘Plaut’ rather than ‘Lukas’). 
They used first names throughout their correspondence for most other colleagues.
On October 12 Blaauw wrote (typed and in English):
\begmarg
I have been very busy lately and not given much thought to the Groningen
succession. I just received Plaut’s letter for a short article, which I shall
write.
\endmarg
No further mention of the whole business occurs in subsequent letters
between Oort and Blaauw. From this correspondence it appears that Lukas
Plaut played a definite 
role in setting up the tribute on the occasion of van Rhijn’s  imminent
retirement and inviting persons 
to contribute to a small publication with collected contributions by
colleagues and friends. Plaut edited the booklet and in his modesty
felt it inappropriate for 
him to add an introduction, or even mention his role.

The 13 contributors comprise van Rhijn’s students
Jan Schilt, Jan Oort, Bart Bok,
Jean-Jacques Raimond, Peter van de Kamp, and Adriaan Blaauw, and also 
Jan Borgman. Dirk Brouwer was a Leiden PhD but
collaborated in the {\it Plan of Selected Areas}. Lukas Plaut was a colleague
and collaborator, Henk van de Hulst and Piet Oosterhoff were important
colleagues from Leiden, while Herman Zanstra  and Marcel Minnaert
were directors of the astronomy institutes in Amsterdam and Utrecht. All of them
described their {\it own} work or that of their Institutions,  usually, but not
even always, concluding with a sentence or two of personal appreciation of van
Rhijn, but nowhere is anything included like a summary of van Rhijn’s
contributions or a statement on the importance of his work. Those persons that
had emigrated to the USA and established themselves there permanently (Bok,
Brouwer and van de Kamp) wrote in English, the rest in Dutch. It seems
altogether not a volume intended for a wide audience. 

The Oort Archives (No. 149) contain a text of the speech Oort gave at
the special NAC  meeting, presenting the booklet. He acknowledged Plaut’s
editing work, regretting he was not present. He described the set-up as
letting each contributor address what interested them at the time. Oort
does review van Rhijn’s work in some detail and noted (my translation)
\begmarg
Prof. van Rhijn saw, as it were, a great task before him when he was appointed
to the Groningen Laboratory, and he has been completely faithful to this task
during the long period he has been attached to it. This commands admiration.
It is clear from the scope of this work, but also from its nature. It is a
task that only someone capable of self-sacrifice can accomplish.
\endmarg

\noindent
And Oort finished by saying
\begmarg
Both you and I will be reminded on this occasion of a day when, I believe
almost ten years ago, in more intimate circle, we also faced each other in
much the same way. What I remember most about that day, Piet, is how
beautifully your friend van der Leeuw, who died so much too soon, spoke to you
then.

Van der Leeuw then spoke to you. He more or less said that the main reason why
he felt such a warm friendship for you was your aversion to making too much
fuss and your simplicity, which springs from a deep-rooted sense of truth.

It is not only van der Leeuw who saw this in you. We all appreciate you for
this.

It is also because of this that today, instead of the frills of an official
tribute,  we only wanted to give you something of
that which ultimately affects us most deeply: our own work.
\endmarg

\noindent
It would have been fitting had Oort provided an introduction along these lines
or an advance summary of this speech to include in the booklet.
\bigskip

In his final years before retirement and actually also beyond van Rhijn
published three more papers in the {\it Publications of the Astronomical
Laboratory at Groningen}, the final one even the last volume of that series.
The institute publications in the Netherlands existed alongside the {\it
Bulletin of the Astronomical Institutes of the Netherlands}, which had been
founded as a national journal in 1921. The {\it Supplement Series} of the
BAN was instituted in the 1960s, the first volume starting in 1966 and the
institute series were already in anticipation phased out then. Between 1960
and 1966 there had been no need for a Supplement publication from Groningen,
but eventually Plaut used the new series and published his measurements for
the {\it Palomar-Groningen Survey} in the Supplements of the BAN.

Van Rhijn’s final research was quite original and timely. It actually resulted
also from the Vosbergen Conference, where the local structure in the Galaxy was
seen as an important and urgent subject of research.  He started
determining the space distribution of A0-stars near the Sun from low latitude
Selected Areas and for higher latitudes for A0- to A5-stars, also in
Selected Areas, but other sources as well as in cases, where
the number of A0-stars in
the Areas is too small, see \cite{PJvR1955}. It was suggested at
Vosbergen to use A0-stars because of their small spread in intrinsic luminosity.
The spectral types were
taken from the {\it Henri Draper Catalogue}, {\it Bergedorf(-Groningen-Harvard)
Spectral-Durchmusterung} and the Potsdam southern equivalent, and the magnitudes
were corrected for absorption from the color excesses, which could be determined
since the spectral types were known. The densities then follow from the
distribution of corrected apparent magnitudes and the luminosity function,
which he assumed to be a Gaussian around a mean value with 0.5 magnitudes
dispersion. Within 200 pc from the plane of the Galaxy he found that the
A-star density correlated with the hydrogen density derived from the 21-cm
surveys at Kootwijk by Leiden observers.

Next van Rhijn turned to K-giants, \cite{PJvR1956}, doing essentially the same
thing. The percentages of giants among K-stars was taken to be the same as in
various detailed  studies in the literature. Using new determinations for the
hydrogen densities derived in Leiden, he found, while  for the A-stars that
were closer than
50 pc from the Galactic plane their density correlated with that of
hydrogen, for K-giants there was no evidence for such a correlation.  Above
50 pc neither the A-star nor the K-star distributions correlated with that of
the hydrogen. Typically for van Rhijn, there is no discussion on what this
would mean for understanding Galactic  structure or the presence of spiral arms.

Van Rhijn’s final research paper appeared after his retirement, and concerned
{\it The ellipsoidal velocity distribution of the A stars as a function of the
distance from the Galactic plane}, \cite{PJvR1960}. He extended his study of
A0 stars to A5 at low latitudes
using proper motions to derive the distribution of their
random motions, and from this the properties of the Schwarzschild distribution,
the dispersions in the principal directions and the vertex deviation (see
Section~\ref{sect:extinction}), at various distances from the plane of the
Galaxy. Solving for the axes of the velocity ellipsoid for a sample of stars
can be done from their distribution of proper motions (the components parallel
and perpendicular to the Galactic equator) using a method developed in 1948 in
Leiden by Coert Hendrik Hins (1890--1951) and Adriaan Blaauw. Van Rhijn found
the dispersion of the motions to increase with vertical distance, while their
ratio’s remained more or less constant.
These and the vertex deviation of 26\degs\
were not unreasonable. He attributed the change in mean vertical velocities with
height above the plane to a mixture of two components with a different velocity
dispersion. Although he investigated the possibility of a difference in spectral
types among the two distributions and presented for this a table, but he
failed to discuss this table and to draw any conclusions from it, nor did he
discuss the implication for our understanding of the structure of the Galaxy.
Note that this was published in 1960,
when the concept of Stellar Populations was widely
accepted and a discussion of the results in this context would have been highly
appropriate. The two components had dispersions of about 10 and 25 km/sec (the
smaller one contributing some 90\%), so could in terms
of the scheme of the Vatican Symposium of 1957
be interpreted as young population I and the (older) disk population
\cite{AB1965}. In  that context it would have been no
surprise that A-stars, being relatively young, would overwhelmingly
be part of the young Population I.

These three investigations, comprising van Rhijn’s final research, stemmed
from the Vosbergen meeting. The results were interesting and provided new
insight; yet, van Rhijn presented them without any real discussion on what the
impact on our understanding of
Galactic structure at a fundamental level would be. Unfortunately,
the impact as revealed by citations cannot be evaluated with the ADS. During
this research I discovered that the last six volumes of the {\it Publications
of the Astronomical Laboratory at Groningen} were not included in ADS. It
turned out that the scans were in the system, but for some reason the
corresponding records were not created. I was promised this would be corrected. 

\section{Van Rhijn's succession}\label{sect:PJvRsuccession}

Van Rhijn’s retirement would be at the end of of the academic year 1955-1956.
 The University of Groningen had the
choice between on the one hand continuing the Laboratorium and try and raise
its prominence to what it had been in Kapteyn’s days by appointing a prominent
astronomer or on the other close it and arrange for a physics professor to take
care of astronomy teaching as a task secondary to his main activities or
appoint an astronomer from Leiden (Utrecht and Amstardam were
possibly even too small in staff for this)
to come on a part-time basis to teach. In the
end they chose to pursue the first option. But it was clear that there was a
need for not only  a change in leadership but in addition a modernization of
the way the laboratory was
organized and associated with that new investments and
an increase in staff were required.
The complete reliance on other observatories providing
the plate material, to be measured and reduced in Groningen, was no longer a
guarantee for success.

Van Rhijn was very much concerned with the issue of his succession and the
future of his Laboratorium. Already on November 24, 1952, he wrote to Oort
(from the Oort Archives Nr. 149, my translation):
\begmarg
\noindent
Dear Jan,\\
I would like to write to you about a matter which has been occupying me for
some time now and which also gives me concern, namely the future development
of the Kapteyn Laboratorium.

As you know, the possibilities for work here are very limited, Kapteyn has
tried to establish a proper possibility for scientific research in Groningen,
which he has been refused. The only possibility, which I saw to change this,
was the purchase of a second-hand mirror, which happened to be offered to me
at the time. I have had the greatest difficulty in obtaining permission to
install the instrument on the roof of the laboratory from money from private
funds, that is how strongly `Den Haag' [the Minister] was opposed to expansion.
The instrument is very useful […] But the weather in Groningen is too
bad for photometric work. A few years ago I applied to Z.W.O. for a
spectrograph, which of course can be used even if the sky is not too clear.
The application was turned down. And so we are still stuck with  very limited
possibilities of research. Many times I have received a refusal on a request
to take plates and even if they were granted, the work of the Groningen
laboratory was subordinated to that of the particular
Observatory itself. It is not my
intention to reproach anyone for this. But it does make research in Groningen
very difficult.

In a few years time I will have to resign and of course I am thinking
about the future of the Kapteyn Laboratorium. It has been suggested a few
times to close the laboratory as a financial cut-back measure. I would regret
this very much, not only because the University of Groningen is already
disadvantaged in several respects compared to Leiden and Utrecht, but also
because a concentration of astronomy does not seem to me to be in the best
interest of the enterprise. Just think of the time of Bakhuyzen in Leiden,
Nijland in Utrecht and Kapteyn in Groningen. Leiden and Utrecht were of little
importance for the development of astronomy at that time. Especially in such
times the importance of an independent institute, which goes its own way,
becomes apparent.

If, however, Groningen is to continue to exist, there will have to be other
possibilities for research here in the future than there have been so far.
But I have given this matter some thought and one obvious possibility is radio
research. In the decisions to be made concerning this research I would ask you
to consider the possibility that in the future the Groningen astronomers will
wish to participate in this research. I would also be glad to attend the
meetings of the Foundation, since the development of radio research in the
Netherlands is, in my opinion, also very important for the future of the
Kapteyn Laboratorium. What do you think of this?
\endmarg
Oort replied on  29 Noyember, 1952, saying among others  (Oort Archives Nr. 149; my translation):
\begmarg
The starting point is who you will be able to get as your successor upon
retirement. Should Blaauw by then be available for that, I would not be the
least bit worried about the future prosperity of the lab. 
\endmarg
He then went into a long argument why it is difficult to add another astronomer
to the Board of the Foundation for Radioastronomy, in the end suggesting 
\begmarg 
[...] that you attend a meeting occasionally, when an issue in which you have
a special interest is on the agenda? Another possibility would be to have
somewhat alternating meetings to discuss the technical issues
of the instruments and
the like and meetings to discuss the general scientific side, which could then
include a few more people.
\endmarg

The issue of Oort keeping van Rhijn out of the Board of the radio astronomy
foundation is outside the present scope, but has been treated in detail in
Astrid Elbers’ thorough study of the history of Dutch radio astronomy
\cite{Elbers} and need not be covered here. In the end the outcome was
that van Rhijn joined the board. On September 14, 1956, according to him a few
days before his formal retirement, van Rhijn wrote to Oort, saying he was
grateful the latter considered including others than Leiden astronomers in
the board of the Kerkhoven-Bosscha Fund, but complained rather bitterly that he
(and Kapteyn) had been denied an observatory with the argument that two in
Leiden and Utrecht was enough, but now a very expensive third one was being
built (in Dwingeloo) without involving the director in Groningen in this. 

Returning to the future of astronomy in Groningen, I quote from a letter of
van Rhijn to Oort of December 15,  1952 (Oort Archives, Nr. 149; my
translation). On the issue of appointing a successor or closing the
Laboratorium, he described the position of the Faculty:
\begmarg
In such a Faculty [covering natural sciences] astronomy
should, in the opinion of our Faculty, be represented. Astronomy is too
important to leave the teaching of that subject to a lecturer who comes to
Groningen once a fortnight from the west of the Netherlands,
\endmarg
This shows that there was a firm decision in Groningen to continue with a
professor and an institute in astronomy.  He continued:
\begmarg
If one develops the practice of astronomy at an institute to a very high degree
compared to others, then there is a danger that the smaller institutes will be
closed down as prosperity declines in the Netherlands.  Thus in 1932 the plan
was to close down the Observatory in Utrecht and the Kapteyn Laboratory was in
those days also on a proposal to disappear. All the more reason not to make
the difference in size and importance of the various institutes too great.

It is undoubtedly true that a competent astronomer, appointed as my successor,
will be able to bring the Kapteyn Laboratory to a favorable development.  But
it is precisely the question of how to make the appointment to Groningen
plausible for this man, and to this end new possibilities for scientific work
will have to be created here. Otherwise he will be grateful for the honor, but
look for a job elsewhere.

Sending a Groningen astronomer abroad to collect material could indeed
partially solve the difficulties. The Kapteyn Laboratorium would have to have
a workshop for the production of instruments and tools, which are not available
at the observatory. If one does not have access to a workshop, it will be very
difficult to undertake something new and one is dependent upon work of the same
nature as is already done at that observatory.

In this way one remains dependent upon the directors of other institutes. This
remains a major obstacle, as I have painfully experienced on several occasions.
\endmarg

Van Rhijn added to the letter:
\begmarg
\noindent
P.S. I am convinced of Blaauw's abilities. His articles on the O-associations
I have read. It is excellent work. It seems unlikely to me that he would be
willing to come from Yerkes to Groningen in a few years time. If he feels
somewhat at home in America he will stay there because of the greater
possibilities for research.
\endmarg

So, Jan Oort had been involved from an early stage on. It was obvious that his
favorite was Adriaan Blaauw. The problem was, of course, the latter's move to
Yerkes Observatory.

In the Oort Archives we can find some correspondence between Blaauw and Oort on
this issue. On March 8, 1956, Blaauw wrote (handwritten, in Dutch) to Oort
(my translation).
\begmarg
\noindent
Dear Jan,\\
As you will of course know, van Rhijn has indeed asked me, on behalf of the
Groningen Faculty, to succeed him. I have already replied to him that I am
seriously considering the offer, and have exchanged views with him about
certain conditions which I believe must be met in any case. However, I have
also made it clear that my decision will also depend on the future character
of my position here, which, now that the first three-year period ends soon,
will have to be reviewed in any case. On this last point I have not yet heard
a decision. Nevertheless, I am resolved to make the decision within a few weeks
whether or not I desire the Groningen position, at least in principle. Van
Rhijn would like a speedy decision.

I have already informed van Rhijn, among other things, that I believe that the
special character of the laboratory as a place for processing material obtained
elsewhere could be fruitfully maintained if the necessary arrangements were made
so that regularly Groningen observers could personally collect the material
elsewhere. This requires, among other things, a guarantee of a relatively high
annual travel fund, and an expansion of the staff with at least one young and
energetic astronomer.

A subject to which van Rhijn alluded but which I do not yet clearly understand
concerns the part that Groningen should play in radio astronomy. Van Rhijn did
mention that Groningen has a seat on the board of the Radio Foundation, but I
would like to be clear first of all of the view that you and Henk van de Hulst
have on this. You may remember that we touched on this point during our
conversation in Hamburg last summer, and my impression was that you were rather
reserved about the participation of other institutes in the 21 cm work.
[...] Would you write me what you
think now of a possible Groningen share in Dutch research? Of course I would
also like to hear your views on the Groningen research work in general.
\endmarg

He added that he will drop the plan to visit the Netherlands that
summer partly in view of his possibly going to Groningen,
but mainly because of a serious backlog of work.

This apparently crossed a letter by Oort to Blaauw on March 12, in which Oort
mentioned that he had already
heard from Maarten Schmidt that  Blaauw would not visit the
Netherlands that summer, but invited him to come in the fall for a few months
as lecturer on a temporary position. 

When Oort received Blaauw’s letter he answered `immediately' (March 14, 1956;
my translation):
\begmarg
\noindent
Dear Adriaan,\\
\noindent
[...]
I am glad that now Groningen has taken the first step and congratulate you with
the fact đat you have been placed as Number 1 on the list. I have not seen van
Rhijn for a long time, it seems as if his already very limited appetite for
travel has become even less lately.

I don't have to tell you how much I would enjoy your coming to
Groningen. There is not the slightest doubt in my mind that we would work
together pleasantly and fruitfully in all the areas you would like. I hope
that you feel the same way and that you would never feel that the larger
Sterrewacht in Leiden is drawing too much in for itself. This is certainly not
our ideal. It is our ideal to build a useful  collaboration for all concerned.
Groningen, Leiden, Dwingeloo, Hartebeestpoort and Pretoria could form a
wonderful set of institutes for research of the Milky Way system, interstellar
media and evolution.

I meanwhile agree with you that making ample funds available for travel to
America and South Africa and expanding the staff with a scientific staff member
for Groningen are the necessary conditions for healthy development. In most
cases this can at present  be done incidentally through applications to Z.W.O.
Although the acceptance of such requests will depend on the resources available
at Z.W.O. and the number of other requests, in my experience if the request is
well-founded we can almost count on it being accepted [...]

But now to the actual question. I can answer without reservation that there
will certainly be very much room for Groningen to participate in the work of
the radio observatory, at least if someone like you were to come and put it
in good shape, [...]
\endmarg

As remarked above already,
it is remarkable that both Blaauw and Oort kept referring to `van Rhijn'
without using his first name, while they used first names addressing each
other, and in their correspondence used first names when referring to e.g.
Henk van de Hulst or Pieter Oosterhoff (although in correspondence above they
refer to ‘Plaut’ without using his first name).
Both Oort and Blaauw had obtained
their PhD’s with van Rhijn as their ‘promotor’, but were senior now
themselves. Oort addressed van Rhijn as `Piet' when
writing him, and Blaauw probably did the same.

The reference to `Dwingeloo' of course is to the 25-meter Dwingeloo radio
telescope,  about to become operational, and `Hartebeespoortdam' to the Leiden
Southern Station. This latter facility resulted from the condition for
Hertzsprung to accept his appointment in 1918 that there should be a southern
telescope for his research. Willem de Sitter then had arranged access for
Leiden to the facilities of the Unie Sterrewag in Johannesburg, including the
Franklin-Adams Telescope (referred to as F.A. below). Later this was
extended with the Rockefeller Telescope,  a twin 40-cm astrograph, that Leiden
had built from funds provided by the Rockefeller Foundation. After the war
Oort had furthermore arranged funding from Leiden University for an additional
90-cm `light-collector', which would become operational in Hartebeespoortdam
near Pretoria in 1958. It was to be
outfitted with a five-channel photometer built by Th\'eodore (Fjeda) Walraven
(1916--2008). The Leiden station (and the Unie Sterrewag telescopes) had been
moved  from Johannesburg to Hartebeespoortdam in 1957.
\bigskip

Oort followed his letter up on April 16, 1956 (my translation;
`Pieter' is Pieter Oosterhoff):
\begmarg 
In this connection I would like to write to you that during the NAC meeting in
Groningen, where van Rhijn was honored, Borgman gave a lecture with
demonstration on the electronic blink-machine. I was very impressed by the
skill and complete mastery of the subject which he demonstrated. He has thought
the problems through exceptionally well and also has great skill in putting
things together so that they work effectively. He reminded me very much of
Walraven in this respect, only he is much calmer and more balanced, If you
could keep him in Groningen you would, I believe, have a worker of the very
highest quality.

Indeed I have written to the Groningen Faculty that we would also like to
foster a collaboration at our Southern Station. Pieter and I have reviewed
this after receiving your letter. We are of the opinion that there will be
room enough to also have Groningen observers there at times, when we have both
the light-collector and the Rockefeller there as well as part of the F.A.
Moreover, Walraven is going to build another telescope for continuous
observation of special variable stars [...]

Fee for the use of our instruments we would rather not want to ask. Only if
you occasionally want to take over part of our time with the Radcliffe
telescope would it be reasonable if Groningen would bear the cost.
\endmarg

However, on April 20, 1956 Adriaan Blaauw wrote (typed, in English):
\begmarg
\noindent
Dear Jan,\\
I have, after long hesitation, decided to decline the offered position at
Groningen, It was a most difficult decision to make, and several times during
the past two months I felt strongly inclined to go back to [the Netherlands].
But I was, after all, not really convinced that it would be worth discontinuing
my present position and work, and transferring my family again, especially not
because I believe there are other astronomers who may be interested in this
position and do as good a job -- and who at present work under less favorable
circumstances than I do.
\endmarg

He must have
had a particular person in mind, but did not specify who it was. Oort urged
him to reconsider, or at least postpone a final decision, until after the summer
when he might spend some time in Leiden and could visit Groningen.
In a handwritten PS to an (otherwise missing) letter to Oort on
May 29, 1956, Blaauw mentioned that van Rhijn had asked him who he had had in
mind and that he had told him it was Bruno van Albada.

Gale Bruno van Albada (1912--1972; for an obituary see \cite{JHO1973}), uncle
of at that time  student Tjeerd van Albada in Groningen, had studied in
Amsterdam and obtained a PhD under Pannekoek in 1945. The thesis was in
Dutch but concerned pioneering methods of analysis of spectral lines to
obtain chemical element
abundances that would come to fruition only after electronic computer were
used, and he worked on theories of the origin of the chemical elements. After
a short stay in the USA van Albada moved to Indonesia in 1949 to become
director of the Bosscha Sterrenwacht, which he rebuilt into a working
observatory under extremely difficult (`less favorable'  in Adriaan Blaauw's
words is a euphemism) circumstances.  

From the correspondence between Oort and  van Rhijn it can be learned that
Oort suggested all activities in Groningen related to van Rhijn's succession
should be postponed. Blaauw was to
visit Leiden in the fall and Oort seemed hopeful he could persuade Adriaan
Blaauw to visit Groningen and reconsider. It turned out, according to this exchange,
that in addition to
van Albada also Peter van de Kamp was reconsidered an option. He was a student
of van Rhijn and worked in the field he had always worked in, astrometry. Van
Rhijn asked Oort on his opinion on these two persons
without giving his own. He did apparently not
know van Albada very well. He wrote to Oort  (May 7, 1956, mt translation):
\begmarg
We have heard that he is stubborn and finds it  difficult to accept the opinions of
others. These traits would make it difficult to work with the Faculty.
\endmarg
Oort replied that he never had any problems with van Albada and if he had a
difficult character that was ‘more appearance than reality’. He did think
van de Kamp would certainly do good work, but Groningen needed someone with
the ‘enthusiasm and talent to start something new’. Of the two he
definitely preferred van Albada.

Blaauw in the mean time had also himself
suggested to postpone any decision until September
when he was to visit the Netherlands. As far as I am aware, van Albada and van
de Kamp were never approached.

Through Oort's intervention (as seems the case according to the exchamge of
letters) Groningen University postponed the decision and
decided not to approach van Albada, and await
Blaauw’s reconsideration.  Eventually, but a few years later,
Bruno van Albada felt forced to leave
Indonesia in view of the growing discontent towards Dutchmen as  former
colonizing nationals. He moved to Amsterdam in 1958, where he became director
in 1959. 

Blaauw indeed spent a few months in Leiden in the fall of 1956, eventually
October and November, on a temporary appointment as lecturer arranged by Oort.
The visit had the effect Oort had
hoped for. On December 10, 1956, Adriaan Blaauw’s wife Atie wrote to Oort that
the day after his return Adriaan had fallen ill with what had been diagnosed
in the meantime as  jaundice and was unable to write himself (my translation): 
\begmarg
\noindent
In principle A. has decided to accept the position, but he needs to write to
Groningen about further details. [...]\\
Prof. van Rhijn does not yet know all this; A. has not talked to him about this.
What do your think of this in principle?
\endmarg 

Oort responded on December 17 that ‘Mieke and I are very much delighted’.

Blaauw accepted the position in Groningen with some special agreed conditions.
These conditions were a new staff position, travel fund for
observing and provision to build an observatory with a workshop to build
instrumentation outside Groningen. Hugo van
Woerden was appointed at the new staff position to start radio astronomy as part
of the research effort. Jan Borgman and student Tjeerd van Albada did site
surveying in the area around Groningen and found the village of Roden, some
20 km to the southwest of Groningen, to be best in terms of darkness at night
and accessibility. The Kapteyn Sterrenwacht has been inaugurated there in 1965
and operated a 62-cm telescope.
\bigskip

Blaauw had on September 1, 1956, become adjunct-director of Yerkes and McDonald
Observatories, so his future in the USA seemed very bright. Yet, he decided
to accept the Groningen directorship. In 1984 Adriaan Blaauw celebrated his
70-th birthday.
At this occasion in September 1984 the Dutch astronomical community organized
a symposium to honor him. At this meeting Hugo van Woerden, who was 
the first appointment Blaauw had made
in Groningen and his first PhD student, gave an extensive
overview of his career \cite{HvW1985} and I quote from this as follows:
\begmarg 
Groningen had acquired world fame under Kapteyn in 1900-1920. But
while van Rhijn, as successor to Kapteyn (1921–1956), had continued the
latter's work with distinction [...], the Kapteyn Laboratory had remained small
and poorly equipped, and its research no longer drew much attention.

Within the Netherlands, astronomy was dominated by Leiden under Oort's
directorship. Leiden Observatory was ‘big’ and famous; it had 3 full
professors (Oort, Oosterhoff and van de Hulst), a tenured academic  staff of
about 10 plus a dozen or more graduate students; its personnel totaled about
40. Its pre-war fame due to de Sitter, Hertzsprung and Oort had been strongly
enhanced by post-war developments, notably the 21-cm line studies of rotation
and spiral structure of our Galaxy, but also research on interstellar matter,
stellar associations, comets and variable stars. Utrecht, with Minnaert as
professor, and Houtgast and de Jager as associate professors, was smaller but
growing rapidly; it was famous for its solar work and developing in stellar
astrophysics. Am\-ster\-dam, under Zanstra, was small but had many good students.
Groningen, with van  Rhijn retiring, had one staff member (Plaut), one
scientific assistant (Borgman), 3 computing assistants, one draftsman, one
junior technician, no secretary, and ... no students. Borgman's doctorate in
1956 had been  the first in 10 years.
\endmarg
That is to say, no {\it graduate} students. In 1957 Tjeerd van Albada and Harm
Habing entered their fourth and third year respectively as undergraduate
students.
\begmarg
\noindent
[...] By
early 1968, when Blaauw accepted the Scientific Directorship of ESO, the total
number of people working in the Department amounted to about 40 — a growth by
a factor of five in ten years.
\endmarg
\bigskip

Now the question is, what made Adriaan Blaauw accept the appointment at the
Kapteyn Laboratorium? And why he did he go to Yerkes in the
first place? For this it is important to find out what kind of person
he was. First note that Blaauw was a person that liked adventures,
challenges and exploring new grounds. But he was also unconventional and could
be annoyed and amused at the same time
by bureaucracy. To illustrate this consider
the following anecdote. In the central building of the
University of Groningen (the Academy Building) many rooms are decorated with
paintings of well-known (emeritus) professors, dressed in gown, jabot and
sometimes barret
(as Kapteyn in Fig.~2). Now Blaauw was not regarded an emeritus professor, as
he had {\it resigned} in 1975 and therefore never {\it retired} from a Groningen
professorship. And the administrators adored bureaucracy,
as they were cautious not to set any precedents.
After his retirement from Leiden Blaauw had been given the status of guest
researcher in Groningen, because he lived nearby, but that would not qualify
him to have his portrait added. It meant he also
did not receive announcements and invitations other emeriti did. As others I
was annoyed by this and thought it was thankless and unfair, and  I appealed a
few times to the upper officials, trying to use my weight as director of the
Kapteyn Institute or dean of the Faculty of Mathematics and Natural Science,
but to no avail. Blaauw was on the one hand a bit piqued by this bureaucratic
nitpicking, but foremost amused. We were unable to break this deadlock for a
long time. With Adriaan’s upcoming 90-th birthday in 2004 we wanted to add a
painting of him to the gallery in the Academy Building.
In the end Rector Magnificus Simon Kuipers intervened and Blaauw
was appointed for one year as honorary professor so that he would qualify for
the status of Groningen emeritus professor after that and a painting of him was added.
Typically for Adriaan Blaauw he selected without hesitation the by far most
unconventional painter of the lot he could choose from. This was Adriana Engelina
Maria (Janneke) Vieger. It was displayed on
the cover of the Institute's 2004 Annual Report \cite{AnnRep2004}, see
Fig.~\ref{fig:AB2004}.

\begin{figure}[t]
\sidecaption[t]
\includegraphics[width=0.64\textwidth]{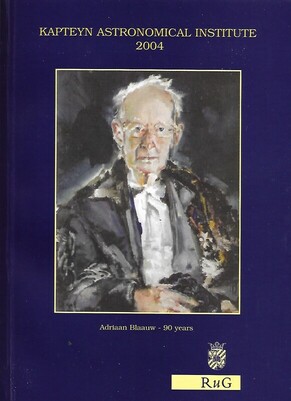}
\caption{\normalsize The cover of the 2004 Annual Report of the Kapteyn
  Astronomical Institute \cite{AnnRep2004} with Adriaan Blaauw's painting
  by Janneke Viegens 
  for the gallery in the Academy Building. University of Groningen.}
\label{fig:AB2004}
\end{figure}

That Blaauw was also adventurous can be illustrated with his travels when a
young staff member in Leiden. He has told on many occasions that he very much
enjoyed his participation of the second Kenya expedition of the Sterrewacht
Leiden. These two expeditions (in 1931-1933 and 1947-1951) aimed at measuring
absolute declinations, that is to say free from corrections for flexure in the
telescope and atmospheric refraction. The principle is that at the equator
declinations can be measured as the angle along the horizon (azimuth)
between the positions where a star did rise and set or 
between one of these and that of the north (see \cite{JKM91} or
\cite{JHObiog}). Blaauw was at the
Kenya site for five months in 1949-1950. His description of the extensive
travel involved, full of detours through Africa,
in his autobiographical contribution to the {\it Annual Review},
\cite{Blaauw2004}, illustrates very well his adventurous character. 

With this  in mind we can 
understand that Blaauw gave up a permanent
position in Leiden to accept a temporary one at Yerkes and McDonald with less
pay. Of this he said in an unpublished interview with Jet Katgert-Merkelijn
some time in the 1990s (my translation):
\begmarg
But then you see, I did decide to take that job in America. Of course, that was
also a decision of, um.... Well, on the one hand it was quite nice in Leiden.
On the other hand, I also had the feeling that others in Leiden also had,
of .... a certain stabilized situation. Oort was the boss and there was, eh
.... Oosterhoff, who directed things in that other building which we then
called the `Astro'. And then you had the feeling of `Yes, you could sit
here for another 20 or 25 years, but the world is so big and there's so much
more interesting to look at, so, eh .... let's do that, so I just resigned as
a lecturer then. And I accepted a temporary job at the University of Chicago
at the Yerkes Observatory. In retrospect, actually a curious gamble. And I
think few people at present would still do that, that you give up a
permanent job for ..... And that was also a job in America in which I was, I
think, paid less than I was in Leiden, but there was apparently so much that
feeling of either you stay here and then you're a bit walled in, although,
walled in in a nice community, or you go and have another adventure. Well,
that's what I did. Look, others have done it too of course, Gart Westerhout
with the same feeling. And there have been others like that.
\endmarg

Adriaan Blaauw saw a lack of room to develop and start new initiatives in
Leiden where Oort dominated and everything was set and determined by him. 

Yerkes Observatory of
the University of Chicago had been founded just across the state border
from Illinois in Williamsbay, Wisconsin, by George Hale in 1897, who erected
there the largest refracting telescope with an aperture lens of 40 inches.
Yerkes had become one of the leading centers in astronomy in the United States,
especially under the directorship of Otto Struve (1897–1963), descendent of a
long von Struve  line of famous Russian astronomers from St. Petersburg (Otto
had dropped the ‘von’ when he became a USA citizen). Blaauw twice spent
extensive periods at Yerkes of nine and six months (1947-1948 and 1952) --
as was
the case for Kenya without his family. Fig.~\ref{fig:Yerkes} shows him with
Jan Oort, director Struve and later director Gerard Kuiper in 1947.

\begin{figure}[t]
\sidecaption[t]
\includegraphics[width=0.64\textwidth]{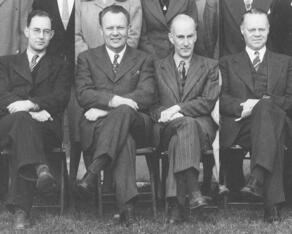}
\caption{\normalsize Adriaan Blaauw (left)
at Yerkes Observatory with from left to
right Gerard Kuiper, Jan Oort, and Otto Struve. These frames are cut
from a group photograph of staff and visitors at Yerkes Observatory in 1947.
Oort Archives.}
\label{fig:Yerkes}
\end{figure}

Yerkes employed some first rate astronomers, including in addition to Struve,
Jesse Greenstein (after his work  referred to above
on the wavelength dependence of interstellar extinction he moved to the
California Institute of Technology), Surahmanyan Chandrasekhar (1920--1995;
known e.g.
from his work on white dwarfs and degeneration pressure), William (Bill)
Morgan, William Albert Hiltner (1914--1991; another pioneer of photo-electric
photometry), Bengt Str\"omgren (wel-known for his work chemical composition of
stars and physics of the interstellar medium), and others.
Yerkes was responsible for the
operation of the 84-inch telescope on McDonald Observatory in Texas, one of
the most productive American telescopes, and Yerkes astronomers had easy access
to it. In 1947 Struve left for Berkeley and then Kuiper had been director until
Str\"omgren was appointed  in 1950; the latter stayed until 1957 when he went to
Princeton and Kuiper took over again. Blaauw had published important papers
with Struve and Morgan.
\bigskip

Coming back to the Groningen appointment the question is: why then did he
elect to leave the USA and become director of a place that was in poor
shape with little promise to flourish again? By that time he was 43 years of
age. This question was not covered in the
Katgert-Merkelijn interview (which concerned Jan Oort primarily), but it
was in the AIP interview. Here are some excerpts:
\begmarg
So I said, ‘Well, certainly I want to stay at Yerkes till somewhere in 1957.’

\noindent
[...] It was to complete things I had been working on, and also because we were
not quite sure that we wanted to go back to [the Netherlands]. We had gone to
the United States, as I said, more or less as emigrants from [the Netherlands],
and we liked it very much there. Where we lived at William's Bay was a very
pleasant place and we had very nice friends there. The educational situation
for the children was not bad.  [...]

I think we all felt that the possibilities for the future of children in [the
Netherlands] were rather limited. [...] 

On the other hand, I felt attracted to the job in [the Netherlands], and there
was a circumstance -- the general situation at Yerkes that had developed at
that time. [...]
And while people have sometimes said that you can have at an opera only one
prima donna, and a lot of good players, but you cannot have six prima donnas.
It doesn't last for long, unless you have one person who very strongly keeps
it together. And maybe one should say that there were all these prima donnas,
but not one who kept the whole thing together strongly enough, by one common
motivation, you see. I think one should look at it in that way. Maybe not the
fact that there were frictions. There are frictions almost everywhere. I don't
know of any observatory where there have not been frictions. […]

[...], the catalogue work [at the Kapteyn Laboratorium] had been practically
finished by that time. On the other hand, radio astronomy was developing in
[the Netherlands]. There was the Dwingeloo Radio Observatory that was dedicated
in 1956. I was still away then. It really opened the future for participation
in that work. So then when I went back to [the Netherlands], I said of course,
`Now, this place has to be, in a way, rebuilt, modernized, and we want to
participate in the radio work.'
\endmarg

Blaauw saw it as a challenge and an opportunity to bring the laboratory in
Groningen back to prominence, while at the same offering him the possibility
to gather his own research group around him and lead it to become a significant
factor in astronomy. The complex composition and balance between personalities
on the  Yerkes staff with many `prima donna’s' was not the environment he was
looking for. You either strive to be a prima donna too or settle with being one
of the smaller players. This was to some extent the same as working in Leiden
under the dominance of Oort had been, which made him leave Leiden.
Groningen meant working next to Oort,
not under him. And, of course a challenge and an adventure were opportunities
never to let pass.

\begin{figure}[t]
\sidecaption[t]
\includegraphics[width=0.64\textwidth]{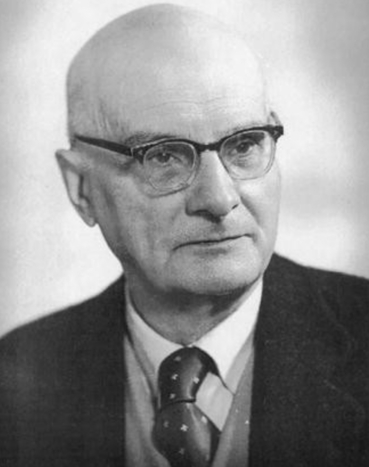}
\caption{\normalsize Pieter Johannes van Rhijn in 1957. From the collection of
Adriaan Blaauw and Kapteyn Astronomical Institute.}
\label{fig:portret}
\end{figure}

\section{The Laboratorium after van Rhijn}
At the end of the academic year 1955-56 van Rhijn had been given honorable
discharge since he had become 70 years of age earlier during that year. Like the
appointment of professors this was by Royal Decree. However, since he had no
successor he was appointed, also by Royal Decree and on the same date of July
30, on a temporary basis to be responsible for education in astronomy. He
remained listed in the {\it Yearbook} \cite{Yearbooks}  of the University of
1956 as the director of the Kapteyn Laboratorium. The next {\it Yearbook}
mentions the appointment of Blaauw, but no discharge of van Rhijn. 

When Blaauw had taken over, van Rhijn prepared the publishing of his final
research as detailed above: his last publication \cite{PJvR1960} constituted the
final volume of the {\it Groningen Publications}, number 61 in 1960. He also
died that same year, on May 9, 1960 at the age of 74. In the {\it Yearbook}
of the University \cite{Yearbooks} for 1960 he is remembered in an {\it In
Memoriam} by Adriaan Blaauw, of which I will quote a part (my translation).
\begmarg
When van Rhijn accepted his position in 1921 as Kapteyn's successor with
a lecture on `The Gravitation Problem', he consciously followed in his teacher's
footsteps. He considered it his task to carry out the grand plan designed by
Kapteyn. This plan envisaged an extensive investigation of the structure of
the Galaxy. At the time of Kapteyn's resignation, important preparatory research
had been completed; some highly significant results had already been obtained.
However, only the foundations had been laid for the broad scope actually
intended. A start had been made on international cooperation, with the aim of
systematically collecting all kinds of observation data. Throughout his career,
van Rhijn has been the driving force behind this `Plan of Selected Areas'.
With enthusiasm and unyielding perseverance he managed to obtain the
cooperation of observatories all over the world and, what was certainly not
the easiest thing, he managed to persuade almost all collaborators to complete
their often time-consuming work. The arsenal of data thus created is a lasting
monument to van Rhijn's efforts.[...]

\noindent	
Van Rhijn had a, for astronomy, large number of students. They can be found in
important posts at observatories abroad — in Australia and in the United States
-- and in the Netherlands. In all their work the imprint of their teacher can
be detected. Many of them -- both students and colleagues-- are now mourning
the passing of a friend whose exemplary rectitude they admired. Because his
thinking and his actions, both in his work and in his daily life, were
dominated by sincerity and modesty, those who came into contact with him on a
superficial level were under the impression that he barely cared about what
was going on in the world around him. But those who knew him better know that
he had a profound interest in and a sharp judgment of what was going on in the
world.

The War years were very difficult for van Rhijn: his rectorate ended
under the German occupation and then he suffered
from extended illness. With great willpower
he resigned himself to the therapy that demanded so much patience in
sanatorium and hospital. Fortunately, after the war, he was given many more
years of fruitful scientific work.

Van Rhijn was averse to insincere honors. He could speak with glowing
indignation about speeches at special occasions in which merits were
exaggerated and shortcomings concealed. He lived in the deep awareness that
there is a judgment of man and his work, higher and purer than that which can
be pronounced by fellow men.   
\endmarg

Van Rhijn left behind his twenty years younger wife after 28 years of marriage
and their two children, who by now had grown up and were in their twenties.
Regnera L.G.C. de Bie survived him by 37 years. After she died in 1997 she
was buried next to her husband in a cemetery in the north of the city of
Groningen (see Fig.~\ref{fig:graf}). 
\bigskip

\begin{figure}[t]
\sidecaption[t]
\includegraphics[width=0.64\textwidth]{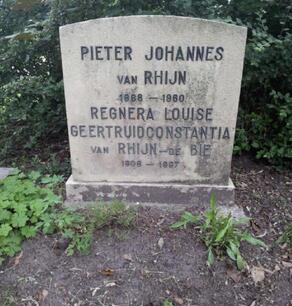}
\caption{\normalsize Tombstones on the graves of van Rhijn and his wife on the
cemetery `Selwerderhof' in the city of Groningen. From
`Online begraafplaatsen' \cite{Graven}.}
\label{fig:graf}
\end{figure}

It would carry us beyond the scope of this article to go into the development
of the Kapteyn Laboratorium under the directorship of Blaauw in the same
detail as done here for the long period of van Rhijn’s leadership. Yet in
order to contrast  how it to some extent regained its prominence, a short
summary of the first decade or so is appropriate. 

Adriaan Blaauw assumed his position
in Groningen on September 1, 1957.  He immediately
started to expand the Laboratorium by hiring a young radio-astronomer from
Leiden, Hugo van Woerden (1926--2020). Van Woerden had studied astronomy in
Leiden and had been involved in the research with the Kootwijk radiotelescope,
a refurbished German radar antenna of the Atlantikwall, a 7.5 meter
`W\"urzburg Riese'. In 1955 he had started a PhD protect that involved the
25 meter Dwingeloo radiotelescope that subsequently had become operational in
1956. This work on the structure and motion in the interstellar gas in the
Orion region was transferred to Groningen and resulted in the first PhD
thesis completed and defended under Blaauw in 1962. 

Blaauw’s next hire was Andreas Bernardus (Andr\'e) Muller (1918--2006). He also
came from Leiden after having obtained his PhD in 1953 with Oosterhoff as
supervisor on a thesis on the variable star XZ Cygni, an RR Lyrae type variable
star. He spent a few years at the Leiden Southern Station, as main observer
in long-term photometric programs and oversaw the move
of the Rockefeller Telescope from Johannesburg to Hartebeespoortdam. At the
new site Walraven was busy erecting the Light-collector. Two Leiden astronomers
in South-Africa was too much of a luxury, and Muller and Walraven did not get
along very well, so one of the two had to come back to the Netherlands. In his
interview with Jet Katgert-Merkelijn, Blaauw recalls:
\begmarg
Then there are persons like Andr\'e [Muller], who were not treated very nicely
by Jan [Oort]. Andr\'e had been in South Africa with his family. It must have
been around '57, when Andr\'e came back with his wife and he said to me that
he had been told to look for a job elsewhere. That was very painful and had
to do with Oort feeling that Walraven was the better man to be there. Andr\'e
and his wife Louise had a large family. They had 6 children, although I don't
know they had already 6 then. [...] In any case André had been in Africa for
quite some time and looked after things very well, but he was simply dismissed
and I felt that was very unjust.
\endmarg

Now, the developments in the project to found a large European Observatory in
the south, for which Oort and Baade had taken the initiative had developed to
the stage that there had been an `ESO Committee' of which by then Oort was
permanent President. With the limited funding that was provided by the
members of the informal organization, site testing campaigns had been organized.
Blaauw was very much involved in all this and wanted to hire Muller to
contribute (my translation):
\begmarg 
Well, Andr\'e then was given a job in Groningen. I had convinced people to
accept him, since the situation in Leiden was very poor. And then the
developments in the framework of ESO demanded that someone would supervise
things that developed and Andr\'e with his experience in South Africa was the
right person. And slowly he shifted from a job in Groningen to one at ESO. It
all worked out well for him, but he and his family sacrificed a lot when he
worked in Chile. That has been very difficult.’
\endmarg

The sequence of events was as follows:
In the same year 1957, Oort in this capacity as President of the ESO Committee
had approached the Ford Foundation for funding to proceed to found the
European Southern Observatory ESO. This was successful in 1959.
In that year Blaauw took up the role of Secretary of the ESO Committee and
was keen on playing a major role in the setting up of this organization.
In January 1964, four of the five countries (the Netherlands, Sweden, Germany and
France) had signed and ratified a convention, the Ford funds
became available. Belgium ratified only in 1967, a few
months after Denmark had joined as the sixth member state.
The appointment of Andr\'e Muller in 1959 ensured a major involvement of the
Kapteyn Laboratorium in ESO \cite{ABESO}.
Muller turned out instrumental for ESO; a
well-written article was published by Richard West at the time of Muller's
death, which describes his work and personality very well \cite{WestAM}.

In 1961 a young student Martien de Vries was added to the personel
of the laboratorium to work with
Muller and Blaauw on stellar photometry using the Leiden Southern Station.
Unfortunately de Vries never finished his PhD thesis and after he
participated setting up the Kapteyn Observatory in
Roden under Borgman, he went on long
term sick-leave and eventually was dismissed from the staff late in the 1970s.

In 1968 Tjeerd van Albada and Harm Habing obtained their Candidaats
degrees (Bachelor). In
respectively February 1962 and March 1963 they passed their Doctoraal exams
(Masters) and started PhD research. 

The radio group under van Woerden was extended  in 1962 with the appointment of
Swiss born Ulrich J. Schwarz (1932--2018), who joined the Kapteyn Laboratory in 1962 to become its second radio astronomer. For an {\it in memoriam} see
\cite{UJS}. He had studied physics in Bern, but did spend a year (1957--1958)
in Leiden working on radio astronomy and receivers (with American Charles
Seeger preparing for the Benelux Cross Antenna projects, see e.g.
\cite{JHObiog} or \cite{JHOEng}), where he had met Hugo van Woerden's sister
Elisabeth, whom he married in 1961. Van Woerden left after obtaining his
PhD for a two-year fellowship of the Carnegie Institution at the Mount Wilson
and Palomar Observatories and Schwarz took over the coordination of radio
astronomical work in Groningen 

And in 1960 a professorship of astrophysics had been allocated by Groningen
University, which after a long search was filled in 1963 by appointing
Stuart Robert  Pottasch (1932--2018) \cite{SRP}. He studied at  Harvard and
Colorado and had been a postgraduate fellow at Leiden and postdoctoral fellow
at Utrecht and Paris before joining the Kapteyn Laboratorium. His field was
stellar atmospheres and interstellar physics.

So within six years the Kapteyn Laboratorium had grown from three staff
members (Blaauw, Plaut and Borgman) and no students to seven staff members
(plus van Woerden, Muller, Schwarz and Pottasch) and (as it turned out)
three students! 
\bigskip

\begin{table}[t]
\begin{center}
  \caption{\normalsize
    Number of articles in the \textit{Bulletin of the Astronomical Institutes
in the Netherlands} and its \textit{Supplement Series} (BAN) and the share
of the Kapteyn Laboratorium Groningen (RUG).}
\begin{tabular}{crcccr|crccr}
\toprule
Years & BAN & RUG & \% &\ \ \ &\ \ \ & Years & BAN & RUG & \% \\
\toprule
1921-1925  &163\ \    & 7 &   4 &&& 1946-1950 &  97 &  \ \ 2 &   2 \\
1926-1930   & 215\ \  & 7   &   3 &&& 1951-1955 &  59 & \ \ 2 & 3 \\
1931-1935  &  104\ \   & 0   &  0 &&& 1956-1960  &  85 &  \ \ 5 & 6 \\
1936-1940  & 142\ \   & 2 &   1 &&& 1961-1965   &  97 & 12 & 12 \\
1941-1945 &  82\ \  & 1 &   1 &&& 1966-1969   & 99 &  21 &   21 \\
\midrule
Main 1921-1969   &1104 & 59  &5 &&& Suppl. 1966-1969 &  34\ \    & 5 &  15 \\
\bottomrule
\end{tabular}
\label{table:BAN}
\end{center}
\end{table}

The growth was also reflected in the publications output. Most, but not all
papers appeared in the {\it Bulletin of the Astronomical Institutes of the
Netherlands} and extensive data in tabular form in the  the {\it Publications of
  the Astronomical Laboratory at Groningen},
the latter having been terminated with
van Rhijn’s final publication and was replaced by the {\it Supplement Series}
of the {\it BAN}, of which the first volume started in 1966.
In 1969 both the main and supplements editions were terminated
and incorporated in {\it Astronomy \&\ Astrophysics}.
Table~\ref{table:BAN} shows the number of papers in the {\it BAN} over the
full period it appeared in five-year totals. The {\it Supplement Series} was
published only during the final interval (1966-1969)
and since of these papers an abstract
was published in the main journal and in listings counted as an article, the
numbers for the main journal have been corrected for this. The overall
Groningen share is 5.3\%, but there is a pronounced change from a few percent
before 1955 under van Rhijn
to 20\%\ towards the end. It should be noted that up to the Second
World War the numbers are significantly distorted by Ejnar Hertzsprung’s
publication record, who published large numbers of short,
(often single page,) single author
papers often on ‘elements’ (periods, amplitudes, etc.) of variable stars
and other objects. In the period 1921-1942, he published 123 papers of the
total number of 638 articles in the journal, which amounts to 19\%. 
Keeping this in mind, even then the effect of the change of the directorship
from van Rhijn to Blaauw is very large.

There was of course also the budget increase for travel.
According to van Woerden
\cite{HvW1985} it amounted to the sizable sum of 10,000 guilders per year,
which in current times would be around 35,000 \textgreek{\euro}. It was
probably used among other things to pay for part of
Plaut's long stays at Palomar Observatory
for the {\it Palomar-Groningen Variable-Star Survey} and Borgman's observing and
research trips to McDonald, Lowell and Yerkes Observatories.
It is very likely that this budget has been used also for
some of the expenses related to the sight-testing activities for ESO in South
Africa under Andr\'e Muller. 

The Groningen Laboratory did not contribute any further data sets to
the {\it Plan of Selected Areas} after van Rhijn finished his research.
With van Rhijn no longer present
as the driving force it also disappeared from the research program of
Kapteyn's laboratory, which badly needed a new focus and new inspiration.
\bigskip

It would carry too far to go into details of the new research projects that
were initiated under Blaauw. In addition to his own research, which concerned
young stars, stellar associations, interstellar medium and star formation, the
fields opened were Galactic structure using pulsating stars under Lukas
Plaut, Galactic radio astronomy with the large Dwingeloo dish under Hugo van
Woerden \cite{HvWS} and under Jan Borgman the founding of the Kapteyn
Observatory in Roden with a 62-cm telescope and a well-equipped workshop to
develop instrumentation, infrared photometry from balloons
and ultraviolet photometry in the first Astronomical Netherlands Satellite ANS.
It is fair to say that the prominent position the Kapteyn Laboratorium had
under Kapteyn, and had to a major extent lost under van Rhijn, was largely
recovered in Adriaan Blaauw’s term as director.

\section{Discussion}

I want to start this discussion with a few numbers that illustrate the enormous
amount of work that the Kapteyn Astronomical Laboratorium has devoted to
provide fundamental data. Here are the numbers of stars of the Groningen shares
for which positions and magnitudes have been derived in catalogues produced as
part of the {\it Plan of Selected Areas}:

\begin{tabular}{lr}
{\it Harvard(-Groningen) Durchmusterung}: &  231,981,\\
{\it Mt. Wilson(-Groningen) Catalogue}: & 45,448,\\
{\it Durchmusterung of the Special Plan}: & 140,082,\\
{\it Bergedorf(-Groningen-Harvard) Spektral Durchmusterung}: & 173,559.\\
\end{tabular}

These in total 591,070 position and brightness determinations were performed
roughly between 1910 and 1950, so 40 years of work
(not allowing for termination and slowing down of work during the Depression
and World War II). Now the
workweek usually was 6 days and since 1919 the workweek was by law set at
45 hours. 
Allowing for closure during public holidays, we may take a year as 51 workweeks
or 2295 working hours. With these assumptions this means that during
working hours another star with measured coordinates and apparent magnitude
was added to the data base on average every 9 to 10 minutes. And even
less if allowance is made for times of fewer operating hours.  An incredible
accomplishment! Of course, at any time a number of persons were working on
this.

Although remarkable it should be said that it is not unprecedented.
For example the initial version of the {\it Bonner
Durchmusterung} was produced by Friedrich Argelander and his assistants
between 1846 and 1863 (observations
ran from 1853 to 1859, see \cite{Charlier21}) and this resulted in
a catalogue of the positions and apparent
magnitudes of approximately 325,000 stars. And Annie J. Cannon
classified some 350,000 stellar spectra for the {\it
Henri Draper Catalogue} and extensions between about 1900 and
1940, with a reported peak frequency of about
three per minute \cite{Sheis}! 

The production of positions and magnitudes started even earlier
in Groningen with 
the {\it Cape Photographic Durchmusterung} or {\it CPD}, which also concerned
positions and magnitudes. This involved a total of 454,875 stars and was
achieved in 12 years, so repeating the calculation with these numbers yields a
rate of adding another star with measured and reduced coordinates and apparent
magnitude to the set of completed measurements on average every 4 or 5 minutes
for this period.
This is even faster, but it should be realized that the {\it CPD} plates were
measured with Kapteyn’s very clever ‘parallactic method’. In this method a
small telescope was placed at a distance from the plate exactly the same as
the focal distance of the telescope, so that in a correct orientation the two
perpendicular angular distances of a star from the plate center that were read
from dials on the axes of the small telescope were direct measurements of the
differences in right ascension and declination of the star and the plate center.

The Kapteyn Laboratorium between 1888 and 1957 (the year of van Rhijn's
retirement) measured 1.045,945 positions and magnitudes of stars in the
context of the Cape and Selected Areas Durchmusterungs. A quick scan
of the {\it Groningen Publications} shows that in addition, over this period,
there have been measurements of over 27,500 parallaxes (often of course upper
limits) and/or proper motions from plates provided by Anders Donner at
Helsingfors Observatorion, Ejnar Hertzsprung at Potsdam Observatorium, the
Cape Royal Observatory, Radcliffe Observatory, l'Observatoire Bouzareah
(Algers), and a number of other places. Of these about 7800 were obtained in
Kapteyn’s time and the Algers collaboration accounted for
the largest faction 
(41\%). So between 1888 and 1957, the Kapteyn Laboratorium produced a new
position/magnitude combination every 9 minutes during working hours and a
proper motion and/or parallax every $5\H$ hours.
After the War there have been
identifications and magnitude measurements of thousands of variable stars in
Lukas Plaut's research. This of course continued further in the latter's
variable star survey on Palomar plates up to about 1970. When Kapteyn in 1885
wrote David Gill to take on the measurements for the CPD, which he estimated
would be an effort of six or seven years, and a little later developed his
concept of an astronomical laboratory, he could not have foreseen the
enormous production this would lead to.
\bigskip

Before addressing the questions I formulated in the Introduction I quote
Adriaan Blaauw, who has expressed himself more than
others on van Rhijn's personality
and style of research. In his interview with Jet Katgert-Merkelijn
he said about van Rhijn (my translation):
\begmarg
Yes, he must have regretted seeing that the great development in the field of
Galaxy research went a little bit past him. But at the same time, if you were
to say, that was also kind of his own fault, you're right about that too. And
I think he also sensed that a bit. But in his last years he had, I think,
enough problems to deal with: his health, his family, and yes... the cuts from
the University and the staff. He also wasn't a.... He was a man who very
conscientiously carried out a program, but with little.... what can I say...?
Not very inspired, I would say.  
\endmarg

I now return to the questions posed in the Introduction:
\bigskip

\noindent
{\it Was van Rhijn the successor Kapteyn had wished for or had he preferred someone else?}

We have no written statement or account by Kapteyn, but we can say a few things 
about this. It should be seen in a
national context.  He had been offered the directorship of the Sterrewacht
Leiden at the retirement of Gerardus van de Sande Bakhuyzen in 1908, which was
untimely as at that time he started his Research Associate appointment with the
Carnegie Institution. So he had pushed for his student Willem de Sitter for
this. Against his advice 
Leiden had appointed Ernst van de
Sande Bakhuyzen as director and de Sitter in the professorship.
This first action was indeed a failure due to the
first’s conservative nature and when in 1918 he unexpectedly died before
retirement, de Sitter and Kapteyn had already considered a course of action,
which resulted in a new structure with three departments with renowned leaders
(like Hertzsprung). From this it is clear that Kapteyn must have viewed the
future of Dutch astronomy to be the much better staffed and equipped
Sterrewacht in Leiden and not his much smaller Groningen establishment. It is
also important to realize that in the proposal for his {\it Plan of Selected
Areas}, Kapteyn had presented it in addition to the science as a means to
secure the near future of the Laboratorium.

So, Kapteyn would have looked for someone that would be dedicated to executing
the Plan, rather than someone to start new initiatives before
this Plan had
been completed. Someone to follow ‘in his footsteps’, we may say. Van Rhijn
was just that and must have been seen also by Kapteyn as the right person. He
obviously expected a great future for Jan Oort, who studied with him, and
recommended him strongly to de Sitter for a future job in Leiden (along with
Jan Schilt, but Oort with preference). The conclusion would be that Kapteyn
might have been content with van Rhijn taking over his position in Groningen.
Van Rhijn had carried out, one
would think to Kapteyn’s satisfaction, the preparatory work for the star count
analysis in 1920 \cite{KvR1920a}, which would lead to the Kapteyn Universe and
was the basis for his ‘first attempt’ \cite{JCK1922}.
\bigskip

\noindent
{\it What future might Kapteyn have had in mind for his Laboratorium?}

At the opening of his Astronomical Laboratorium, still on a temporary location,
on January 16, 1896, Kapteyn said in his lecture (see \cite{JCKbiog}, p.228; my translation):
\begmarg
[...] we need to look at the work that needs to be carried out at a
photographic observatory. It consists of two parts: $1^{\rm st}$. The work
of the photographer. His part ends with the completed photographic plate.
$2^{\rm nd}$. The work of the astronomer. His part starts with the completed
photographic plate. […] 

Eliminate the photographer and all you need is a building with some six rooms,
for which no very special requirements exist. Still we cut out what seems to
us the least vital, the lack of which would present an eminent advantage to
most astronomers if they prefer research to teaching, […] No laborious efforts
on the part of the photographer, only concentration on the real astronomical
work! What pleasure, what alleviation!
\endmarg
In his presentation of the {\it Plan of Selected Areas} he wrote in the
Introduction:
\begmarg
\noindent
That such a plan was evolved at the Laboratory of Groningen is only natural.

The nature of our astronomical institution makes our work dependent on that of
other observatories. Every work of a practical nature undertaken here is of
necessity a work of cooperation with another astronomer who possesses the
means of having photographs of the sky taken. [...]

At the same time, in pursuance of the same plan, I devoted all my leisure left,
to the investigation of the structure of the stellar system by means of the
data already available.  […] These labors have now come to that point, that
they ought to enable me to make at last definite plans for the future work of
the laboratory. [...]
\endmarg

Not only was the Plan a means of securing a future for the institution,
it also illustrated that to secure the collaboration of others required a
vision in terms of a plan, the usefulness of which had to convince directors
of observatories. In the 1896 lecture above Kapteyn’s assumption was that the
observatories produced abundantly more plates than they were able to measure,
reduce and analyze. The case of Anders Donner at Helsingfors was a clear
example. Presenting a good proposal was sufficient to obtain the plates. As
long as this situation prevailed the concept of an astronomical laboratory
could survive.

However, in van Rhijn’s days these assumptions started to fail. The work at the
Mount Wilson  telescopes by astronomers like Harlow Shapley, Edwin Hubble,
Walter Baade, etc. was manifestly not intended to provide research material
for others. Even in the collaboration for the {\it Mount Wilson(-Groningen)
Catalogue}, Frederick Seares pursued his own part independently and published
the catalogue without Groningen in its name (but with Kapteyn as co-author).
In the days of Kapteyn,
observatories were often collecting data to produced catalogues for the general
benefit, but these times were the past in van Rhijn's time  and
would never return.
\bigskip

\noindent
{\it Was van Rhijn’s work really so unimaginative and how important has it
turned out to be?}

I start with the second part. This concerns his scientific research, not the
efforts for the {\it Plan of Selected Areas}. It can be maintained, I believe,
that van Rhijn’s work on the Luminosity Function, the provision of tables of
mean parallaxes of stars as a function of position, apparent magnitude, proper
motion and spectral type, etc. were no more than continuing Kapteyn’s approach.
His research work on the distribution of stars in the plane of the Galaxy after
corrections for average extinction, and that of distributions of stars at a
line of sight at higher Galactic latitude for various spectral types in the
1930s were in itself good pieces of research,
but the papers failed to put the results in a
larger context. The PhD theses produced under him were of good or high quality,
especially the one by Broer Hiemstra was ingenious and original. So my
assessment is that van Rhijn’s scientific work in general was solid, but in
general it was straightforward and not following a new,
innovative strategy.

A useful illustration would be to compare van Rhijn’s 1936 study, in which he
studied the distribution of stars in the plane of the Galaxy \cite{PJvR1936},
with Oort’s 1938 study of the distribution of stars in the Galaxy in
directions out of that plane \cite{JHO1938}. In van Rhijn’s paper average
extinctions per kiloparsec were derived and applied to Selected Areas near
latitude zero. He ignored the effects of irregularities in the dust
distribution and in the end found only a local result (of which he seemed not
really convinced) that the Sun is located in a local cluster with a radius of
order 1 kpc and a drop in density of a factor almost two lower at the borders.
Results from larger distances on
the structure of the `larger system’ are ambiguous, uncertain and inconclusive.
As usually the case with van Rhijn papers, he failed to provide a discussion of
the repercussions of his work on the larger picture of the structure of the
Galaxy.

Oort’s 1938 study \cite{JHO1938}  contrast with this in that he is interested
in the larger view on the Sidereal System and attempts to supersede Kapteyn’s
`first attempt’ with an improved version with the inclusion of what had become
known subsequently.  In particular his approach is to limit himself to high
latitudes and at the lowest latitudes only to include those for which Hubble’s
galaxy counts allow an estimate of the corrections involved. Furthermore, he
stays away from very low latitudes, because he then no longer can simplify the
work by assuming the absorption takes place in front of the bulk of the stars.
This leads to a first glimpse at how our Stellar
System compares to external spiral galaxies (from the Summary, p.233):
\begmarg
The structural features indicated are large-scale phenomena, extending on both
sides of the Galactic plane to distances of 500 or 700 parsecs from this plane.
The analogy between this structure and the spiral structure of strongly
flattened extra-galactic systems is discussed, but the data are still too
inaccurate and incomplete to permit any conclusion about the course of the
eventual `arms’, except that the Sun should be between two arms. 
\endmarg

Van Rhijn’s work is not unimaginative and certainly not unimportant. But he
does not take the wider view of how his work fits in with the emerging picture
of the Galaxy we live in. Or at least he does not address this when presenting
results of his investigations.

As it turns out, Adriaan Blaauw also took these same two papers to make a
comparison between van Rhijn and Oort in his interview with Jet
Kargert-Merkelijn (quoted earlier). He took a somewhat different approach and
stressed another fundamental difference in conducting research (my translation):
\begmarg
Look, van Rhijn's working method was, in a way, that of Kapteyn. You make a
plan, you know those and those observational data are all here, this is the
problem we want to solve and we are going to do that step and that step and
you finish that way. But of course when you do that, it happens that at a
certain moment you say: `Hey, there is  something I never thought of.  Now I
will just have to look at the matter again.' Well, that flexibility, he had
very little of. A `serendepity'-development  did not fit in van Rhijn's
schedule. [...]

Van Rhijn’s style of research can be characterized well by comparing two pieces
of research that appeared in 1936 and 1938. […] Here you can see the different
ways of approach: the somewhat rigid, systematic, programmatic one of van Rhijn
and a more flexible one of Oort. [...]

Well, there were two people, both of them had the task of `we must do Milky Way
research and take into account the interstellar matter.' They had the same data
and then you see van Rhijn working according to a certain scheme, he says in
advance, `I am going to go with those  reddening measurements, I am  going to
calculate the absorption per kiloparsec and then I will apply that [...] .'
And Oort leaves things more open in the approach he will take. And at a certain
moment he says: `Well, I will leave the low latitudes for what they are, I can
at least say something about the higher latitudes. Because I know how thick
that layer is and I know that if I look up by a certain amount, I can trust the
star counts. 
\endmarg
\bigskip

\noindent
{\it Was Kapteyn’s concept of an astronomical laboratory, an observatory
without a telescope, a viable one?}

It should be noted that this concept was not Kapteyn’s preferred option; he was
forced to accept it as the only way to get himself involved in astronomical
research at the forefront. I have described this development of  Kapteyn not
obtaining his own observatory and telescope earlier, \cite{JCKbiog} or
\cite{JCKEng}, as a blessing in disguise. Had he been running his own
observatory, he might have been prevented from conducting research, measuring
plates or interpreting the results thereof to the extent he now has been able
to. Add to this the fact that the material he did receive to measure was far
superior to what he could have obtained from Groningen.

The success formula worked for Kapteyn’s lifetime, but would it hold up in the
long run? In the first place it required a personality like him to succeed him
and also that the world would not change.
As mentioned above, the eminent astronomers drawn to
Pasadena to exploit the new Mount Wilson Observatory were not the kind that
were satisfied to provide others with unique  observational material; they
wanted to push astronomy forward themselves and take the credit for the
progress. And this was not limited to Mount Wilson, but probably part of a
much more fundamental development. Blaauw’s story of the `damned van
Rhijn program’, as Harvard astronomers described the obligatory taking of
plates only to be sent to Groningen, illustrates this well.  

Van Rhijn was aware of this and took steps to obtain his own telescope and did
in fact have a viable observing plan for it. The delay due to economic
depression and World War does not invalidate this point. But after the War the
real world moved to a situation where observing facilities of top quality (or
simply superior size) were not possible for the average university institute.
Blaauw realized this and accepted the position of successor of van Rhijn only
with guarantees for access to facilities (Dwingeloo, Hartebeespoortdam, foreign
observatories) and
funds to actually use these. 

So, yes the `observatory without a telescope’ concept worked for the person
Kapteyn in his times. The commitment to the {\it Plan of Selected Areas} worked
for van Rhijn profitability as a work plan to pursue, but without a road-map
how to exploit the  data and synthesize a model for the structure of the
Galaxy it produced little more than catalogues and data bases. Van Rhijn seemed
to have been satisfied with that, but the future of the astronomical laboratory
needed a change in tactics. Having funds to go observe elsewhere solved at least
part of the issue under Blaauw. What remained was the problem of how to be
awarded the necessary observing time, but collaborations helped to solve that
(Plaut at Palomar, Borgman at Yerkes, McDonald, etc.) for the moment. The
existence of a nationally funded foundation to run the Dwingeloo dish solved
the radio astronomical needs.  The initiative to build ESO would do the same in
the optical.

So, is the concept of an astronomical laboratory still a viable one?
Maybe paradoxically I would say: Yes, I would
think so.  After all, in a sense the current Kapteyn Astronomical Institute
is not fundamentally different from the laboratory of Kapteyn’s days.
Astronomers obtain their data from elsewhere, often without the need to travel
and observe personally. We sometimes indeed do
travel to observe at observatories the Netherlands
participates in such as on La Palma, La Silla or at Paranal,
but in many cases we can use service observing and this
is increasingly an option. We use other facilities through collaborations
with astronomers that have access to facilities, in which we are
not a stakeholder, such as in my case Palomar or Siding Spring, sometimes
traveling there, sometimes not.
In all cases material comes to Groningen for analysis
much like the plates that reached Kapteyn and van
Rhijn. Or we use space facilities, where data arrive at
your doorstep. As a PhD student in Leiden in
the 1960s I did not have to go to the
Dwingeloo for radio observations (although I did for the larger part of my
observations) and only had to fill in a form to ‘order’ observations.
The same has
been the normal modus operandi  in Westerbork from the start in 1970.
Obviously, satellites and observatories in space by definition are operated
this way. Dedicated staff ensures efficient use of the telescope or satellite.

The problem van Rhijn faced was  the reliance on other observatories to
provide date without having the authority Kapteyn had. The concept failed if
it {\it relied entirely} on the generosity of directors of observatories with
telescopes, but  with funding to arrange access to facilities it is a model
working well in many places. In fact, going to the telescope oneself is seen
more and more as a poor and inefficient use of manpower and budgets. Kapteyn
could concentrate on research without having the burden of running an
observatory. We still conduct our research that way.
\bigskip

\noindent
{\it What precisely caused the Kapteyn Laboratorium to loose its prominent
      place in the international context?}

There is no single cause. One simple answer is that van Rhijn was not in the
class of Kapteyn. Indeed, Kapteyn really was a though act to follow.
It would be difficult to rival the discovery of the Star
Streams, the vision of the {\it Plan of Selected Areas}, the two concepts of
statistical astronomy and galactic dynamics that Kapteyn had developed. But
there must have been more to it. Part of that was van Rhijn’s attitude to give
data collection in the context of the {\it Plan of Selected Areas} priority
over analysis to develop an overall picture of Galactic structure. And there
of course was Leiden, which with no less than three world class leaders in
Willem de Sitter, Ejnar Hertzsprung and Jan Oort was  impossible to better by
any single person. And then there is the lack of funding, resulting among
others from the remoteness to the seat of
government. This affected all of the University of Groningen but more so
astronomy, where in addition
funding the Sterrewacht in Leiden was more opportune
than the small scale Laboratorium in distant Groningen.

On the other hand, one should not forget that the Kapteyn Laboratorium and
Pieter van Rhijn did not become insignificant in the global ranking of
astronomy. 
\bigskip

\noindent
{\it How important was the Plan of Selected Areas for the progress of astronomy
and did it ever reach the goals Kapteyn had in mind for it?}

Astronomy profited from the coordinated approach in various ways. First of
course the availability of a database in general. For photometry the existence
of photometric sequences in the Selected Areas, sometimes to very
faint magnitudes has been important, as discussed by Kinman in the Legacy
Symposium \cite{TK2000}. The existence of reasonably consistent and uniform
catalogues of spectral types, color indices, proper motions and radial
velocities must have been of value for many ongoing investigations.

Was this what Kapteyn had in mind? Not in the first place. His focus was an
analysis to determine what we now call the structure and dynamics of the
Galaxy, which he pioneered in his seminal `first attempt’ paper of 1922
\cite{JCK1922}, in which he was the first to apply dynamics to observational
information. The dynamics was studied extensively by,
in particular, James Jeans,
Arthur Eddington, Bertil Lindblad and Jan Oort, so not in van Rhijn's
research establishment. But it seems faur to say that the ultimate goal of
understanding Galactic structure was brought within reach by the
data collecting of the {\it Plan of Selected Areas}. On the other hand, the
Plan did not fundamentally contributed to the discovery
of extinction.

Kapteyn’s original goal had been to use the data collected in the Plan
to determine the distribution of the stars in space as a road-map for the
future of his Laboratorium.
In his presentation of the Plan \cite{SA} he wrote in the Preface:
\begmarg
As will be explained more fully in the introduction, the present plan was
originally meant only as a working plan for the astronomical laboratory of
Groningen. Having first determined on its main lines it was my task to try to
convince astronomers, who were in a situation of executing the necessary
plates and observations, of its urgency and thus win their cooperation.
\endmarg
\noindent
And in the Introduction:
\begmarg
In the following pages a plan is outlined, wich may be realised by the
cooperation of few astronomers in a, relatively speaking, moderate time.

The aim of it is to bring together, as far as is possible with such an effort,
all the elements which at the present time must seem most necessary for a
successful attack on the sidereal problem, that is: the problem of the
structure of the sidereal world.
\endmarg
What he had in mind was a final attack on this problem when all the data would
be in along the lines of his ‘first attempt’. I have argued that Oort’s paper
in 1938 \cite{JHO1938} was just that, except that much of the data used where
not yet in the final form. 

This concerned the {\it Systematical Plan}. The {\it Special Plan}, I stress,
adopted under pressure from Pickering, had a few goals which can be appreciated
from his justification of the choice off the {\it Special Areas} in his
presentation of the Plan. As far as I am aware only the study of those few
Areas that were targeted to dark clouds was actually realized in the
thesis by Broer Hiemstra.
\bigskip

\noindent
{\it Did van Rhijn get the support he deserved from his university and the
government?}

The correspondence between van Rhijn and the Curators of Groningen University
show a healthy level of support for the Kapteyn Laboratorium. The problem is
the bias, at least in the 1930s,
of the Ministry for observational facilities in favor of Leiden and
Utrecht and against Groningen. The statement that computers in Groningen
should not expect to be paid at the same level as in Leiden is curious, to say
the least.  After all, the work is similar and the salary should not depend on
the success or prominence of an institution on an international level
or the fame of the
professor in charge.  It is much to van Rhijn’s credit that he persevered in
his attempts to obtain his own telescope. 

A comment that should be added here is that van Rhijn's
idea in the 1930s to address
the important and urgent issue of the wavelength dependence of interstellar
extinction by comparing similar stars with strong effects of reddening and
insignificant amounts, was timely. The difficulties to obtain the funding and
the crippling effects of the economic depression and the War delayed the
project by about two decades, so that by that time photoelectric techniques
had made photographic approaches obsolete, while the results obtained by van
Rhijn and Borgman have shown the program to be viable and capable of providing
definitive answers. The lack of support from the Ministry in funding the
telescope and refusal to
fund at least the dome, has been a major factor in the delay and has therefore
been a major cause of van Rhijn missing this important opportunity.
\bigskip 

\noindent
{\it How influential was van Rhijn and what was his role in the prominence of
Dutch astronomy that followed during the twentieth century?}

Obviously, in the field of stellar and Galactic
astronomy he was overshadowed by de Sitter,
Hertzsprung and Oort in Leiden, but probably not by Pannekoek in
Amsterdam. The latter’s legacy after all is in the area of astrophysics and
stellar atmospheres. Minnaert in Utrecht was prominent in solar research.
Van Rhijn kept Kapteyn’s legacy alive, but the dominance
of Dutch galactic and stellar astronomy in the twentieth century resulted 
from the work of the trio from Leiden and constitutes a major part of
Kapteyn's legacy.
Van Rhijn’s influence on the
development of Dutch astronomy was limited, but his coordination of the work
on the Selected Areas remained important for astronomy as a whole.

That is not to say, van Rhijn did not contribute.
He did produce — or at least
supervised in the final stages — a number of excellent PhD theses and
students, that later
would occupy very prominent positions in the Netherlands and abroad,
such as Jan Schilt, Jan Oort, Peter van de Kamp, Bart Bok and Adriaan Blaauw.
In the same period there had been in Leiden
Dirk Brouwer (with de Sitter), and Pieter Oosterhoff,
Gerrit (Gerard) Kuiper and Adriaan Wesselink (with Hertzsprung). 
In this respect van Rhijn did extremely well, also
compared to Leiden.
Of course there have been more outstanding astronomers in the Dutch school,
but these had their PhDs only after
van Rhijn’s active period: Oort’s most
prominent students Maarten Schmidt and Lodewijk Woltjer and van de Hulst’s
Gart Westerhout followed in 1956 and 1958. Utrecht with Marcel Minnaert
contributed with Henk van de Hulst and Cees de Jager, and Pannekoek in
Amsterdam with Bruno van Albada, but only after WWII.  This selection is of
course a subjective inventory
and somewhat arbitrary, but does show van Rhijn did
contribute very significantly to the reputation of Dutch astronomy in the first
half and a bit of the twentieth century.
\bigskip

\noindent
{\it What made Adriaan Blaauw accept his appointment as van Rhijn’s successor?}

I have argued that a significant factor was Blaauw’s appetite for adventure and
challenges. This made him accept Oort’s request to join the Kenya expedition
and give up leave his secure, tenured position in Leiden for an
uncertain, temporary one at Yerkes. But adventure or challenge is not enough.
The advent of radio astronomy and the allocation upon his request for a new
staff position and of funds for travel abroad to collect
new observational
material together made an effort  to effectuate a revival of Groningen
astronomy one with a reasonable chance of success. I suppose that had it
failed, Blaauw would easily have found
some position elsewhere. What helped him to pursue the
kind of career he desired,  was the attitude of his wife to agree to
let him go away from his family for often long periods and her consent to move
the family across the Atlantic twice in only a few years. Adriaan Blaauw was
exactly what Groningen needed after van Rhijn.

Of course, the University of Groningen had been keen as well on attracting
Blaauw and even provide opportunities for growth of the astronomy
laboratory. I have not been able to find any written statement on this in the
Archives, but to back this up
I would think the university leaders and administrators were
still sufficiently aware of Kapteyn’s significance and his stature as one of
the greatest scientists in its history to keep his legacy intact, even though
the focus of Dutch astronomy had moved with Kapteyn’s prot\'eg\'es de Sitter,
Hertzsprung and Oort to Leiden. 
\bigskip

In conclusion,  Pieter van Rhijn may not have been
of the same caliber as Kapteyn (who could possibly blame him for
that?), and his work and leadership may have been not highly imaginative,
but still an enormous amount of
important and useful work has been done at the
Laboratorium under his directorship to the great benefit of astronomy.
And an admirable number of extraordinary students wrote their theses under his
supervision. He deserves much credit for keeping the
{\it Plan of Selected Areas} alive and
coordinating the enormous effort that was not likely to lead to much fame
or praise for himself. Van
Rhijn disliked travel and attending large meetings, and it was his misfortune
to have much of his directorship coincide with the Great
Depression, World War II
and his own protracted suffering of tuberculosis, while
the Minister denied him funding at the level he was
prepared to supply Leiden, yet under these unfavorable circumstances,
in the end in his modest and unassuming manner
his efforts left the spirit and heritage of his inspiring predecessor  intact
until new leadership took over. 
\bigskip

\noindent
{\bf Acknowledgments} 
I thank my colleague astronomer Jan Willem Pel and historian Klaas van Berkel,
for a critical reading of a draft of this paper and making many detailed and
very useful comments and suggestions. David Baneke and  Ton Schoot Uiterkamp
also read the manuscript and made a number of important remarks.
I am grateful to Kalevi Mattila of Helsinki Observatory for providing the
information on the correspondence of Kapteyn with Donner on the issue of 
Kapteyn's succession.
I am especially grateful to Gert Jan van Rhijn,
grandson of van Rhijn’s brother Maarten,
who maintains the van Rhijn Website \cite{PJgenea}, for good quality
photographs and much extensive biographical information on
Pieter van Rhijn. 
I thank the staff of the Kapteyn Astronomical Institute, in particular the
secretariat and computer group,  for support and help
and the director, Prof. L\'eon Koopmans, for hospitality extended to an
emeritus professor as guest scientist. 
\bigskip
}

\newpage

\end{document}